\documentclass[10pt]{revtex4}
\usepackage{amssymb,amsmath,amscd}
\usepackage{graphicx,calc,epsfig,pstricks,bbm}
\usepackage{tikz}

\newcommand {\be}{\begin{equation}}
\newcommand {\ee}{\end{equation}}
\newcommand{\ba}{\begin{array}{c}}
\newcommand{\ea}{\end{array}}

\renewcommand{\exp}[1]{\operatorname{exp}\left(#1\right)}

\newcommand{\sixj}[6]{\begin{Bmatrix}  #1&#2&#3\\#4&#5&#6\end{Bmatrix}}
\newcommand{\tinysixj}[6]{\{\begin{smallmatrix}  #1&#2&#3\\#4&#5&#6\end{smallmatrix}\}}

\newcommand{\tinyninej}[9]{\left\{\begin{smallmatrix}  #1&#2&#3\\#4&#5&#6\\#7&#8&#9\end{smallmatrix}\right\}}

\newcommand{\scr}{\scriptscriptstyle}
\newcommand{\vgraph}{\mathfrak{n}}
\newcommand{\cube}{\ba
 \begin{tikzpicture}
\pgfmathsetmacro{\cubex}{0.15}
\pgfmathsetmacro{\cubey}{0.15}
\pgfmathsetmacro{\cubez}{0.15}
\draw (0,0,0) -- ++(-\cubex,0,0) -- ++(0,-\cubey,0) -- ++(\cubex,0,0) -- cycle;
\draw (0,0,0) -- ++(0,0,-\cubez) -- ++(0,-\cubey,0) -- ++(0,0,\cubez) -- cycle;
\draw (0,0,0) -- ++(-\cubex,0,0) -- ++(0,0,-\cubez) -- ++(\cubex,0,0) -- cycle;
\end{tikzpicture}
\ea}

\begin{document}

\title{Quantum Reduced Loop Gravity: Semiclassical limit}
\author{Emanuele Alesci}
\email{emanuele.alesci@fuw.edu.pl}
\affiliation{Instytut Fizyki Teoretycznej, Uniwersytet Warszawski, ul. Ho{\.z}a 69, 00-681 Warszawa, Poland, EU}
\author{Francesco Cianfrani}
\email{francesco.cianfrani@ift.uni.wroc.pl}
\affiliation{Institute
for Theoretical Physics, University of Wroc\l{}aw, Pl.\ Maksa Borna
9, Pl--50-204 Wroc\l{}aw, Poland.}

\begin{abstract}
We discuss the semiclassical limit of Quantum Reduced Loop Gravity, a recently proposed model to address the quantum dynamics of the early Universe. 
We apply the techniques developed in full Loop Quantum Gravity to define the semiclassical states in the kinematical Hilbert space and evaluating the expectation value of the 
euclidean scalar constraint we demonstrate that it coincides with the classical expression, {\it i.e.} the one of a local Bianchi I dynamics. The result holds as a leading order expansion in the scale factors of the Universe and opens the way to study the subleading corrections to the semiclassical dynamics. We outline how by retaining a suitable finite coordinate length for holonomies our effective Hamiltonian at the leading order coincides with the one expected from LQC. This result is an important step in fixing the correspondence between LQG and LQC.
\end{abstract}

\maketitle

\section{Introduction}

A viable Quantum Gravity model must reduce to General Relativity (GR) in the proper semiclassical limit. Although this is a quite natural requirement, nevertheless it can be a far-from-trivial issue. This is the case also for the approaches as Loop Quantum Gravity (LQG) in its canonical \cite{Rovelli:2004tv,Thiemann:2007zz} or covariant formulation (Spinfoam Models) \cite{Perez:2003vx,Rovelli:2011eq}. In fact, while going from the classical to the quantum realm is a well-settled procedure, going back is much more complicated since it involves the construction of a proper semiclassical limit.  
The definition of semiclassical states for a quantum theory of the geometry has been given in \cite{Thiemann:2000bw, Thiemann:2002vj} in the kinematical Hilbert space via the application of the complexifier technique.
At the end one can define states peaked around a given set of classical holonomies and fluxes, but these have to be tested against the dynamics. This can be done looking at the graviton propagator in the Spinfoam setting \cite{propagatore1, propagatore2, propagatore3, propagatore4} or looking at the expectation value of the Hamiltonian \cite{Thiemann:1996aw} or the Master Constraint \cite{master1,master2} in canonical LQG. The difficulties with finding an analytic expression for the scalar constraint matrix elements \cite{Gaul:2000ba, io e antonia, Alesci:2013kpa} in the spin network basis forbids a direct computation of the dynamic behavior of semiclassical states. Only the Master Constraint operator in the context of Algebraic Quantum Gravity \cite{Giesel:2006uj} has been shown to convergence to the right classical expression in the semiclassical limit \cite{Giesel:2006uk,Giesel:2006um} under the simplifying replacing of the gauge group $SU(2)$ with $U(1)^3$.

The situation is quite different in Loop Quantum Cosmology (LQC) \cite{Bojowald:2011zzb,Ashtekar:2011ni}, the standard cosmological implementation of LQG (other cosmological models related with LQG are given in \cite{Rovelli:2008aa,Bianchi:2010zs}, \cite{Borja:2010gn} and \cite{Gielen:2013kla}). In LQC, the quantization is performed in minisuperspace, {\it i.e.} after reducing the phase space according with the homogeneity requirement for Bianchi models. All the kinematical symmetries ($SU(2)$ gauge symmetry and background independence) are fixed on a classical level, such that quantum states are described by quasi periodic functions of the three independent connection components $c_a$. The semiclassical states are naturally defined by peaking around classical trajectories. The dynamic issue is greatly simplified and an analytic expression for the scalar constraint is obtained. A crucial point is the regularization, which is realized by fixing non vanishing polymeric parameters $\bar\mu_a$, such that the momenta operators have a discrete spectrum, whose eigenvalues $\propto\mu_a$. The expectation value of the scalar constraint in the presence of a clock like scalar field reproduces the classical expression as soon as the energy density of the field $\rho>>\rho_{cr}$, $\rho_{cr}$ being a critical energy density related with $\bar\mu_a$. For $\rho\sim\rho_{cr}$ quantum effects are not negligible and they induce a bouncing scenario replacing the inital singularity \cite{Ashtekar:2006wn}.    

In \cite{Alesci:2013xya} we proposed a new loop quantum model, namely Quantum Reduced Loop Gravity (QRLG), in which the dynamic issue is simplified with respect to the full theory thanks to the restriction to a diagonal metric tensor (see also \cite{Alesci:2012md,Alesci:2013xd} for a local Bianchi I space). The idea of QRLG is to implement such a restriction directly in the kinematical Hilbert space of LQG. This allows to retain the basic structure of the full theory, such as graphs and intertwiner structures, but in a simplified framework. As a consequence, the (reduced) graphs have only a cuboidal structure, while intertwiners are only complex numbers. In the limit of the Belinski Lipschitz Kalatnikov (BKL) conjecture \cite{bkl1,Belinsky:1982pk}, the dynamics preserves the metric to be diagonal and it locally coincides with those of the Bianchi I model. The associated scalar constraint can be defined along the lines developed in the full theory. The volume operator turns out to be diagonal in the (reduced) spin network basis and the matrix elements of the scalar constraint can be analytically evaluated. The possibility to apply LQG techniques in a computable model makes QRLG a tantalizing subject of investigation.  

In this work we investigate the semiclassical limit of QRLG. We will outline how the construction of semiclassical states can be done as in \cite{Thiemann:2000bw, Thiemann:2002vj, Thiemann:2000ca, Bahr:2007xn, Bianchi:2009ky, Magliaro:2010qz}, the only difference being that the sum over spin quantum numbers is replaced by the one over the maximun/minimum magnetic indexes. Then, we will evaluate explicitly the expectation value of the non-graph changing  euclidean scalar constraint on such states. We will carry on in details all the calculations. By using the asymptotic expansion of the Clebsch-Gordan coefficients entering the final expression we will compute the leading order contribution in the limit of high spin quantum numbers. This way, we demonstrate how at each node the expectation of the euclidean scalar constraint reproduces  the analogous expression for the Bianchi I model and, in the continuum limit, the corresponding classical expression, {\it i.e.} a local Bianchi I dynamics.  

Hence, \emph{the semiclassical dynamics in QRLG coincides with the one a local Bianchi I model}. This means that the quantum restriction we performed to simplify the dynamic problem is well-grounded, since the resulting quantum system approaches in the classical limit GR within the proper approximation scheme.
   
QRLG can thus be used to realize a viable quantum description for the Universe, in which all the prediction of the Standard Cosmological Model are safe, while we can get some hints on the fate of the initial singularity in LQG. 

At the same time, even if in this work we investigate only the leading order term in the semiclassical expansion, we will set up all the techniques to evaluate the corrections, which can provide non trivial modification with respect to the classical behavior. We want to stress how among such corrections there are the ones related with the fundamental SU(2) structure, which have no counterpart in LQC and come from the next-to-leading order expansion of the 3j, 6j and 9j symbols entering the euclidean scalar constraints. 

The expectation value of the Hamiltonian equals the anologous expression for the quantum Hamiltonian used in LQC \cite{MartinBenito:2008wx,Ashtekar:2009vc}, the relevant difference being due to subleading corrections (to be computed in upcoming works) and the explicit presence of the coordinate length in the semiclassical expression, which plays the role of $\bar{\mu}_a$. In our analysis this parameter is not entering at all in the quantum theory, it's only an artifact of the semiclassical construction the requires the use of kinematical states; it can be removed with the same considerations that allow to remove it in the full theory. The final expression we find open the way to properly relate LQC and LQG (see also \cite{Brunnemann:2010qk,Engle:2013qq}) and eventually address the role of the holonomic and triad corrections to LQC \cite{Bojowald:2011aa,Bojowald:2012qq}.

The article is organized as follows: in section \ref{LQG}, LQG is reviewed, focusing our attention on the construction of semiclassical states. In section \ref{QRLG}, the framework of QRLG is introduced, while semiclassical states are defined in section \ref{semiQRLG} along the lines of the full theory. The action of the Euclidean scalar constraint is evaluated on basis states based at dressed nodes in section \ref{Eucl}, such that we can compute the expectation value of the scalar constraint on semiclassical states in section \ref{Scal}. Concluding remarks follow in section \ref{concl}.

\section{Loop Quantum Gravity}\label{LQG}

Gravity phase space in LQG is described by the holonomies of Ashtekar-Barbero connections $A^i_a$ along curves and the fluxes of inverse densitized triads $E^i_a$ across surfaces. The corresponding kinematical Hilbert space $\mathcal{H}$ is the direct sum over all graph $\Gamma$ of the single Hilbert spaces $\mathcal{H}_\Gamma$ associated with each graph. The elements of $^G\mathcal{H}_\Gamma$ are gauge-invariant functions of $L$ copies of the $SU(2)$ group, $L$ being the total number of links in $\Gamma$. Basis vectors are given by invariant spin networks
 \be
<h |\Gamma,\{j_l\},\{x_n\}>=\prod_{n\in\Gamma} {x_{n}}\cdot   \prod_{l}D^{j_{l}}(h_{l}),
\label{spinnet solite} 
\ee
$D^{j_{l}}(h_{l})$ and ${x_{n}}$ being Wigner matrices in the representation $j_{l}$ and invariant intertwiners, respectively, while the products extend over all the nodes $n$ in $\Gamma$ and all the links $l$ emanating from $n$. The symbol $\cdot$ means the contraction between the indexes of intertwiners and Wigner matrices.

Fluxes $E_i(S)$ across a surface $S$ realize a faithful representation of the holonomy-flux algebra and they act as left (right)-invariant vector fields of the SU(2) group. In particular, given a surface $S$ having a single intersection with $\Gamma$ in a point $P\in l$, such that $l=l_1\bigcup l_2$ and $l_1\cap l_2=P$, the operator $\hat{E}_i(S)$ is given by 
\be
\hat{E}_i(S)D^{(j_l)}(h_l)
=8\pi\gamma l_P^2 \; o(l,S) \; D^{j_l}(h_{l_1})\,\tau_{i}\,D^{j_l}(h_{l_2}),\label{Eop}
\ee  
$\gamma$ and $l_P$ being the Immirzi parameter and the Planck length, respectively, while $o(l,S)$ is equal to $0,1,-1$ according with the relative sign of $l$ and the normal to $S$, and $\tau_i$ denotes the SU(2) generator in $j_l$-dimensional representation.  

Indeed, one still has to impose background independence and this can be done in the dual space $\mathcal{H}^*$ via s-knots, which are equivalence class of spin networks under diffeomorphisms. 

Finally, the last constraint to implement is the scalar one $\hat{H}$, for which a regularized expression can be given in \cite{Thiemann:1996aw} by a graph-dependent triangulation of the spatial manifold. This triangulation $T$ contains the tetrahedra $\Delta$ obtained by considering all the incident links at a given node and all the possible nodes of the graph $\Gamma$ on which the operator acts. For each pair of links $l_i$ and $l_j$ incident at a node $n$ of  $\Gamma$ we choose a semi-analytic arcs $a_{ij}$  whose end points $s_{l_i},s_{l_j}$ are interior points of $l_i$ and $l_j$, respectively, and  $a_{ij}\cap\Gamma=\{s_{l_i},s_{l_j}\}$. The arc $s_i$ ($s_j$) is the segment of $l_i$ ($l_j$) from $n$ to $s_{l_i}$ ($s_{l_j}$), while $s_{i}$, $s_{j}$ and $a_{ij}$ generate a triangle $\alpha_{ij} := s_{i} \circ a_{ij} \circ s_j^{-1}$. 
The Euclidean  and Lorentian parts of the scalar constraint can then be promoted to operators replacing the classical holonomies and fluxes entering the regularized expression with their quantum expression. In this process one can fix an arbitrary representation $(m)$ for the holonomies contained in the regularized constraint. The final expression for the Euclidean part is then
 \be
   \label{Hm_delta:quantum}
 \hat{H}_E=\sum_{\Delta\in T} \hat{H}^m_{\Delta}[N]:= \sum_{\Delta\in T}\, N(n)C(m) \,  \, \epsilon^{ijk} \,
   \mathrm{Tr}\Big[\hat{h}^{(m)}_{\alpha_{ij}} \hat{h}^{(m)-1}_{s_{k}} \big[\hat{h}^{(m)}_{s_{k}},\hat{V}\big]\Big] .
 \ee
$V$ being the volume operator and $C(m)=\frac{-i}{8\pi \gamma l^2_P N^2_m}$ denotes a normalization constant depending on the representation $(m)$ chosen for the holonomy operators where $N^2_m=- d_m\, m (m+1)$.

The lattice spacing $\epsilon$ of the triangulation $T$ (which here acts as a regulator) can be removed in a suitable operator topology in the space of s-knots. 

Even though one can write formal solutions to the constraint in terms of graphs based at ``dressed'' nodes \cite{Thiemann:2007zz,Thiemann:1996aw}, these solutions are only formal since an analytical expression for the matrix elements of the volume $V$, thus of the whole scalar constraint, is missing. 

The same strategy can be adopted to build regularizations using different decompositions, for example in terms of cubulations and can be extended to be graph changing or not using loops to regularize the curvature that belong or not to the underlying spin networks \cite{Thiemann:2007zz,Giesel:2006uk}.

\subsection{Semiclassical limit of LQG}
The development of semiclassical states in LQG is based on the application of the complexifier technique to a Hilbert space made of functions of copies of the SU(2) group. Let us consider a single link $l$ and a dual surface $S$ and let us suppose that we want to peak around the classical configuration, {\it i.e.} an holonomy $h'$ along $l$ and a flux $E'_i$ across $S$. These two quantities can be combined to form the complexifier $H'$ 
\be
H'=h'\exp{\frac{\alpha}{8\pi\gamma l_P^2}E'_i\tau_i},\label{Hlqg}
\ee
$\alpha$ being a parameter, which is an element of $SL(2,C)$, the complexification of the original $SU(2)$ group. Following \cite{Thiemann:2000ca}, a state $\psi^\alpha_{H'}(h_l)$ peaked around such a classical configuration can be constructed from the heat-kernel of the Laplace-Beltrami operator $\Delta_{h_l}$ applied to the $\delta$-function over the group elements $h_l$,  {\it i.e.}
\be
K_\alpha(h_l,h')=e^{-\frac{\alpha}{2}\Delta_{h_l}}\delta(h_l,h'),
\ee
and the explicit form reads
\be
K_\alpha(h_l,h')=\sum_{j_l}(2j_l+1)e^{-j_l(j_l+1)\frac{\alpha}{2}}Tr(D^{j_l}(h^{-1}_lh')).\label{hklqg}
\ee
The semiclassical state is obtained via analytic continuation from $h'\in SU(2)$ to $H'\in SL(2,C)$ as follows
\be
\psi^\alpha_{H'}(h_l)=K_\alpha(h_l,H').\label{sclqg}
\ee
These states are eigenfunctions of the operator $\hat{H}_l=h_l\exp{\frac{\alpha}{8\pi\gamma l_P^2}E_i(S)\tau_i}$ with eigenvalue $H'$ and, just like the usual coherent states in quantum mechanics, they are peaked around $h_l=h'$ and $E_i(S)=E'_i$, while fluctuations are controlled by the parameter $\alpha$. 

By repeating this construction for several copies of the $SU(2)$ group one can define coherent states for a generic graph, so finding 
\begin{equation}
\Psi_{H',\Gamma}(\{h_l\})=\prod_{l\in\Gamma}\psi^\alpha_{H'}(h_l).
\end{equation}
The main difficulty is to reconcile such a construction with $SU(2)$ gauge-invariance. In fact the expression above behavess as follows under a SU(2) transformation
\begin{equation}
\Psi'_{H',\Gamma}(\{h_l\})=\prod_{l\in\Gamma}\psi^\alpha_{H'}(g_{t_l}\,h_l\,g^{-1}_{s_l}),
\end{equation}
$s_l$ and $t_l$ being the source and target points of $l$. In order to define gauge-invariant coherent states one must average $\Psi'_{H',\Gamma}(\{h_l\})$ over $g_{s_l}$ and $g_{t_l}$ for the links of the graph. The resulting expression can be expanded in terms of invariant spin networks. However, while by construction the gauge variant coherent state exibiths the right peakness properties, this is not necessary the case for gauge-invariant ones. In fact, it has been verified only by explicit calculation that gauge invariant coherent states are proper semiclassical states in $^G\mathcal{H}$  \cite{Thiemann:2000ca}.

\section{Quantum reduced Loop Gravity}\label{QRLG}
QRLG realizes the quantum reduction of the full LQG kinematical Hilbert space down to a proper reduced space $\mathcal{H}^R$ capturing the relevant degrees of freedom of a system with a diagonal metric tensor \cite{Alesci:2013xya} (see also \cite{Alesci:2012md,Alesci:2013xd} for early attempts restricted to the Bianchi I model). Such a projection has been perfomed by 
\begin{enumerate}
{\item the implementation of the partial gauge fixing condition of diffeomorphisms invariance restricting to a diagonal metric tensor: this implies a truncation of the admissible graphs to reduced graphs, which are the union of some links which are parallel to one of the three fiducial vectors $\omega_i$ (we denote these links as being of the kind $l_i$ for some $i$),}  
{\item the implementation of a $SU(2)$ gauge-fixing condition: this is realized via the restriction to those functions of $SU(2)$ group elements based at $l_i$ which are entirely determined by their restriction to some functions of $U(1)_i$ group elements. Such $U(1)_i$ are the $U(1)$ subgroup obtained by stabilizing the $SU(2)$ group around the internal direction $\vec{u}_i$ 
\be
\vec{u}_1=(1,0,0),\qquad \vec{u}_2=(0,1,0),\qquad \vec{u}_3=(0,0,1).
\ee
.}
\end{enumerate}   

These two steps affect the kind of symmetries we have on a kinematical level. The former implies that not full background independence is realized. In fact, the only kind of diffeomorphisms which survives after the truncation to reduced graphs are those mapping reduced graphs among themeselves (and on a classical level preserving the diagonal form of the metric). We call these transformations reduced diffeomorphisms. 
As for the $SU(2)$ gauge-fixing, it makes $SU(2)$ gauge invariance not manifest anymore. Nevertheless, since the $U(1)_i$ groups are not independent, some reduced intertwiners arise as a relic of the original $SU(2)$ gauge invariance.

Finally, the kinematical Hilbert space $\mathcal{H}^R$ is the direct sum over all reduced graphs $\Gamma$ of the ones based on a single reduced graph $\mathcal{H}^R_{\Gamma}$,  
\be
\mathcal{H}^R=\oplus_{\Gamma}\mathcal{H}^R_{\Gamma}.
\ee
A generic element $\psi_{\Gamma}\in\mathcal{H}^R_{\Gamma}$ is a proper function of $L_1+L_2+L_3$ copies of $SU(2)$ group elements $h_{l}$, $L_i$ being the total number of the links of the kind $l_i$ in $\Gamma$. Given a link $l$, let us denote by $u_l$ the internal direction corresponding to it (if $l$ is the kind $l_i$ then $\vec{u}_l=\vec{u}_i$), the functions of $h_l$ group elements can be expandend in terms of the following projected Wigner matrices 
\be
\;{}^l\!D^{j_{l}}_{j_{l} j_{l}}(h_{l})= \langle j_{l},\vec{u}_l|D^j(g)|j_{l},\vec{u}_l\rangle,\qquad \;{}^l\!D^{j_{l}}_{-j_{l} -j_{l}}(h_{l})= \langle j_{l},-\vec{u}_l|D^j(g)|j_{l},-\vec{u}_l\rangle \label{reduced basis elements}
\ee
$|j,\vec{u}_l\rangle$ and $|j,-\vec{u}_l\rangle$ being the basis of $SU(2)$ irreducible representations with spin number $j$ and magnetic components along the direction $\vec{u}_l$ equal to $j$ and $-j$, respectively. We will denote them by ${}^l\!D^{j_{l}}_{m_{l} m_{l}}(h_{l})$ with $m_{l}=\pm j_{l}$. Here for the first time we will consider also reduced states with minimum magnetic numbers. The projected Wigner matrices are entirely determined by their restriction the the $U(1)_i$ subgroup.   

The whole basis state in the gauge-invariant reduced space ${}^G\mathcal{H}^{R}$ is obtained by inserting at each node $n$ the reduced intertwiners $\langle{\bf j_{l}}, {\bf x}_n|{\bf m_{l}},  \vec{{\bf u}}_l \rangle$, which are constructed from the $SU(2)$ intertwiner basis ${\bf x}_n$. At the end, one gets 
\be
\langle h|\Gamma, {\bf m_l, x_n \bf}\rangle= \prod_{n\in\Gamma}\langle{\bf j_{l}}, {\bf x}_n|{\bf m_{l}},  \vec{{\bf u}}_l \rangle 
\prod_{l} \;{}^l\!D^{j_{l}}_{m_{l} m_{l}}(h_{l}),
\label{base finale}
\ee
where $\prod_{n\in\Gamma}$ and $\prod_{l}$ extend over all the nodes $n\in\Gamma$ and over all the links $l$ emanating from $n$, respectively. Henceforth, each basis element is labeled by the reduced graph $\Gamma$, the spin quantum numbers $m_l$ associated with each link and the $SU(2)$ intertwiners ${\bf x}_n$ used to construct the reduced ones ($j_l=|m_l|$).

These basis states are not orthogonal with respect to ${\bf x}_n$, since the scalar product is given by 
\be
<\Gamma, {\bf m_l, x_n }|\Gamma', {\bf m'_l, x'_n } >=\delta_{\Gamma,\Gamma'}\; \prod_{n\in\Gamma}\,\prod_{l\in\Gamma} \delta_{m_l,m'_l}\;  <{\bf m_l}, \vec{{\bf u}}_l | {\bf j_l} , {\bf x_n}><{\bf j_l}, {\bf x'_n}|{\bf m_l}, \vec{{\bf u}}_l >.
\label{not orto}
\ee
The reduced fluxes ${}^{R}E_i$ are defined only across the surfaces $S^i$ dual to $\omega_i$ and their action is non-vanishing only on those states based at links $l_i$, in which case it reads 
\be
{}^{R}\hat{E}_i(S^i){}^l\!D^{j_{l}}_{m_{l} m_{l}}(h_{l})= 8\pi\gamma l_P^2\, m_{l}\,{}^l\!D^{j_{l}}_{m_{l} m_{l}}(h_{l}) \qquad l_i\cap S^i\neq \oslash .\label{redei}
\ee
As a consequence the reduced volume operator is diagonal in the basis (\ref{base finale}). For instance the volume of a region $\omega$ containing the node $n$ acts as follows on basis vectors based at the links $l_i$ emanating from $n$  
\be\label{volume ridotto}
{}^{R}\hat{V}(\omega)\prod_{l} 
\langle{\bf j_{l}}, {\bf x}_n|{\bf m_{l}},  \vec{{\bf u}}_l \rangle  
\cdot \;{}^l\!D^{j_{l}}_{m_{l} m_{l}}(h_{l})=(8\pi\gamma l_P^2)^{3/2}V_{{\bf m_l}}\prod_{l} 
\langle{\bf j_{l}}, {\bf x}_n|{\bf m_{l}},  \vec{{\bf u}}_l \rangle  
\cdot \;{}^l\!D^{j_{l}}_{m_{l} m_{l}}(h_{l}).
\ee%%controlla
where
\be
V_{{\bf m_l}}=\sqrt{\prod_{i}|\sum_{l_i}m_{l_i}|}
\ee
The sum inside the square root extends over the links of the kind $l_i$ emanating from $n$, thus generically it is the sum 
of two terms (based at the links incoming and outcoming in $n$).

The invariance under reduced diffeomorphisms can be implemented on a quantum level according with standard LQG techniques, {\it i.e.} by defining reduced s-knots
\be
<s,{\bf j_l},{\bf x_n}|h>=%\sum_{\Gamma\in s}<h |P|\Gamma,{\bf j_e},{\bf x_v}>^*=
\sum_{\Gamma\in s}<\Gamma,{\bf j_l},{\bf x_n}|h>,
\ee    
where the sum is over all the reduced graphs related by a reduced diffeomorphism.

The scalar constraint can be implemented in ${}^G\mathcal{H}^{R}$ by taking the expression of the full theory and substituting the elements of the reduced Hilbert space as in \cite{Alesci:2013xd}. 
This procedure provides a quantum operator acting in the reduced Hilbert space describing a diagonal metric tensor. 
Hence, it is well-grounded only if the classical action of the scalar constraint preserves the gauge condition on the metric tensor. This is not generically the case, since after a finite transformation generated by $H$ the metric is diagonal only modulo a diffeomorphism (which is not a reduced one). 
 The definition of the modified constraint preserving the diagonal form of the metric and its quantization will be discussed elsewhere.
Here we return back to the first application to QRLG, the inhomogeneous extension of the Bianchi I model, in which case the dynamics is entirely determined by the reduced Euclidean scalar constraint, which preserves the diagonal form of the metric.

\subsection{Inhomogeneous extension Bianchi I model}

The Bianchi I model is the anisotropic extension of the flat FRW space-time. The spatial sections are still flat and the fiducial one forms, whose dual $\omega_i$ are Killing vectors, can be taken as $\omega^i=\delta^i_adx^a$, $x^a$ being Cartesian coordinates. The line element reads
\be
ds^2_I=N^2(t)dt^2-a_1^2(t)dx^1\otimes dx^1-a_2^2(t)dx^2\otimes dx^2-a_3^2(t)dx^3\otimes dx^3,\label{BI}
\ee
$N=N(t)$ being the lapse function, while $a_i$ $(i=1,2,3)$ denote the three scale factors, all depending on the time variable only. 

We considers the following inhomogeneous extension of the line element (\ref{BI})
\be
ds^2_I=N^2(x,t)dt^2-a_1^2(t,x)dx^1\otimes dx^1-a_2^2(t,x)dx^2\otimes dx^2-a_3^2(t,x)dx^3\otimes dx^3,
\ee
in which each scale factor $a_i$ is a function of time and of the spatial coordinates. By fixing the group of internal rotations \cite{Cianfrani:2011wg,Cianfrani:2012gv} the densitized inverse 3-bein vectors can be taken as
\be
E^a_i=p^i(t,x)\delta^a_i,\qquad p^i=\frac{a_1a_2a_3}{a_i},\label{Ein}
\ee
where the index $i$ is not summed. In the following, repeated indexes will not be summed. As for Ashtekar connections, we get a similar expression, {\it i.e.}
\be
A^i_a(t,x)= c_i(t,x)\delta^i_a,\qquad c_i(t,x)=\frac{\gamma}{N}\dot{a_i}.\label{Ain}
\ee  
in the two relevant cases of i) reparametrized Bianchi I model (in which each scale factor $a_i$ is a function of the corresponding Cartesian coordinate $x^i=\delta^i_ax^a$ only) and  ii) the generalized Kasner solution within a fixed Kasner epoch (in which spatial gradients are negligible with respect to time derivatives). It is worth noting how the expression for $A^a_i$ (\ref{Ain}) is exact in the former case, which is equivalent to the homogeneous Bianchi I model, while it holds only approximatively in the latter by assuming the BKL conjecture \cite{Belinsky:1982pk}. In which case, the inhomogeneous model is made of a collection of homogeneous patches, one for each point. 

In reduced phase space the $SU(2)$ Gauss constraint and the vector constraint do not vanish but they generate $U(1)_i$ gauge transformations and reduced diffeomorphisms. The Lorentzian part of the scalar constraint is proportional to the Euclidean one, such the sum is $1/\gamma^2$ times the latter and the explicit expression reads
\be
H[N]=\frac{1}{\gamma^2}H_E[N]=\frac{1}{\gamma^2}\int d^3x N\left[\sqrt{\frac{p^1p^2}{p^3}}c_1c_2+\sqrt{\frac{p^2p^3}{p^1}}c_2c_3+\sqrt{\frac{p^3p^1}{p^2}}c_3c_1\right],\label{HbI}
\ee
which can be seen as the sum of local Bianchi I patches, {\it i.e.}
\be\label{localBI}
H[N]=\frac{1}{\gamma^2}\sum_{x} V(x)N(x)\left[\sqrt{\frac{p^1p^2}{p^3}}c_1c_2+\sqrt{\frac{p^2p^3}{p^1}}c_2c_3+\sqrt{\frac{p^3p^1}{p^2}}c_3c_1\right](x),
\ee
$V_x$ being the volume of the homogeneous patch based at the point $x$, where all the $c_i$ and $p^i$ variables are evaluated. This the kind of classical dynamics we are going to compare with the semiclassical limit of QRLG, since it preserves the diagonal form of the metric.

\section{Semiclassical states in QRLG}\label{semiQRLG}
Let us define semiclassical states in QRLG by projecting the expression (\ref{sclqg}) down to $^{red}\mathcal{H}$. Hence, let us first define the semiclassical states along a given link $l$. The analogous of the expression (\ref{hklqg}) now reads
\be
K_\alpha(h_{l},h')=\sum_{m_{l}=-\infty}^{+\infty}(2j_{l}+1)e^{-j_{l}(j_{l}+1)\frac{\alpha}{2}}\;{}^l\!D^{j_{l}}_{m_{l} m_{l}}(h^{-1}_{l}h'),\label{hkqrlg}
\ee
where $j_l=|m_{l}|$ and $h'$ is an element of the $SU(2)$ subgroup generated by $\tau_i$ ($U(1)_i$), $i$ being the internal direction associated with the link $l$ ($l$ is of the kind $l_i$), {\it i.e.}
\be
h'=e^{i\theta_l\tau_i},
\ee
$\theta_l$ being the parameter along the group, which can be determined from the explicit expression of the holonomy along the links $l$. In the limit in which $c_i$ is constant along $l$, there is the direct identification $\theta_l=\pm\epsilon_{l}\,c_i$, $\epsilon_{l}$ being the length of $l$ and the $+$(-) signs is for positive(negative)- oriented $l$.

The complexification of $h'$ is given by 
\be
H'=h'e^{\frac{\alpha}{8\pi\gamma l_P^2}E'_i\tau_i},\label{Hqrlg}
\ee
which differs from the expression (\ref{Hlqg}) because the indexes $i$ in the exponent are not summed and $h'$ is a $U(1)_i$ group element. We can rewrite the expression (\ref{Hqrlg}) as follows
\be
H'=R(\vec{u}_l)e^{i\theta_l\tau_3+\frac{\alpha}{8\pi\gamma l_P^2}E'_i\tau_3}R^{-1}(\vec{u}_l),\label{Hqrlg2}
\ee
$R(\vec{u}_l)$ being the rotation sending the direction $\vec{u}_l$ into the direction $\vec{u}_3$. A clear interpretation of the classical data we are peaking on can now be given in terms of a cellar decomposition. In fact, we can compare Eq.(\ref{Hqrlg2}) with the expression of the coherent states for a homogeneous model defined in \cite{Bianchi:2009ky,Magliaro:2010qz}. These coherent states are defined via a geometrical parametrization of the phase space in terms of twisted geometries \cite{Freidel:2010aq},\cite{Rovelli:2010km}, in which two $SU(2)$ rotations are inserted at the target and source points. By comparing Eq.(\ref{Hqrlg2}) with Eq.(52) in \cite{Magliaro:2010qz}, one sees how in our case, in which the intrinsic curvature of the spatial section vanishes, these two rotations coincides and they are given by $R(\vec{u}_l)$. 

The semiclassical state for QRLG takes the following expression 
\be   
\psi^\alpha_{H'}(h_{l})=K_\alpha(h_{l},H'),\label{scqrlg}
\ee
$K_\alpha(h_{l},H')$ and $H'$ given by Eq.(\ref{hkqrlg}) and (\ref{Hqrlg}), respectively. We can write an explicit expression for $\psi^\alpha_{H'}(h_{l})$ in terms of basis vectors (\ref{reduced basis elements}), thanks to the fact that the $SU(2)$ representation $\;{}^l\!D^{j_{l}}_{m n}(h_{l})$ of $U(1)_i$ group elements is diagonal, {\it i.e.} 
\be
\;{}^l\!D^{j_{l}}_{m_{l} m_{l}}(h^{-1}_{l}H')=\sum_{n=-j_{l}}^{j_{l}}\;{}^l\!D^{j_{l}}_{m_{l} n}(h^{-1}_{l})\;{}^l\!D^{j_{l}}_{n m_{l}}(H')=\;{}^l\!D^{j_{l}}_{m_{l} m_{l}}(h^{-1}_{l})\;{}^l\!D^{j_{l}}_{m_{l} m_{l}}(H').\label{d1}
\ee  
The last factor on the right-hand side of the equation above can be easily evaluated, so getting
\be
\;{}^l\!D^{j_{l}}_{m_{l} m_{l}}(H')=e^{i\theta_lm_{l}}e^{\frac{\alpha}{8\pi\gamma l_P^2}E'_im_{l}}.\label{d2}
\ee
By collecting together all the equations of this section one finds the following expression for the semiclassical states in QRLG
\be
\psi^\alpha_{H'}(h_{l})=\sum_{m_{l}=-\infty}^\infty \psi^\alpha_{H'}(m_{l})\;{}^l\!D^{j_{l}}_{m_{l} m_{l}}(h^{-1}_{l}),
\label{finscqrlg}
\ee
with 
\begin{equation}
\psi^\alpha_{H'}(m_{l})=(2j_{l}+1)e^{-j_{l}(j_{l}+1)\frac{\alpha}{2}}e^{i\theta_lm_{l}}e^{\frac{\alpha}{8\pi\gamma l_P^2}E'_im_{l}}.
\end{equation}
where $j_l=|m_l|$.
It is worth noting how in the limit $\frac{E'}{8\pi\gamma l_P^2}>>1$ one has
\begin{equation}
-j(j+1)\frac{\alpha}{2}+m\frac{\alpha E'}{8\pi\gamma l_P^2}=-m(m\pm 1)\frac{\alpha}{2}+m\frac{\alpha E'}{8\pi\gamma l_P^2}\sim
-\frac{\alpha}{2}\left(m -\frac{E'}{8\pi\gamma l_P^2}\right)^2+\alpha\left(\frac{E'}{8\pi\gamma l_P^2}\right)^2\,.
\end{equation} 
Hence the coefficients $\psi^\alpha_{H'}(m_{l})$ modulo a factor not depending on $m_{l}$ become Gaussian weights and $\psi^\alpha_{H'_l}(h_{l})$ can be written as
\be
\psi^\alpha_{H'}(h_{l})\sim\sum_{m_{l}=-\infty}^\infty(2j_{l}+1)e^{-\frac{\alpha}{2}\left(m_{l}-\frac{E'_i}{8\pi\gamma l_P^2}\right)^2}e^{i\theta_lm_{l}}\;{}^l\!D^{j_{l}}_{m_{l} m_{l}}(h^{-1}_{l})=\sum_{m_{l}=-\infty}^\infty \psi^{\alpha}_{H'}(m_{l})\;{}^i\!D^{j_{l}}_{m_{l} m_{l}}(h^{-1}_{l})
,\label{finscqrlg2}
\ee
which outlines that the state is peaked around $\bar{m}_{l}=\frac{E'_l}{8\pi\gamma l_P^2}$. Such a value corresponds to the following momenta $\bar{p}^l$
\be
\bar{p}^l\delta_l^2=8\pi\gamma l_P^2\bar{m}_l\,, \label{delta}
\ee
$\delta_l^2$ being the area of the surface across which $E'_l$ is smeared in the fiducial metric.
Similarly it can be shown that the state is also peaked around the classical holonomy $h'$.

For multi-link states, one simply has to consider the direct product of states of the kind (\ref{finscqrlg}) and to insert invariant intertwiners at nodes. We remember that in QRLG the invariant intertwiners are merely coefficients, so the extension of the expression (\ref{finscqrlg2}) to the gauge invariant Hilbert space ${}^G\mathcal{H}^{R}$ can be done straightforwardly  by inserting reduced intertwiners both in basis elements and in the coefficients, so finding
\begin{equation}
\psi^{{\bf\alpha}}_{\Gamma{\bf H'}}=\sum_{{\bf m_{l}}}\prod_{n\in\Gamma} \langle{\bf j_{l}}, {\bf x}_n|{\bf m_{l}},  \vec{{\bf u}}_l \rangle^*\;\prod_{l\in\Gamma} \psi^\alpha_{H'_{l}}(m_{l})\;\langle h|\Gamma, {\bf m_l, x_n \bf}\rangle\,,
\label{semiclassici ridotti inv}
\end{equation}
where $\sum_{{\bf m_{l}}}=\prod_{l\in\Gamma}\sum_{m_{l}}$.

\section{The Hamiltonian on basis states}\label{Eucl}
We are now interested in implementing the action of the Hamiltonian ${}^{R}\hat{H}$ as
\be\label{hamEc}
{}^{R}\hat{H}=\frac{1}{\gamma^2}{}^{R}\hat{H}_E, 
\ee
via an operator ${}^{R}\hat{H}_E$ defined on $^{\mathcal{G}} \mathcal{H}^{R}$: a convenient way of constructing it is to replace in the expression (\ref{Hm_delta:quantum}), regularized via a cubulation $C$ adapted to the reduced spinnetwork graph, as explained in \cite{Alesci:2013xd}, quantum holonomies and fluxes with the ones acting on the reduced space as follows
\be
{}^{R}\hat{H}_E[N]=\sum_{\cube} {}^{R}\hat{H}^m_{E\cube}[N]
\ee
where 
\be
{}^{R}\hat{H}^m_{E\cube}[N]:= N(\vgraph) C(m) \,  \, \epsilon^{ijk} \,
   \mathrm{Tr}\Big[{}^{R}\hat{h}^{(m)}_{\alpha_{ij}} {}^{R}\hat{h}^{(m)-1}_{s_{k}} \big[{}^R\hat{h}^{(m)}_{s_{k}},{}^{R}\hat{V}\big]\Big]. 
   \label{Hridotto}
\ee
 The reduced holonomy operators ${}^{R}\hat{h}$ are obtained by projecting the $SU(2)$ ones on the projected Wigner matrices \eqref{reduced basis elements}, while the reduced volume operator is the one given in \eqref{volume ridotto}. 
%The lattice spacing of the cubulation $C$ that acts as a regularization parameter and it can be changed via a reduced diffeomorphisms. Hence, on reduced s-knot states there exists a 4suitable operator topology in which the regulator can be safely removed as in full LQG. Therefore, \emph{the action of the superHamiltonian operator can be regularized in the reduced Hilbert space}. 

The action of this operator has already been computed in \cite{Alesci:2013xd}, however there we allowed only states of the kind $^{j}\!D^j_{jj}(h)$ for the holonomies contained in the expresssion (\ref{Hridotto}), but here we consider general states in $^{\mathcal{G}} \mathcal{H}^{R}$ of the form $^{j}\!D^j_{nn}(h)$ with $n=\pm j$ %for the $h$'s
. This implies that in the regularized Hamiltonian the intertwiners between two different directions will have the possibility to connect holonomies projected on maximum and minimum magnetic number running on different segments $s_i$. The practical rule is then to connect in the Hamiltonian \eqref{Hridotto} objects of the kind:
\be
{}^{R} D^j_{mn}(h_l)=\sum_{\lambda=\pm1} <j,m | j,\lambda\vec{u_l}><j,\lambda\vec{u_l}| D^j(h_l)| j,\lambda\vec{u_l}><j,\lambda\vec{u_l}|j,n>.
\label{reduced with free indexes, doppio}
\ee
to the standard intertwiners.

In view of the application of the non graph changing version of the hamiltonian we consider the simplest state on which the action is non trivial, namely\footnote{ We take the coordinates $x,y,z$ along the fiducial directions $i=1,2,3$}:
\be
|\vgraph^z\rangle_R=
\ba
\ifx\JPicScale\undefined\def\JPicScale{0.6}\fi
\psset{unit=\JPicScale mm}
\psset{linewidth=0.3,dotsep=1,hatchwidth=0.3,hatchsep=1.5,shadowsize=1,dimen=middle}
\psset{dotsize=0.7 2.5,dotscale=1 1,fillcolor=black}
\psset{arrowsize=1 2,arrowlength=1,arrowinset=0.25,tbarsize=0.7 5,bracketlength=0.15,rbracketlength=0.15}
\begin{pspicture}(0,0)(101,143)
\psline{|*-}(19,78)(21.83,80.83)
\rput(17,81){$\scr{j_x}$}
\rput{45}(28.5,87.5){\psellipse[](0,0)(2.21,-2.12)}
\rput(24,91){$\scr{h_{x}}$}
\psline[fillstyle=solid]{-|}(42.83,101.83)(40,99)
\pspolygon[](46,103)(44,101)(42,103)(44,105)
\psline[fillstyle=solid](51,110)(44.84,103.84)
\rput(40,104){$\scr{R_x}$}
\psline[fillstyle=solid]{|*-}(16.83,75.83)(14,73)
\rput(10,74){$\scr{R^{-1}_x}$}
\rput(21,85){$\scr{R_x}$}
\psline[fillstyle=solid](26.83,85.83)(24,83)
\psline(30,89)(32.82,91.82)
\pspolygon[](36,93)(34,91)(32,93)(34,95)
\rput(32,97){$\scr{R^{-1}_x}$}
\psline[fillstyle=solid]{|-}(38,97)(35.17,94.17)
\rput(11,77){$\scr{j_x}$}
\psline{|*-}(51,137)(51,133)
\rput(44,129){$\scr{j_z}$}
\rput{0}(51,130){\psellipse[](0,0)(3,-3)}
\rput(51,130){$\scr{h_{z}}$}
\psline[fillstyle=solid]{-|}(51,110)(51,120)
\psline[fillstyle=solid]{|*-}(51,139)(51,143)
\psline{-|}(51,127)(51,123)
\rput(46,141){$\scr{j_z}$}
\pspolygon[](25,82)(23,80)(21,82)(23,84)
\psline[fillstyle=solid](12,71)(6,65)
\pspolygon[](15,72)(13,70)(11,72)(13,74)
\psline{|*-}(83,78)(80.17,80.83)
\rput(84,82){$\scr{j_y}$}
\rput(76,91){$\scr{h_{y}}$}
\psline[fillstyle=solid]{-|}(59.17,101.83)(62,99)
\pspolygon[](58,105)(60,103)(58,101)(56,103)
\rput(62,104){$\scr{R_y}$}
\psline[fillstyle=solid]{|*-}(85.17,75.83)(88,73)
\rput(95,73){$\scr{R^{-1}_y}$}
\rput(79,86){$\scr{R_y}$}
\psline[fillstyle=solid](75.17,85.83)(78,83)
\psline(72,89)(69.18,91.82)
\pspolygon[](68,95)(70,93)(68,91)(66,93)
\rput(72,96){$\scr{R^{-1}_y}$}
\psline[fillstyle=solid]{|-}(64,97)(66.83,94.17)
\rput(90,77){$\scr{j_y}$}
\pspolygon[](79,84)(81,82)(79,80)(77,82)
\psline[fillstyle=solid](90,71)(96,65)
\pspolygon[](89,74)(91,72)(89,70)(87,72)
\rput{43.83}(73.44,87.44){\psellipse[](0,0)(2.21,-2.12)}
\rput(14,60){$\scr{j_l}$}
\rput(80,64){$\scr{j_l}$}
\psline[border=0.3,fillstyle=solid](51,110)(57.16,103.84)
\rput(47,120){$\scr{j_z}$}
\psline{|*-}(42,39)(39.17,41.83)
\rput(43,43){$\scr{j_l}$}
\rput(35,52){$\scr{h_{l_y}}$}
\psline[fillstyle=solid]{-|}(18.17,62.83)(21,60)
\pspolygon[](17,66)(19,64)(17,62)(15,64)
\rput(21,65){$\scr{R_y}$}
\psline[fillstyle=solid]{|*-}(44.17,36.83)(47,34)
\rput(44,30){$\scr{R^{-1}_y}$}
\rput(38,47){$\scr{R_y}$}
\psline[fillstyle=solid](34.17,46.83)(37,44)
\psline(31,50)(28.18,52.82)
\pspolygon[](27,56)(29,54)(27,52)(25,54)
\rput(31,57){$\scr{R^{-1}_y}$}
\psline[fillstyle=solid]{|-}(23,58)(25.83,55.17)
\rput(49,38){$\scr{j_l}$}
\pspolygon[](38,45)(40,43)(38,41)(36,43)
\psline[fillstyle=solid](49,32)(51.83,29.17)
\pspolygon[](48,35)(50,33)(48,31)(46,33)
\rput{43.83}(32.44,48.44){\psellipse[](0,0)(2.21,-2.12)}
\psline[fillstyle=solid](11,70)(16.16,64.84)
\psline{|*-}(80,57)(77.17,54.17)
\rput(64,38){$\scr{j_l}$}
\rput{45}(70.5,47.5){\psellipse[](0,0)(2.21,-2.12)}
\rput(75,46){$\scr{h^{-1}_{l_x}}$}
\psline[fillstyle=solid]{-|}(56.17,33.17)(59,36)
\pspolygon[](53,32)(55,34)(57,32)(55,30)
\psline[fillstyle=solid](52,29)(54.16,31.16)
\rput(58,28){$\scr{R_x}$}
\psline[fillstyle=solid]{|*-}(82.17,59.17)(85,62)
\rput(88,58){$\scr{R^{-1}_x}$}
\rput(80,52){$\scr{R_x}$}
\psline[fillstyle=solid](72.17,49.17)(75,52)
\psline(69,46)(66.18,43.18)
\pspolygon[](63,42)(65,44)(67,42)(65,40)
\rput(71,40){$\scr{R^{-1}_x}$}
\psline[fillstyle=solid]{|-}(61,38)(63.83,40.83)
\pspolygon[](74,53)(76,55)(78,53)(76,51)
\psline[fillstyle=solid](87,64)(92,69)
\pspolygon[](84,63)(86,65)(88,63)(86,61)
\rput(41,108){$\scr{j_x}$}
\rput(58,108){$\scr{j_y}$}
\rput(101,67){$\scr{j'_y}$}
\rput(3,68){$\scr{j'_x}$}
\end{pspicture}
\ea
\label{v3}
\ee
Note that here and in the following in the graphical formulas the 3-valent nodes represent Clebsh-Gordan coefficients, while in the previous papers \cite{Alesci:2013xd,Alesci:2012md} instead, we were using the same symbol for 3-j symbols. This is just to avoid the presence of dimension factors that would appear in subsequent recouplings, and if one keeps track of the direction of the holonomies, there is not difference between the two choices.
Note also that the quantum numbers $j_l$ here properly speaking are the $m_l$ of the previous section (the magnetic numbers), but here we consider states with positive magnetic number and to keep the notation analogous to the $SU(2)$ one we set $j_l=m_l$. 
This state $|\vgraph^z\rangle_R=|l_x,l_y,l_z,l_l, j_x, j_y, j_z ,j_l, x_n \rangle_R$ is based on a  dressed node $\vgraph$ with three non coplanar outgoing links $l_x, l_y, l_z$ in the directions $x,y,z$ respectively and an arc $l_l$ lying in the plane orthogonal to the direction $z$ formed by two links $l_{l_x}$ and $l_{l_y}$ respectively parallel to $l_x$ and $l_y$ and closing a squared loop with them as in \eqref{v3}.

The operator ${}^{R}\hat{H}^m_{E\cube} |\vgraph^z \rangle_R$ acting at the node $\vgraph$ is the sum of three terms  ${}^{R}\hat{H}^m_{E\cube} |\vgraph^z \rangle_R=\sum_{k}{}^{R}\hat{H}^{m,k}_{E\cube} |\vgraph^z \rangle_R$ where $k=x,y,z$ for $s_k\in l_x, l_y, l_z$ respectively.

Now we restrict our attention to ${}^{R}\hat{H}^{m,z}_{E\cube} |\vgraph^z \rangle_R$ because for an appropriate choice of coherent states (based on $\alpha_{ij}=l_x\circ l_{l_y}\circ l_{l_x}\circ l^{-1}_y$) this will be the only operator that matters.
As noted several times \cite{Borissov:1997ji,Gaul:2000ba,Alesci:2013kpa} only the term in the commutator of \eqref{Hridotto} with the holonomy ${}^R\hat{h}^{(m)}_{s_{z}}$ on the right contributes. This holonomy produces (from now on we focus to the central 3-valent node with links in the three orthogonal direction, we will analyze the remaining nodes in the following):

\be
^{R}\hat{h}^{(m)}_{s_{z}} |\vgraph^z\rangle_R\; =^{R}\!\hat{h}^{(m)}_{s_{z}}
\ba
\ifx\JPicScale\undefined\def\JPicScale{0.7}\fi
\ifx\JPicScale\undefined\def\JPicScale{1}\fi
\psset{unit=\JPicScale mm}
\psset{linewidth=0.3,dotsep=1,hatchwidth=0.3,hatchsep=1.5,shadowsize=1,dimen=middle}
\psset{dotsize=0.7 2.5,dotscale=1 1,fillcolor=black}
\psset{arrowsize=1 2,arrowlength=1,arrowinset=0.25,tbarsize=0.7 5,bracketlength=0.15,rbracketlength=0.15}
\begin{pspicture}(0,65)(90,140)
\psline{|*-}(19,78)(21.83,80.83)
\rput(17,81){$\scr{j_x}$}
\rput{44.4}(28.5,87.5){\psellipse[](0,0)(2.21,-2.12)}
\rput(24,91){$\scr{h_{x}}$}
\psline[fillstyle=solid]{-|}(42.83,101.83)(40,99)
\pspolygon[](46,103)(44,101)(42,103)(44,105)
\psline[fillstyle=solid](51,110)(44.84,103.84)
\rput(40,104){$\scr{R_x}$}
\psline[fillstyle=solid]{|*-}(16.83,75.83)(14,73)
\rput(10,74){$\scr{R^{-1}_x}$}
\rput(21,85){$\scr{R_x}$}
\psline[fillstyle=solid](26.83,85.83)(24,83)
\psline(30,89)(32.82,91.82)
\pspolygon[](36,93)(34,91)(32,93)(34,95)
\rput(32,97){$\scr{R^{-1}_x}$}
\psline[fillstyle=solid]{|-}(38,97)(35.17,94.17)
\rput(43,108){$\scr{j_x}$}
\rput(13,77){$\scr{j_x}$}
\psline{|*-}(51,137)(51,133)
\rput{0}(51,130){\psellipse[](0,0)(3,-3)}
\rput(51,130){$\scr{h_{z}}$}
\psline[fillstyle=solid]{-|}(51,110)(51,120)
\psline[fillstyle=solid]{|*-}(51,139)(51,143)
\psline{-|}(51,127)(51,123)
\rput(46,141){$\scr{j_z}$}
\pspolygon[](25,82)(23,80)(21,82)(23,84)
\psline[fillstyle=solid](12,71)(9.17,68.17)
\pspolygon[](15,72)(13,70)(11,72)(13,74)
\psline{|*-}(83,78)(80.17,80.83)
\rput(84,82){$\scr{j_y}$}
\rput(76,91){$\scr{h_{y}}$}
\psline[fillstyle=solid]{-|}(59.17,101.83)(62,99)
\pspolygon[](58,105)(60,103)(58,101)(56,103)
\rput(62,104){$\scr{R_y}$}
\psline[fillstyle=solid]{|*-}(85.17,75.83)(88,73)
\rput(95,73){$\scr{R^{-1}_y}$}
\rput(79,86){$\scr{R_y}$}
\psline[fillstyle=solid](75.17,85.83)(78,83)
\psline(72,89)(69.18,91.82)
\pspolygon[](68,95)(70,93)(68,91)(66,93)
\rput(72,96){$\scr{R^{-1}_y}$}
\psline[fillstyle=solid]{|-}(64,97)(66.83,94.17)
\rput(57,107){$\scr{j_y}$}
\rput(90,77){$\scr{j_y}$}
\pspolygon[](79,84)(81,82)(79,80)(77,82)
\psline[fillstyle=solid](90,71)(92.83,68.17)
\pspolygon[](89,74)(91,72)(89,70)(87,72)
\rput{44.4}(73.44,87.44){\psellipse[](0,0)(2.21,-2.12)}
\psline[fillstyle=solid](51,110)(57.16,103.84)
\rput(48,115){$\scr{j_z}$}
\end{pspicture}
\ea
=
\ee
\be
= 
\ba
\ifx\JPicScale\undefined\def\JPicScale{0.7}\fi
\psset{unit=\JPicScale mm}
\psset{linewidth=0.3,dotsep=1,hatchwidth=0.3,hatchsep=1.5,shadowsize=1,dimen=middle}
\psset{dotsize=0.7 2.5,dotscale=1 1,fillcolor=black}
\psset{arrowsize=1 2,arrowlength=1,arrowinset=0.25,tbarsize=0.7 5,bracketlength=0.15,rbracketlength=0.15}
\begin{pspicture}(0,70)(90,138)
\psline{|*-}(19,78)(21.83,80.83)
\rput(17,81){$\scr{j_x}$}
\rput{44.4}(28.5,87.5){\psellipse[](0,0)(2.21,-2.12)}
\rput(24,91){$\scr{h_{x}}$}
\psline[fillstyle=solid]{-|}(42.83,101.83)(40,99)
\pspolygon[](46,103)(44,101)(42,103)(44,105)
\psline[fillstyle=solid](51,110)(44.84,103.84)
\rput(40,104){$\scr{R_x}$}
\psline[fillstyle=solid]{|*-}(16.83,75.83)(14,73)
\rput(10,74){$\scr{R^{-1}_x}$}
\rput(21,85){$\scr{R_x}$}
\psline[fillstyle=solid](26.83,85.83)(24,83)
\psline(30,89)(32.82,91.82)
\pspolygon[](36,93)(34,91)(32,93)(34,95)
\rput(32,97){$\scr{R^{-1}_x}$}
\psline[fillstyle=solid]{|-}(38,97)(35.17,94.17)
\rput(43,108){$\scr{j_x}$}
\rput(13,77){$\scr{j_x}$}
\psline{|*-}(51,137)(51,133)
\rput{0}(51,130){\psellipse[](0,0)(3,-3)}
\rput(51,130){$\scr{h_{z}}$}
\psline[fillstyle=solid]{-|}(51,110)(51,120)
\psline[fillstyle=solid]{|*-}(51,139)(51,145)
\psline{-|}(51,127)(51,123)
\rput(46,144){$\scr{j_z}$}
\pspolygon[](25,82)(23,80)(21,82)(23,84)
\psline[fillstyle=solid](12,71)(9.17,68.17)
\pspolygon[](15,72)(13,70)(11,72)(13,74)
\psline{|*-}(83,78)(80.17,80.83)
\rput(84,82){$\scr{j_y}$}
\rput(76,91){$\scr{h_{y}}$}
\psline[fillstyle=solid]{-|}(59.17,101.83)(62,99)
\pspolygon[](58,105)(60,103)(58,101)(56,103)
\rput(62,104){$\scr{R_y}$}
\psline[fillstyle=solid]{|*-}(85.17,75.83)(88,73)
\rput(95,73){$\scr{R^{-1}_y}$}
\rput(79,86){$\scr{R_y}$}
\psline[fillstyle=solid](75.17,85.83)(78,83)
\psline(72,89)(69.18,91.82)
\pspolygon[](68,95)(70,93)(68,91)(66,93)
\rput(72,96){$\scr{R^{-1}_y}$}
\psline[fillstyle=solid]{|-}(64,97)(66.83,94.17)
\rput(57,107){$\scr{j_y}$}
\rput(90,77){$\scr{j_y}$}
\pspolygon[](79,84)(81,82)(79,80)(77,82)
\psline[fillstyle=solid](90,71)(92.83,68.17)
\pspolygon[](89,74)(91,72)(89,70)(87,72)
\rput{44.4}(73.44,87.44){\psellipse[](0,0)(2.21,-2.12)}
\psline[fillstyle=solid](51,110)(57.16,103.84)
\rput(46,112){$\scr{j_z}$}
\psline(51,115)(55,115)
\psline(51,142)(55,142)
\rput(46,118){$\scr{j_z+\mu}$}
\rput(57,117){$\scr{\mu}$}
\rput(58,141){$\scr{\mu}$}
\rput(45,139){$\scr{j_z+\mu}$}
\end{pspicture}.
\ea
\ee
with the magnetic index $\mu=\pm m$ (remember that in the reduced case in the recoupling rules  are the relative magnetic numbers that determine the resulting representation see appendix \ref{appendix}).

Then the Volume in \eqref{Hridotto} acts diagonally multiplying by $(8\pi\gamma l_P^2)^{3/2}\sqrt{j_x\;j_y\;(j_z+\mu)}$; considering then the inverse holonomy along z we have:
\be
{}^{R}h^{(m)\,-1}_{s_{z}}\;\; {}^{R}\hat{V}\;\; {}^{R}h^{(m)}_{s_{z}} |\vgraph^z\rangle_R\; = (8\pi\gamma l_P^2)^{3/2}\sqrt{j_x\;j_y\;(j_z+\mu)}
\ba
\ifx\JPicScale\undefined\def\JPicScale{0.7}\fi
\psset{unit=\JPicScale mm}
\psset{linewidth=0.3,dotsep=1,hatchwidth=0.3,hatchsep=1.5,shadowsize=1,dimen=middle}
\psset{dotsize=0.7 2.5,dotscale=1 1,fillcolor=black}
\psset{arrowsize=1 2,arrowlength=1,arrowinset=0.25,tbarsize=0.7 5,bracketlength=0.15,rbracketlength=0.15}
\begin{pspicture}(0,0)(106.29,89)
\psline{|*-}(95.69,20.31)(92.86,23.14)
\rput(87.2,35.87){$\scr{j_y}$}
\rput{112.48}(83.67,32.34){\psellipse[](0,0)(3,-3)}
\rput(83.67,32.33){$\scr{h_{y}}$}
\psline[fillstyle=solid]{-|}(67.4,48.6)(70.23,45.77)
\pspolygon[](65.28,54.96)(69.52,50.72)(65.28,46.48)(61.04,50.72)
\psline[fillstyle=solid](57,59)(63.16,52.84)
\rput(63.87,50.72){$\scr{R_y}$}
\pspolygon[](102.05,18.19)(106.29,13.95)(102.05,9.71)(97.81,13.95)
\psline[fillstyle=solid]{|*-}(97.1,18.9)(99.93,16.07)
\rput(100.64,13.95){$\scr{R^{-1}_x}$}
\pspolygon[](90.74,29.51)(94.98,25.26)(90.74,21.02)(86.5,25.27)
\rput(89.32,25.26){$\scr{R_y}$}
\psline[fillstyle=solid](85.79,30.21)(88.62,27.38)
\psline{|*-}(81.54,34.46)(78.72,37.28)
\pspolygon[](76.59,43.65)(80.84,39.41)(76.59,35.16)(72.34,39.4)
\rput(75.18,39.41){$\scr{R^{-1}_y}$}
\psline[fillstyle=solid]{|-}(71.64,44.36)(74.47,41.53)
\rput(68.82,54.25){$\scr{j_y}$}
\psline{|*-}(18.31,20.31)(21.14,23.14)
\rput(33.87,28.8){$\scr{j_x}$}
\rput{22.48}(30.34,32.33){\psellipse[](0,0)(3,-3)}
\rput(30.33,32.33){$\scr{h_{x}}$}
\psline[fillstyle=solid]{-|}(46.6,48.6)(43.77,45.77)
\pspolygon[](52.96,50.72)(48.72,46.48)(44.48,50.72)(48.72,54.96)
\psline[fillstyle=solid](57,59)(50.84,52.84)
\rput(48.72,52.13){$\scr{R_x}$}
\psline[fillstyle=solid](9.83,11.83)(7,9)
\pspolygon[](16.19,13.95)(11.95,9.71)(7.71,13.95)(11.95,18.19)
\psline[fillstyle=solid]{|*-}(16.9,18.9)(14.07,16.07)
\rput(11.95,15.36){$\scr{R^{-1}_x}$}
\pspolygon[](27.51,25.26)(23.26,21.02)(19.02,25.26)(23.27,29.5)
\rput(23.26,26.68){$\scr{R_x}$}
\psline[fillstyle=solid](28.21,30.21)(25.38,27.38)
\psline{|*-}(32.46,34.46)(35.28,37.28)
\pspolygon[](41.65,39.41)(37.41,35.16)(33.16,39.41)(37.4,43.66)
\rput(37.41,40.82){$\scr{R^{-1}_x}$}
\psline[fillstyle=solid]{|-}(42.36,44.36)(39.53,41.53)
\rput(52.25,47.18){$\scr{j_x}$}
\rput(15.49,10.41){$\scr{j_x}$}
\psline{|*-}(57,83)(57,79)
\rput(50,75){$\scr{j_z}+\mu$}
\rput{0}(57,76){\psellipse[](0,0)(3,-3)}
\rput(57,76){$\scr{h_{z}}$}
\psline[fillstyle=solid]{-|}(57,59)(57,67)
\psline[fillstyle=solid]{|*-}(57,85)(57,89)
\psline{-|}(57,73)(57,69)
\rput(51,64){$\scr{j_z}+\mu$}
\rput(52,87){$\scr{j_z}+\mu$}
\psline(57,63)(62,63)
\psline(57,87)(62,87)
\psline{|*-}(69,83)(69,79)
\rput{0}(69,76){\psellipse[](0,0)(3,-3)}
\rput(69,76){$\scr{h_{z}^{-1}}$}
\psline[fillstyle=solid]{-|}(69,59)(69,67)
\psline[fillstyle=solid]{|*-}(69,85)(69,87)
\psline{-|}(69,73)(69,69)
\psline(69,87)(62,87)
\rput(71,64){$\scr{\mu'}$}
\end{pspicture}
\ea.
\ee
Before attaching the last part of the action of ${}^{R}\hat{H}^{m,3}_{E\cube}$ it's convenient so simplify the $U(1)_z$ group elements and separate the projected lines obtaining:
\be
{}^{R}h^{(m)\;-1}_{s_z} {}^{R}\hat{V} {}^{R}h^{(m)}_{s_z} |\vgraph^z\rangle_R\; = (8\pi\gamma l_P^2)^{3/2}\sqrt{j_x\;j_y\;(j_z+\mu)}
\ba
\ifx\JPicScale\undefined\def\JPicScale{0.7}\fi
\psset{unit=\JPicScale mm}
\psset{linewidth=0.3,dotsep=1,hatchwidth=0.3,hatchsep=1.5,shadowsize=1,dimen=middle}
\psset{dotsize=0.7 2.5,dotscale=1 1,fillcolor=black}
\psset{arrowsize=1 2,arrowlength=1,arrowinset=0.25,tbarsize=0.7 5,bracketlength=0.15,rbracketlength=0.15}
\begin{pspicture}(0,0)(106.29,89)
\psline{|*-}(95.69,20.31)(92.86,23.14)
\rput(87.2,35.87){$\scr{j_y}$}
\rput{112.48}(83.67,32.34){\psellipse[](0,0)(3,-3)}
\rput(83.67,32.33){$\scr{h_{y}}$}
\psline[fillstyle=solid]{-|}(67.4,48.6)(70.23,45.77)
\pspolygon[](65.28,54.96)(69.52,50.72)(65.28,46.48)(61.04,50.72)
\psline[fillstyle=solid](57,59)(63.16,52.84)
\rput(63.87,50.72){$\scr{R_y}$}
\pspolygon[](102.05,18.19)(106.29,13.95)(102.05,9.71)(97.81,13.95)
\psline[fillstyle=solid]{|*-}(97.1,18.9)(99.93,16.07)
\rput(100.64,13.95){$\scr{R^{-1}_x}$}
\pspolygon[](90.74,29.51)(94.98,25.26)(90.74,21.02)(86.5,25.27)
\rput(89.32,25.26){$\scr{R_y}$}
\psline[fillstyle=solid](85.79,30.21)(88.62,27.38)
\psline{|*-}(81.54,34.46)(78.72,37.28)
\pspolygon[](76.59,43.65)(80.84,39.41)(76.59,35.16)(72.34,39.4)
\rput(75.18,39.41){$\scr{R^{-1}_y}$}
\psline[fillstyle=solid]{|-}(71.64,44.36)(74.47,41.53)
\rput(68.82,54.25){$\scr{j_y}$}
\psline{|*-}(18.31,20.31)(21.14,23.14)
\rput(33.87,28.8){$\scr{j_x}$}
\rput{22.48}(30.34,32.33){\psellipse[](0,0)(3,-3)}
\rput(30.33,32.33){$\scr{h_{x}}$}
\psline[fillstyle=solid]{-|}(46.6,48.6)(43.77,45.77)
\pspolygon[](52.96,50.72)(48.72,46.48)(44.48,50.72)(48.72,54.96)
\psline[fillstyle=solid](57,59)(50.84,52.84)
\rput(48.72,52.13){$\scr{R_x}$}
\psline[fillstyle=solid](9.83,11.83)(7,9)
\pspolygon[](16.19,13.95)(11.95,9.71)(7.71,13.95)(11.95,18.19)
\psline[fillstyle=solid]{|*-}(16.9,18.9)(14.07,16.07)
\rput(11.95,15.36){$\scr{R^{-1}_x}$}
\pspolygon[](27.51,25.26)(23.26,21.02)(19.02,25.26)(23.27,29.5)
\rput(23.26,26.68){$\scr{R_x}$}
\psline[fillstyle=solid](28.21,30.21)(25.38,27.38)
\psline{|*-}(32.46,34.46)(35.28,37.28)
\pspolygon[](41.65,39.41)(37.41,35.16)(33.16,39.41)(37.4,43.66)
\rput(37.41,40.82){$\scr{R^{-1}_x}$}
\psline[fillstyle=solid]{|-}(42.36,44.36)(39.53,41.53)
\rput(52.25,47.18){$\scr{j_x}$}
\rput(15.49,10.41){$\scr{j_x}$}
\psline{|*-}(57,83)(57,79)
\rput(50,75){$\scr{j_z}$}
\rput{0}(57,76){\psellipse[](0,0)(3,-3)}
\rput(57,76){$\scr{h_{z}}$}
\psline[fillstyle=solid]{-|}(57,59)(57,67)
\psline[fillstyle=solid]{|*-}(57,85)(57,89)
\psline{-|}(57,73)(57,69)
\rput(51,64){$\scr{j_z}$}
\rput(52,87){$\scr{j_z}$}
\psline[fillstyle=solid]{-|}(69,59)(69,67)
\psline[fillstyle=solid]{|*-}(69,85)(69,87)
\psline(69,87)(63,87)
\rput(71,64){$\scr{\mu'}$}
%\rput(51,74){$\scr{j_z}$}
\psline[fillstyle=solid]{-|}(63,59)(63,67)
\rput(65,64){$\scr{\mu}$}
\psline[fillstyle=solid]{|*-}(63,85)(63,87)
\rput(65,84){$\scr{\mu}$}
\rput(71,84){$\scr{\mu'}$}
\end{pspicture}
\ea
\ee

The next step is to compute the action of $h_{\alpha_{ij}}-h_{\alpha_{ji}}$ contained in \eqref{Hridotto}; to this aim we can use a recoupling identity (the loop trick, see \cite{Alesci:2013kpa}) namely:
\be
\label{eqn:looptrick}
h_{\alpha_{[ij]}}=(h_{\alpha_{ij}}-h_{\alpha_{ji}})= \sum_{\tilde{m}\in\left(2\mathbb{N}+1\right)}(-)^{2m}
\ba
\includegraphics[scale=0.8]{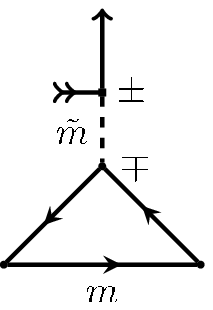}
\ea
\ee	
where all the lines are $SU(2)$ objects with the arrows in the upper part of the diagram representing a recoupling of the free indexes while the arrows in the loop represent the group element $h_{\alpha_{ij}}$, and the sum extends over all the odd value for $\tilde{m}$ compatible with recoupling theory.
Applying this identity, using the reduced recoupling and summing over $\mu$ and $\mu'$ we get
\be
\begin{split}
&\mathrm{Tr}\Big[\left({}^{R}\hat{h}^{(m)}_{\alpha_{xy}}-{}^{R}\hat{h}^{(m)}_{\alpha_{yx}}\right) {}^{R}\hat{h}^{(m)-1}_{s_{z}} {}^{R}\!\hat{V} {}^R\hat{h}^{(m)}_{s_{z}}\Big] |\vgraph^z\rangle_R=\\
&=
(8\pi\gamma l_P^2)^{3/2}\sum_{\tilde{m}}\sum_{\mu,\mu'} \sqrt{j_x\;j_y\;(j_z+\mu)} 
\ba
\ifx\JPicScale\undefined\def\JPicScale{0.7}\fi
\psset{unit=\JPicScale mm}
\psset{linewidth=0.3,dotsep=1,hatchwidth=0.3,hatchsep=1.5,shadowsize=1,dimen=middle}
\psset{dotsize=0.7 2.5,dotscale=1 1,fillcolor=black}
\psset{arrowsize=1 2,arrowlength=1,arrowinset=0.25,tbarsize=0.7 5,bracketlength=0.15,rbracketlength=0.15}
\begin{pspicture}(0,40)(95,143)
\psline{|*-}(19,78)(21.83,80.83)
\rput(17,81){$\scr{j_x}$}
\rput{45}(28.5,87.5){\psellipse[](0,0)(2.21,-2.12)}
\rput(24,91){$\scr{h_{x}}$}
\psline[fillstyle=solid]{-|}(42.83,101.83)(40,99)
\pspolygon[](46,103)(44,101)(42,103)(44,105)
\rput(40,104){$\scr{R_x}$}
\psline[fillstyle=solid]{|*-}(16.83,75.83)(14,73)
\rput(10,74){$\scr{R^{-1}_x}$}
\rput(21,85){$\scr{R_x}$}
\psline[fillstyle=solid](26.83,85.83)(24,83)
\psline(30,89)(32.82,91.82)
\pspolygon[](36,93)(34,91)(32,93)(34,95)
\rput(32,97){$\scr{R^{-1}_x}$}
\psline[fillstyle=solid]{|-}(38,97)(35.17,94.17)
\rput(43,108){$\scr{j_x}$}
\rput(13,77){$\scr{j_x}$}
\psline{|*-}(51,137)(51,133)
\rput(44,129){$\scr{j_z}$}
\rput{0}(51,130){\psellipse[](0,0)(3,-3)}
\rput(51,130){$\scr{h_{z}}$}
\psline[fillstyle=solid]{-|}(51,110)(51,120)
\psline[fillstyle=solid]{|*-}(51,139)(51,143)
\psline{-|}(51,127)(51,123)
\rput(46,141){$\scr{j_z}$}
\pspolygon[](25,82)(23,80)(21,82)(23,84)
\psline[fillstyle=solid](12,71)(9.17,68.17)
\pspolygon[](15,72)(13,70)(11,72)(13,74)
\psline{|*-}(83,78)(80.17,80.83)
\rput(84,82){$\scr{j_y}$}
\rput(76,91){$\scr{h_{y}}$}
\psline[fillstyle=solid]{-|}(59.17,101.83)(62,99)
\rput(62,104){$\scr{R_y}$}
\psline[fillstyle=solid]{|*-}(85.17,75.83)(88,73)
\rput(95,73){$\scr{R^{-1}_y}$}
\rput(79,86){$\scr{R_y}$}
\psline[fillstyle=solid](75.17,85.83)(78,83)
\psline(72,89)(69.18,91.82)
\pspolygon[](68,95)(70,93)(68,91)(66,93)
\rput(72,96){$\scr{R^{-1}_y}$}
\psline[fillstyle=solid]{|-}(64,97)(66.83,94.17)
\rput(57,107){$\scr{j_y}$}
\rput(90,77){$\scr{j_y}$}
\pspolygon[](79,84)(81,82)(79,80)(77,82)
\psline[fillstyle=solid](90,71)(92.83,68.17)
\pspolygon[](89,74)(91,72)(89,70)(87,72)
\rput{43.83}(73.44,87.44){\psellipse[](0,0)(2.21,-2.12)}
\psline{|*-}(42,93)(39.17,90.17)
\rput(26,74){$\scr{m}$}
\rput{45}(32.5,83.5){\psellipse[](0,0)(2.21,-2.12)}
\rput(37,84){$\scr{h_{x}}$}
\psline[fillstyle=solid]{-|}(18.17,69.17)(21,72)
\pspolygon[](15,68)(17,70)(19,68)(17,66)
\psline[fillstyle=solid](14,65)(16.16,67.16)
\rput(21,67){$\scr{R^{-1}_x}$}
\psline[fillstyle=solid]{|*-}(44.17,95.17)(47,98)
\rput(48,94){$\scr{R_x}$}
\rput(42,88){$\scr{R^{-1}_x}$}
\psline[fillstyle=solid](34.17,85.17)(37,88)
\psline(31,82)(28.18,79.18)
\pspolygon[](25,78)(27,80)(29,78)(27,76)
\rput(31,78){$\scr{R_x}$}
\psline[fillstyle=solid]{|-}(23,74)(25.83,76.83)
\rput(18,63){$\scr{m}$}
\rput(51,104){$\scr{m}$}
\pspolygon[](36,89)(38,91)(40,89)(38,87)
\psline[fillstyle=solid](49,100)(53,104)
\pspolygon[](46,99)(48,101)(50,99)(48,97)
\psline{|*-}(60,93)(62.83,90.17)
\rput(74,74){$\scr{m}$}
\rput(63,82){$\scr{h^{-1}_{y}}$}
\psline[fillstyle=solid]{-|}(83.83,69.17)(81,72)
\pspolygon[](85,66)(83,68)(85,70)(87,68)
\psline[fillstyle=solid](89,64)(85.84,67.16)
\rput(81,67){$\scr{R_y}$}
\psline[fillstyle=solid]{|*-}(57.83,95.17)(55,98)
\rput(55,94){$\scr{R_y^{-1}}$}
\rput(60,88){$\scr{R_y}$}
\psline[fillstyle=solid](67.83,85.17)(65,88)
\psline(71,82)(73.82,79.18)
\pspolygon[](75,76)(73,78)(75,80)(77,78)
\rput(70,77){$\scr{R^{-1}_y}$}
\psline[fillstyle=solid]{|-}(79,74)(76.17,76.83)
\rput(86,64){$\scr{j_y}$}
\rput(60,111){$\scr{\tilde{m}}$}
\pspolygon[](64,87)(62,89)(64,91)(66,89)
\pspolygon[](54,97)(52,99)(54,101)(56,99)
\rput{45}(69.56,83.56){\psellipse[](0,0)(2.21,-2.12)}
\psline[fillstyle=solid](53,104)(61,115)
\psline[fillstyle=solid](54,101)(53,104)
\rput(49,112){$\scr{j_z}$}
\rput(47,120){$\scr{j_z}$}
\psline[border=0.3,fillstyle=solid](51,110)(57,104)
\psline[fillstyle=solid](51,110)(44.84,103.84)
\pspolygon[](58,105)(60,103)(58,101)(56,103)
\psline[fillstyle=solid]{-|}(61,115)(58,119)
\psline[fillstyle=solid]{-|}(61,115)(64,119)
\rput(66,121){$\scr{\mu'}$}
\rput(57,120){$\scr{\mu}$}
\psline[fillstyle=solid]{|*-}(64,139)(64,141)
\psline(64,141)(58,141)
\psline[fillstyle=solid]{|*-}(58,139)(58,141)
\rput(61,143){$\scr{m}$}
\rput(58,137){$\scr{\mu}$}
\rput(64,137){$\scr{\mu'}$}
\rput(64,116){$\scr{m}$}
\rput(57,116){$\scr{m}$}
\end{pspicture}
\ea
\end{split}
\label{azione fondam}
\ee
where $\tilde{m}$ is odd.
In this graphical formula we recognize a $\delta_{\mu\mu'}$; this forces the magnetic indexes appearing in the formula to be equal, and this in turns imply that the line in representation $\tilde{m}$ has vanishing magnetic number. From the sum over magnetic indexes along the line $\tilde{m}$ we are then left with a single term

\be
\begin{split}
&\mathrm{Tr}\Big[{}^{R}\hat{h}^{(m)}_{\alpha_{[xy]}} {}^{R}\hat{h}^{(m)-1}_{s_{z}} {}^{R}\!\hat{V} {}^R\hat{h}^{(m)}_{s_{z}}\Big] |\vgraph^z\rangle_R=\\
&=
(8\pi\gamma l_P^2)^{3/2}\sum_{\tilde{m}}\sum_{\mu=\pm m} \sqrt{j_x\;j_y\;(j_z+\mu)}\; s(\mu)  C^{m\,m}_{m\,m \; \tilde{m}0}
\ba
\ifx\JPicScale\undefined\def\JPicScale{0.7}\fi
\psset{unit=\JPicScale mm}
\psset{linewidth=0.3,dotsep=1,hatchwidth=0.3,hatchsep=1.5,shadowsize=1,dimen=middle}
\psset{dotsize=0.7 2.5,dotscale=1 1,fillcolor=black}
\psset{arrowsize=1 2,arrowlength=1,arrowinset=0.25,tbarsize=0.7 5,bracketlength=0.15,rbracketlength=0.15}
\begin{pspicture}(0,50)(95,143)
\psline{|*-}(19,78)(21.83,80.83)
\rput(17,81){$\scr{j_x}$}
\rput{45}(28.5,87.5){\psellipse[](0,0)(2.21,-2.12)}
\rput(24,91){$\scr{h_{x}}$}
\psline[fillstyle=solid]{-|}(42.83,101.83)(40,99)
\pspolygon[](46,103)(44,101)(42,103)(44,105)
\rput(40,104){$\scr{R_x}$}
\psline[fillstyle=solid]{|*-}(16.83,75.83)(14,73)
\rput(10,74){$\scr{R^{-1}_x}$}
\rput(21,85){$\scr{R_x}$}
\psline[fillstyle=solid](26.83,85.83)(24,83)
\psline(30,89)(32.82,91.82)
\pspolygon[](36,93)(34,91)(32,93)(34,95)
\rput(32,97){$\scr{R^{-1}_x}$}
\psline[fillstyle=solid]{|-}(38,97)(35.17,94.17)
\rput(43,108){$\scr{j_x}$}
\rput(13,77){$\scr{j_x}$}
\psline{|*-}(51,137)(51,133)
\rput(44,129){$\scr{j_z}$}
\rput{0}(51,130){\psellipse[](0,0)(3,-3)}
\rput(51,130){$\scr{h_{z}}$}
\psline[fillstyle=solid]{-|}(51,110)(51,120)
\psline[fillstyle=solid]{|*-}(51,139)(51,143)
\psline{-|}(51,127)(51,123)
\rput(46,141){$\scr{j_z}$}
\pspolygon[](25,82)(23,80)(21,82)(23,84)
\psline[fillstyle=solid](12,71)(9.17,68.17)
\pspolygon[](15,72)(13,70)(11,72)(13,74)
\psline{|*-}(83,78)(80.17,80.83)
\rput(84,82){$\scr{j_y}$}
\rput(76,91){$\scr{h_{y}}$}
\psline[fillstyle=solid]{-|}(59.17,101.83)(62,99)
\rput(62,104){$\scr{R_y}$}
\psline[fillstyle=solid]{|*-}(85.17,75.83)(88,73)
\rput(95,73){$\scr{R^{-1}_y}$}
\rput(79,86){$\scr{R_y}$}
\psline[fillstyle=solid](75.17,85.83)(78,83)
\psline(72,89)(69.18,91.82)
\pspolygon[](68,95)(70,93)(68,91)(66,93)
\rput(72,96){$\scr{R^{-1}_y}$}
\psline[fillstyle=solid]{|-}(64,97)(66.83,94.17)
\rput(57,107){$\scr{j_y}$}
\rput(90,77){$\scr{j_y}$}
\pspolygon[](79,84)(81,82)(79,80)(77,82)
\psline[fillstyle=solid](90,71)(92.83,68.17)
\pspolygon[](89,74)(91,72)(89,70)(87,72)
\rput{43.83}(73.44,87.44){\psellipse[](0,0)(2.21,-2.12)}
\psline{|*-}(42,93)(39.17,90.17)
\rput(26,74){$\scr{m}$}
\rput{45}(32.5,83.5){\psellipse[](0,0)(2.21,-2.12)}
\rput(37,84){$\scr{h_{x}}$}
\psline[fillstyle=solid]{-|}(18.17,69.17)(21,72)
\pspolygon[](15,68)(17,70)(19,68)(17,66)
\psline[fillstyle=solid](14,65)(16.16,67.16)
\rput(21,67){$\scr{R^{-1}_x}$}
\psline[fillstyle=solid]{|*-}(44.17,95.17)(47,98)
\rput(48,94){$\scr{R_x}$}
\rput(42,88){$\scr{R^{-1}_x}$}
\psline[fillstyle=solid](34.17,85.17)(37,88)
\psline(31,82)(28.18,79.18)
\pspolygon[](25,78)(27,80)(29,78)(27,76)
\rput(31,78){$\scr{R_x}$}
\psline[fillstyle=solid]{|-}(23,74)(25.83,76.83)
\rput(18,63){$\scr{m}$}
\rput(51,104){$\scr{m}$}
\pspolygon[](36,89)(38,91)(40,89)(38,87)
\psline[fillstyle=solid](49,100)(53,104)
\pspolygon[](46,99)(48,101)(50,99)(48,97)
\psline{|*-}(60,93)(62.83,90.17)
\rput(74,74){$\scr{m}$}
\rput(63,82){$\scr{h^{-1}_{y}}$}
\psline[fillstyle=solid]{-|}(83.83,69.17)(81,72)
\pspolygon[](85,66)(83,68)(85,70)(87,68)
\psline[fillstyle=solid](89,64)(85.84,67.16)
\rput(81,67){$\scr{R_y}$}
\psline[fillstyle=solid]{|*-}(57.83,95.17)(55,98)
\rput(55,94){$\scr{R_y^{-1}}$}
\rput(60,88){$\scr{R_y}$}
\psline[fillstyle=solid](67.83,85.17)(65,88)
\psline(71,82)(73.82,79.18)
\pspolygon[](75,76)(73,78)(75,80)(77,78)
\rput(70,77){$\scr{R^{-1}_y}$}
\psline[fillstyle=solid]{|-}(79,74)(76.17,76.83)
\rput(86,64){$\scr{j_y}$}
\rput(60,111){$\scr{\tilde{m}}$}
\pspolygon[](64,87)(62,89)(64,91)(66,89)
\pspolygon[](54,97)(52,99)(54,101)(56,99)
\rput{45}(69.56,83.56){\psellipse[](0,0)(2.21,-2.12)}
\psline[fillstyle=solid](53,104)(61,115)
\psline[fillstyle=solid](54,101)(53,104)
\rput(49,112){$\scr{j_z}$}
\rput(47,120){$\scr{j_z}$}
\psline[border=0.3,fillstyle=solid](51,110)(57,104)
\psline[fillstyle=solid](51,110)(44.84,103.84)
\pspolygon[](58,105)(60,103)(58,101)(56,103)
\rput(62,116){$\scr{0}$}
\end{pspicture}
\ea
\end{split}
\ee
where  $s(\mu)$ is the sign function with argument $\mu$ and $C^{mm}_{mm \; \tilde{m}0}=2m!\sqrt{\frac{2m+1}{(2m-\tilde{m})! (2m+\tilde{m}+1)!}}$ is a Clebsh Gordan coefficient. The presence of the sign factor follows from the symmetry property of the Clebsh  $(-1)^{a+b-c}C^{c\gamma}_{a\alpha \; b\beta}=C^{c-\gamma}_{a-\alpha \; b-\beta}$ that in our case implies  $C^{mm}_{mm \; \tilde{m}0}=(-1)^{\tilde{m}}C^{m\,-m}_{m\,-m \; \tilde{m}0}=-C^{m\,-m}_{m\,-m \; \tilde{m}0} $ because $\tilde{m}$ is always an odd integer.

Recoupling the rotation matrices $R$ around the central node and multipling the $U(1)$ elements we get :
\be
\begin{split}
&\mathrm{Tr}\Big[{}^{R}\hat{h}^{(m)}_{\alpha_{[xy]}} {}^{R}\hat{h}^{(m)-1}_{s_{z}} {}^{R}\!\hat{V} {}^R\hat{h}^{(m)}_{s_{z}}\Big] |\vgraph^z\rangle_R
=\\
&=(8\pi\gamma l_P^2)^{3/2}\sum_{\mu_x,\mu_y=\pm m}\sum_{k}\sum_{\tilde{m}}\sum_{\mu=\pm m} \sqrt{j_x\;j_y\;(j_z+\mu)}\; s(\mu)  C^{mm}_{mm \; \tilde{m}0}
\ba
\ifx\JPicScale\undefined\def\JPicScale{0.7}\fi
\psset{unit=\JPicScale mm}
\psset{linewidth=0.3,dotsep=1,hatchwidth=0.3,hatchsep=1.5,shadowsize=1,dimen=middle}
\psset{dotsize=0.7 2.5,dotscale=1 1,fillcolor=black}
\psset{arrowsize=1 2,arrowlength=1,arrowinset=0.25,tbarsize=0.7 5,bracketlength=0.15,rbracketlength=0.15}
\begin{pspicture}(0,60)(97,143)
\psline{|*-}(19,78)(21.83,80.83)
\rput(13,81){$\scr{j_x+\mu_x}$}
\rput{45}(28.5,87.5){\psellipse[](0,0)(2.21,-2.12)}
\rput(24,91){$\scr{h_{x}}$}
\psline[fillstyle=solid]{-|}(42.83,101.83)(40,99)
\pspolygon[](46,103)(44,101)(42,103)(44,105)
\psline[fillstyle=solid](51,110)(44.84,103.84)
\rput(40,104){$\scr{R_x}$}
\psline[fillstyle=solid]{|*-}(16.83,75.83)(14,73)
\rput(10,74){$\scr{R^{-1}_x}$}
\rput(21,85){$\scr{R_x}$}
\psline[fillstyle=solid](26.83,85.83)(24,83)
\psline(30,89)(32.82,91.82)
\pspolygon[](36,93)(34,91)(32,93)(34,95)
\rput(32,97){$\scr{R^{-1}_x}$}
\psline[fillstyle=solid]{|-}(38,97)(35.17,94.17)
\rput(43,108){$\scr{j_x}$}
\rput(11,77){$\scr{j_x}$}
\psline{|*-}(51,137)(51,133)
\rput(44,129){$\scr{j_z}$}
\rput{0}(51,130){\psellipse[](0,0)(3,-3)}
\rput(51,130){$\scr{h_{z}}$}
\psline[fillstyle=solid]{-|}(51,110)(51,120)
\psline[fillstyle=solid]{|*-}(51,139)(51,143)
\psline{-|}(51,127)(51,123)
\rput(47,116){$\scr{k}$}
\rput(46,141){$\scr{j_z}$}
\psline(51,114)(53,114)
\psline(51,118)(59,118)
\pspolygon[](25,82)(23,80)(21,82)(23,84)
\psline[fillstyle=solid](12,71)(6,65)
\pspolygon[](15,72)(13,70)(11,72)(13,74)
\psline{|*-}(83,78)(80.17,80.83)
\rput(87,82){$\scr{j_y-\mu_y}$}
\rput(76,91){$\scr{h_{y}}$}
\psline[fillstyle=solid]{-|}(59.17,101.83)(62,99)
\pspolygon[](58,105)(60,103)(58,101)(56,103)
\rput(62,104){$\scr{R_y}$}
\psline[fillstyle=solid]{|*-}(85.17,75.83)(88,73)
\rput(95,73){$\scr{R^{-1}_y}$}
\rput(79,86){$\scr{R_y}$}
\psline[fillstyle=solid](75.17,85.83)(78,83)
\psline(72,89)(69.18,91.82)
\pspolygon[](68,95)(70,93)(68,91)(66,93)
\rput(72,96){$\scr{R^{-1}_y}$}
\psline[fillstyle=solid]{|-}(64,97)(66.83,94.17)
\rput(57,107){$\scr{j_y}$}
\rput(90,77){$\scr{j_y}$}
\pspolygon[](79,84)(81,82)(79,80)(77,82)
\psline[fillstyle=solid](90,71)(96,65)
\pspolygon[](89,74)(91,72)(89,70)(87,72)
\rput{43.83}(73.44,87.44){\psellipse[](0,0)(2.21,-2.12)}
\psline[fillstyle=solid](41,100)(47,98)
\rput(46,94){$\scr{R_x}$}
\rput(51,104){$\scr{m}$}
\psline[fillstyle=solid](49,100)(53,104)
\pspolygon[](46,99)(48,101)(50,99)(48,97)
\psline[fillstyle=solid](61,100)(55,98)
\rput(55,110){$\scr\tilde{{m}}$}
\pspolygon[](54,97)(52,99)(54,101)(56,99)
\psline[fillstyle=solid](53,104)(53,114)
\rput(49,112){$\scr{j_z}$}
\rput(47,120){$\scr{j_z}$}
\rput(33,100){$\scr{j_x+\mu_x}$}
\rput(69,101){$\scr{j_y-\mu_y}$}
\rput(97,67){$\scr{j_y}$}
\rput(5,68){$\scr{j_x}$}
\rput(56,94){$\scr{R^{-1}_y}$}
\psline[fillstyle=solid](54,101)(53,104)
\psline[border=0.3,fillstyle=solid](51,110)(57,104)
\rput(58,120){$\scr\tilde{{m}}$}
\rput(60,117){$\scr{0}$}
\psline[fillstyle=solid]{-|}(19.17,68.17)(22,71)
\pspolygon[](16,67)(18,69)(20,67)(18,65)
\psline[fillstyle=solid](15,64)(17.16,66.16)
\rput(22,66){$\scr{R^{-1}_x}$}
\rput(19,62){$\scr{m}$}
\rput(86,61){$\scr{m}$}
\psline[fillstyle=solid]{-|}(84.83,68.17)(82,71)
\pspolygon[](86,65)(84,67)(86,69)(88,67)
\psline[fillstyle=solid](90,63)(86.84,66.16)
\rput(82,66){$\scr{R_y}$}
\end{pspicture}
\ea
\end{split}
\ee 
with $k$ running from $|j_k-\tilde{m}|$ to $|j_k+\tilde{m}|$ and with $\mu_x=\pm m, \mu_y=\pm m$ being  the magnetic numbers of the reduced holonomy in representation $m$ attached by the hamiltonian in direction $l_x$ and $l_y$, respectively.
The central node can then be simplified obtaining a ${9j}$ symbol as showed in \cite{Alesci:2013kpa}, thus we get

\be
\begin{split}
&\mathrm{Tr}\Big[{}^{R}\hat{h}^{(m)}_{\alpha_{[xy]}} {}^{R}\hat{h}^{(m)-1}_{s_{z}} {}^{R}\!\hat{V} {}^R\hat{h}^{(m)}_{s_{z}}\Big] |\vgraph^z\rangle_R
=(8\pi\gamma l_P^2)^{3/2}\\
&
\sum_{\mu_x,\mu_y=\pm m}\sum_{k}\sum_{\tilde{m}}\sum_{\mu=\pm m} \sqrt{j_x\;j_y\;(j_z+\mu)}\; s(\mu)  C^{mm}_{mm \; \tilde{m}0} \tinyninej {k}{\tilde{m}}{j_z} {j_y-\mu_y} {m} {j_y}  {j_x+\mu_x} {m}  {j_x}
\ba
\ifx\JPicScale\undefined\def\JPicScale{0.7}\fi
\psset{unit=\JPicScale mm}
\psset{linewidth=0.3,dotsep=1,hatchwidth=0.3,hatchsep=1.5,shadowsize=1,dimen=middle}
\psset{dotsize=0.7 2.5,dotscale=1 1,fillcolor=black}
\psset{arrowsize=1 2,arrowlength=1,arrowinset=0.25,tbarsize=0.7 5,bracketlength=0.15,rbracketlength=0.15}
\begin{pspicture}(0,60)(97,143)
\psline{|*-}(19,78)(21.83,80.83)
\rput(13,81){$\scr{j_x+\mu_x}$}
\rput{45}(28.5,87.5){\psellipse[](0,0)(2.21,-2.12)}
\rput(24,91){$\scr{h_{x}}$}
\psline[fillstyle=solid]{-|}(42.83,101.83)(40,99)
\pspolygon[](46,103)(44,101)(42,103)(44,105)
\psline[fillstyle=solid](51,110)(44.84,103.84)
\rput(40,104){$\scr{R_x}$}
\psline[fillstyle=solid]{|*-}(16.83,75.83)(14,73)
\rput(10,74){$\scr{R^{-1}_x}$}
\rput(21,85){$\scr{R_x}$}
\psline[fillstyle=solid](26.83,85.83)(24,83)
\psline(30,89)(32.82,91.82)
\pspolygon[](36,93)(34,91)(32,93)(34,95)
\rput(32,97){$\scr{R^{-1}_x}$}
\psline[fillstyle=solid]{|-}(38,97)(35.17,94.17)
\rput(43,108){$\scr{j_x+\mu_x}$}
\rput(11,77){$\scr{j_x}$}
\psline{|*-}(51,137)(51,133)
\rput(44,129){$\scr{j_z}$}
\rput{0}(51,130){\psellipse[](0,0)(3,-3)}
\rput(51,130){$\scr{h_{z}}$}
\psline[fillstyle=solid]{-|}(51,110)(51,120)
\psline[fillstyle=solid]{|*-}(51,139)(51,143)
\psline{-|}(51,127)(51,123)
\rput(47,114){$\scr{k}$}
\rput(46,141){$\scr{j_z}$}
\psline(51,118)(59,118)
\pspolygon[](25,82)(23,80)(21,82)(23,84)
\psline[fillstyle=solid](12,71)(6,65)
\pspolygon[](15,72)(13,70)(11,72)(13,74)
\psline{|*-}(83,78)(80.17,80.83)
\rput(87,82){$\scr{j_y-\mu_y}$}
\rput(76,91){$\scr{h_{y}}$}
\psline[fillstyle=solid]{-|}(59.17,101.83)(62,99)
\pspolygon[](58,105)(60,103)(58,101)(56,103)
\rput(62,104){$\scr{R_y}$}
\psline[fillstyle=solid]{|*-}(85.17,75.83)(88,73)
\rput(95,73){$\scr{R^{-1}_y}$}
\rput(79,86){$\scr{R_y}$}
\psline[fillstyle=solid](75.17,85.83)(78,83)
\psline(72,89)(69.18,91.82)
\pspolygon[](68,95)(70,93)(68,91)(66,93)
\rput(72,96){$\scr{R^{-1}_y}$}
\psline[fillstyle=solid]{|-}(64,97)(66.83,94.17)
\rput(57,107){$\scr{j_y-\mu_y}$}
\rput(90,77){$\scr{j_y}$}
\pspolygon[](79,84)(81,82)(79,80)(77,82)
\psline[fillstyle=solid](90,71)(96,65)
\pspolygon[](89,74)(91,72)(89,70)(87,72)
\rput{43.83}(73.44,87.44){\psellipse[](0,0)(2.21,-2.12)}
\rput(47,120){$\scr{j_z}$}
\rput(35,100){$\scr{j_x+\mu_x}$}
\rput(67,101){$\scr{j_y-\mu_y}$}
\rput(97,67){$\scr{j_y}$}
\rput(5,68){$\scr{j_x}$}
\psline[border=0.3,fillstyle=solid](51,110)(57,104)
\rput(58,120){$\scr\tilde{{m}}$}
\rput(60,117){$\scr{0}$}
\psline[fillstyle=solid]{-|}(19.17,68.17)(22,71)
\pspolygon[](16,67)(18,69)(20,67)(18,65)
\psline[fillstyle=solid](15,64)(17.16,66.16)
\rput(22,66){$\scr{R^{-1}_x}$}
\rput(19,62){$\scr{m}$}
\rput(87,62){$\scr{m}$}
\psline[fillstyle=solid]{-|}(84.83,68.17)(82,71)
\pspolygon[](86,65)(84,67)(86,69)(88,67)
\psline[fillstyle=solid](90,63)(86.84,66.16)
\rput(82,66){$\scr{R_y}$}
\rput(24,72){$\scr{\mu_x}$}
\rput(79,72){$\scr{\mu_y}$}
\end{pspicture}
\ea
\end{split}
\ee

We can now turn our attention to the remaining nodes, where the loop attached by the Hamiltonian constraint overlaps the existing loop in the state and we have to evaluate the following diagram:

\be
\sum_{\mu'_x, \mu'_y}\quad
\ba
\ifx\JPicScale\undefined\def\JPicScale{0.7}\fi
\psset{unit=\JPicScale mm}
\psset{linewidth=0.3,dotsep=1,hatchwidth=0.3,hatchsep=1.5,shadowsize=1,dimen=middle}
\psset{dotsize=0.7 2.5,dotscale=1 1,fillcolor=black}
\psset{arrowsize=1 2,arrowlength=1,arrowinset=0.25,tbarsize=0.7 5,bracketlength=0.15,rbracketlength=0.15}
\begin{pspicture}(0,0)(104,143)
\psline{|*-}(19,78)(21.83,80.83)
\rput(14,81){$\scr{j_x+\mu_x}$}
\rput{45}(28.5,87.5){\psellipse[](0,0)(2.21,-2.12)}
\rput(24,91){$\scr{h_{x}}$}
\psline[fillstyle=solid]{-|}(42.83,101.83)(40,99)
\pspolygon[](46,103)(44,101)(42,103)(44,105)
\psline[fillstyle=solid](51,110)(44.84,103.84)
\rput(40,104){$\scr{R_x}$}
\psline[fillstyle=solid]{|*-}(16.83,75.83)(14,73)
\rput(10,74){$\scr{R^{-1}_x}$}
\rput(21,85){$\scr{R_x}$}
\psline[fillstyle=solid](26.83,85.83)(24,83)
\psline(30,89)(32.82,91.82)
\pspolygon[](36,93)(34,91)(32,93)(34,95)
\rput(32,97){$\scr{R^{-1}_x}$}
\psline[fillstyle=solid]{|-}(38,97)(35.17,94.17)
\rput(11,77){$\scr{j_x+\mu_x}$}
\psline{|*-}(51,137)(51,133)
\rput(44,129){$\scr{j_z}$}
\rput{0}(51,130){\psellipse[](0,0)(3,-3)}
\rput(51,130){$\scr{h_{z}}$}
\psline[fillstyle=solid]{-|}(51,110)(51,120)
\psline[fillstyle=solid]{|*-}(51,139)(51,143)
\psline{-|}(51,127)(51,123)
\rput(46,141){$\scr{j_z}$}
\pspolygon[](25,82)(23,80)(21,82)(23,84)
\psline[fillstyle=solid](12,71)(-2,57)
\pspolygon[](15,72)(13,70)(11,72)(13,74)
\psline{|*-}(83,78)(80.17,80.83)
\rput(84,82){$\scr{j_y-\mu_y}$}
\rput(76,91){$\scr{h_{y}}$}
\psline[fillstyle=solid]{-|}(59.17,101.83)(62,99)
\pspolygon[](58,105)(60,103)(58,101)(56,103)
\rput(62,104){$\scr{R_y}$}
\psline[fillstyle=solid]{|*-}(85.17,75.83)(88,73)
\rput(95,73){$\scr{R^{-1}_y}$}
\rput(79,86){$\scr{R_y}$}
\psline[fillstyle=solid](75.17,85.83)(78,83)
\psline(72,89)(69.18,91.82)
\pspolygon[](68,95)(70,93)(68,91)(66,93)
\rput(72,96){$\scr{R^{-1}_y}$}
\psline[fillstyle=solid]{|-}(64,97)(66.83,94.17)
\rput(90,77){$\scr{j_y}$}
\pspolygon[](79,84)(81,82)(79,80)(77,82)
\psline[fillstyle=solid](90,71)(104,57)
\pspolygon[](89,74)(91,72)(89,70)(87,72)
\rput{43.83}(73.44,87.44){\psellipse[](0,0)(2.21,-2.12)}
\rput(11,46){$\scr{j_l}$}
\rput(93,50){$\scr{j_l}$}
\psline[border=0.3,fillstyle=solid](51,110)(57.16,103.84)
\rput(47,120){$\scr{j_z}$}
\psline{|*-}(42,26)(39.17,28.83)
\rput(36,25){$\scr{j_l}$}
\rput(26,34){$\scr{h_{l_y}}$}
\psline[fillstyle=solid]{-|}(18.17,49.83)(21,47)
\pspolygon[](17,53)(19,51)(17,49)(15,51)
\rput(9,50){$\scr{R_y}$}
\psline[fillstyle=solid]{|*-}(44.17,23.83)(47,21)
\rput(44,17){$\scr{R^{-1}_y}$}
\rput(33,29){$\scr{R_y}$}
\psline[fillstyle=solid](34.17,33.83)(37,31)
\psline(31,37)(28.18,39.82)
\pspolygon[](27,43)(29,41)(27,39)(25,41)
\rput(20,39){$\scr{R^{-1}_y}$}
\psline[fillstyle=solid]{|-}(23,45)(25.83,42.17)
\rput(52,12){$\scr{j_l}$}
\pspolygon[](38,32)(40,30)(38,28)(36,30)
\psline[fillstyle=solid](49,19)(51.83,16.17)
\pspolygon[](48,22)(50,20)(48,18)(46,20)
\rput{43.83}(32.44,35.44){\psellipse[](0,0)(2.21,-2.12)}
\psline[fillstyle=solid](4,63)(16.16,51.84)
\psline{|*-}(80,44)(77.17,41.17)
\rput(64,25){$\scr{j_l}$}
\rput{45}(70.5,34.5){\psellipse[](0,0)(2.21,-2.12)}
\rput(78,33){$\scr{h^{-1}_{l_x}}$}
\psline[fillstyle=solid]{-|}(56.17,20.17)(59,23)
\pspolygon[](53,19)(55,21)(57,19)(55,17)
\psline[fillstyle=solid](52,16)(54.16,18.16)
\rput(58,15){$\scr{R_x}$}
\psline[fillstyle=solid]{|*-}(82.17,46.17)(85,49)
\rput(88,45){$\scr{R^{-1}_x}$}
\rput(80,39){$\scr{R_x}$}
\psline[fillstyle=solid](72.17,36.17)(75,39)
\psline(69,33)(66.18,30.18)
\pspolygon[](63,29)(65,31)(67,29)(65,27)
\rput(71,27){$\scr{R^{-1}_x}$}
\psline[fillstyle=solid]{|-}(61,25)(63.83,27.83)
\pspolygon[](74,40)(76,42)(78,40)(76,38)
\psline[fillstyle=solid](87,51)(99,62)
\pspolygon[](84,50)(86,52)(88,50)(86,48)
\rput(39,108){$\scr{j_x}+\mu_x$}
\rput(61,108){$\scr{j_y-\mu_y}$}
\rput(104,62){$\scr{j'_y}$}
\rput(-2,62){$\scr{j'_x}$}
\psline[fillstyle=solid]{|*-}(18.83,70.83)(16,68)
\rput(21,67){$\scr{R^{-1}_x}$}
\rput(25,72){$\scr{\mu_x}$}
\psline[fillstyle=solid](14,66)(13,65)
\pspolygon[](17,67)(15,65)(13,67)(15,69)
\psline[fillstyle=solid]{|*-}(83.17,70.83)(86,68)
\rput(80,68){$\scr{R_y}$}
\rput(78,72){$\scr{\mu_y}$}
\psline[fillstyle=solid](88,66)(90,64)
\pspolygon[](87,69)(89,67)(87,65)(85,67)
\rput(24,59){$\scr{\mu'_y}$}
\rput(77,57){$\scr{\mu'_x}$}
\psline{|*-}(42,36)(39.17,38.83)
\rput(45,38){$\scr{\mu'_y}$}
\rput(38,49){$\scr{h_{l_y}}$}
\psline[fillstyle=solid]{-|}(18.17,59.83)(21,57)
\pspolygon[](17,63)(19,61)(17,59)(15,61)
\rput(21,62){$\scr{R_y}$}
\psline[fillstyle=solid]{|*-}(44.17,33.83)(47,31)
\rput(49,34){$\scr{R^{-1}_y}$}
\rput(38,44){$\scr{R_y}$}
\psline[fillstyle=solid](34.17,43.83)(37,41)
\psline(31,47)(28.18,49.82)
\pspolygon[](27,53)(29,51)(27,49)(25,51)
\rput(31,54){$\scr{R^{-1}_y}$}
\psline[fillstyle=solid]{|-}(23,55)(25.83,52.17)
\pspolygon[](38,42)(40,40)(38,38)(36,40)
\psline[fillstyle=solid](49,29)(51.83,26.17)
\pspolygon[](48,32)(50,30)(48,28)(46,30)
\rput{43.83}(32.44,45.44){\psellipse[](0,0)(2.21,-2.12)}
\psline[fillstyle=solid](13,65)(16.16,61.84)
\psline{|*-}(80,54)(77.17,51.17)
\rput(56,37){$\scr{\mu'_x}$}
\rput{45}(70.5,44.5){\psellipse[](0,0)(2.21,-2.12)}
\rput(64,48){$\scr{h^{-1}_{l_x}}$}
\psline[fillstyle=solid]{-|}(56.17,30.17)(59,33)
\pspolygon[](53,29)(55,31)(57,29)(55,27)
\psline[fillstyle=solid](52,26)(54.16,28.16)
\rput(53,34){$\scr{R_x}$}
\psline[fillstyle=solid]{|*-}(82.17,56.17)(85,59)
\rput(80,62){$\scr{R^{-1}_x}$}
\rput(69,53){$\scr{R_x}$}
\psline[fillstyle=solid](72.17,46.17)(75,49)
\psline(69,43)(66.18,40.18)
\pspolygon[](63,39)(65,41)(67,39)(65,37)
\rput(61,43){$\scr{R^{-1}_x}$}
\psline[fillstyle=solid]{|-}(61,35)(63.83,37.83)
\pspolygon[](74,50)(76,52)(78,50)(76,48)
\psline[fillstyle=solid](87,61)(90,64)
\pspolygon[](84,60)(86,62)(88,60)(86,58)
\psline(51,117)(60,117)
\rput(59,119){$\scr\tilde{{m}}$}
\rput(61,116){$\scr{0}$}
\rput(47,114){$\scr{k}$}
\end{pspicture}
\ea
\ee

Recoupling at the nodes produces just a $6j$ symbol per node and we finally find

\be
\begin{split}
&\mathrm{Tr}\Big[{}^{R}\hat{h}^{(m)}_{\alpha_{[xy]}} {}^{R}\hat{h}^{(m)-1}_{s_{z}} {}^{R}\!\hat{V} {}^R\hat{h}^{(m)}_{s_{z}}\Big] |\vgraph^z\rangle_R
= \sum_{\mu'_x \mu'_y\mu_x,\mu_y=\pm m} H^{m\;\,j_xj'_xj_yj'_y}_{\mu_x\mu'_x\mu_y\mu'_y}(j_z,j_l)
 \ba
\ifx\JPicScale\undefined\def\JPicScale{0.7}\fi
\psset{unit=\JPicScale mm}
\psset{linewidth=0.3,dotsep=1,hatchwidth=0.3,hatchsep=1.5,shadowsize=1,dimen=middle}
\psset{dotsize=0.7 2.5,dotscale=1 1,fillcolor=black}
\psset{arrowsize=1 2,arrowlength=1,arrowinset=0.25,tbarsize=0.7 5,bracketlength=0.15,rbracketlength=0.15}
\begin{pspicture}(0,0)(104,143)
\psline{|*-}(19,78)(21.83,80.83)
\rput(14,78){$\scr{j_x+\mu_x}$}
\rput{45}(28.5,87.5){\psellipse[](0,0)(2.21,-2.12)}
\rput(24,91){$\scr{h_{x}}$}
\psline[fillstyle=solid]{-|}(42.83,101.83)(40,99)
\pspolygon[](46,103)(44,101)(42,103)(44,105)
\psline[fillstyle=solid](51,110)(44.84,103.84)
\rput(40,104){$\scr{R_x}$}
\psline[fillstyle=solid]{|*-}(16.83,75.83)(14,73)
\rput(10,74){$\scr{R^{-1}_x}$}
\rput(21,85){$\scr{R_x}$}
\psline[fillstyle=solid](26.83,85.83)(24,83)
\psline(30,89)(32.82,91.82)
\pspolygon[](36,93)(34,91)(32,93)(34,95)
\rput(32,97){$\scr{R^{-1}_x}$}
\psline[fillstyle=solid]{|-}(38,97)(35.17,94.17)
\psline{|*-}(51,137)(51,133)
\rput(44,129){$\scr{j_z}$}
\rput{0}(51,130){\psellipse[](0,0)(3,-3)}
\rput(51,130){$\scr{h_{z}}$}
\psline[fillstyle=solid]{-|}(51,110)(51,120)
\psline[fillstyle=solid]{|*-}(51,139)(51,143)
\psline{-|}(51,127)(51,123)
\rput(46,141){$\scr{j_z}$}
\pspolygon[](25,82)(23,80)(21,82)(23,84)
\psline[fillstyle=solid](12,71)(-2,57)
\pspolygon[](15,72)(13,70)(11,72)(13,74)
\psline{|*-}(83,78)(80.17,80.83)
\rput(76,91){$\scr{h_{y}}$}
\psline[fillstyle=solid]{-|}(59.17,101.83)(62,99)
\pspolygon[](58,105)(60,103)(58,101)(56,103)
\rput(62,104){$\scr{R_y}$}
\psline[fillstyle=solid]{|*-}(85.17,75.83)(88,73)
\rput(95,73){$\scr{R^{-1}_y}$}
\rput(79,86){$\scr{R_y}$}
\psline[fillstyle=solid](75.17,85.83)(78,83)
\psline(72,89)(69.18,91.82)
\pspolygon[](68,95)(70,93)(68,91)(66,93)
\rput(72,96){$\scr{R^{-1}_y}$}
\psline[fillstyle=solid]{|-}(64,97)(66.83,94.17)
\rput(90,79){$\scr{j_y-\mu_y}$}
\pspolygon[](79,84)(81,82)(79,80)(77,82)
\psline[fillstyle=solid](90,71)(104,57)
\pspolygon[](89,74)(91,72)(89,70)(87,72)
\rput{43.83}(73.44,87.44){\psellipse[](0,0)(2.21,-2.12)}
\rput(14,44){$\scr{j_l+\mu'_y}$}
\psline[border=0.3,fillstyle=solid](51,110)(57.16,103.84)
\rput(47,120){$\scr{j_z}$}
\psline{|*-}(42,26)(39.17,28.83)
\rput(38,23){$\scr{j_l+\mu'_y}$}
\rput(26,34){$\scr{h_{l_y}}$}
\psline[fillstyle=solid]{-|}(18.17,49.83)(21,47)
\pspolygon[](17,53)(19,51)(17,49)(15,51)
\rput(9,50){$\scr{R_y}$}
\psline[fillstyle=solid]{|*-}(44.17,23.83)(47,21)
\rput(44,17){$\scr{R^{-1}_y}$}
\rput(33,29){$\scr{R_y}$}
\psline[fillstyle=solid](34.17,33.83)(37,31)
\psline(31,37)(28.18,39.82)
\pspolygon[](27,43)(29,41)(27,39)(25,41)
\rput(20,39){$\scr{R^{-1}_y}$}
\psline[fillstyle=solid]{|-}(23,45)(25.83,42.17)
\pspolygon[](38,32)(40,30)(38,28)(36,30)
\psline[fillstyle=solid](49,19)(51.83,16.17)
\pspolygon[](48,22)(50,20)(48,18)(46,20)
\rput{43.83}(32.44,35.44){\psellipse[](0,0)(2.21,-2.12)}
\psline[fillstyle=solid](4,63)(16.16,51.84)
\psline{|*-}(80,44)(77.17,41.17)
\rput(66,22){$\scr{j_l+\mu'_x}$}
\rput{45}(70.5,34.5){\psellipse[](0,0)(2.21,-2.12)}
\rput(78,33){$\scr{h^{-1}_{l_x}}$}
\psline[fillstyle=solid]{-|}(56.17,20.17)(59,23)
\pspolygon[](53,19)(55,21)(57,19)(55,17)
\psline[fillstyle=solid](52,16)(54.16,18.16)
\rput(58,15){$\scr{R_x}$}
\psline[fillstyle=solid]{|*-}(82.17,46.17)(85,49)
\rput(88,45){$\scr{R^{-1}_x}$}
\rput(80,39){$\scr{R_x}$}
\psline[fillstyle=solid](72.17,36.17)(75,39)
\psline(69,33)(66.18,30.18)
\pspolygon[](63,29)(65,31)(67,29)(65,27)
\rput(71,27){$\scr{R^{-1}_x}$}
\psline[fillstyle=solid]{|-}(61,25)(63.83,27.83)
\pspolygon[](74,40)(76,42)(78,40)(76,38)
\psline[fillstyle=solid](87,51)(99,62)
\pspolygon[](84,50)(86,52)(88,50)(86,48)
\rput(39,108){$\scr{j_x}+\mu_x$}
\rput(61,108){$\scr{j_y-\mu_y}$}
\rput(104,62){$\scr{j'_y}$}
\rput(-2,62){$\scr{j'_x}$}
\rput(87,42){$\scr{j_l+\mu'_x}$}
\psline(51,116)(59,116)
\rput(58,118){$\scr\tilde{{m}}$}
\rput(60,115){$\scr{0}$}
\rput(47,114){$\scr{k}$}
\end{pspicture}
\ea
\end{split}
\ee
with 
\be
H^{m\;\,j_xj'_xj_yj'_y}_{\mu_x\mu'_x\mu_y\mu'_y}(j_z,j_l)=(8\pi\gamma l_P^2)^{3/2}\sum_{k}\sum_{\tilde{m}}\sum_{\mu=\pm m} \sqrt{j_x\;j_y\;(j_z+\mu)}\; s(\mu)  C^{mm}_{mm \; \tilde{m}0} \tinyninej {k}{\tilde{m}}{j_z} {j_y-\mu_y} {m} {j_y}  {j_x+\mu_x} {m}  {j_x} \tinysixj{j_x'}{j_l+\mu'_y}{j_x+\mu_x} {m}{j_x}{j_l}\tinysixj{j_l+\mu'_x}{j_y'}{j_y-\mu_y}{j_y}{m}{j_l}
\label{coefficienti H}
\ee
This is the final form of the hamiltonian action.

\section{Expectation value of the Hamiltonian on coherent states}\label{Scal}
We focus on the action of the Hamiltonian on a coherent state $|\Psi_{H} \; \vgraph^z\rangle$ based on the simple graph on which we computed the action of the hamiltonian in the previous section:

\be
|\Psi_{H} \; \vgraph^z\rangle=\sum_{j_x,j_y,j_z, j_l}
\ba
\ifx\JPicScale\undefined\def\JPicScale{0.6}\fi
\psset{unit=\JPicScale mm}
\psset{linewidth=0.3,dotsep=1,hatchwidth=0.3,hatchsep=1.5,shadowsize=1,dimen=middle}
\psset{dotsize=0.7 2.5,dotscale=1 1,fillcolor=black}
\psset{arrowsize=1 2,arrowlength=1,arrowinset=0.25,tbarsize=0.7 5,bracketlength=0.15,rbracketlength=0.15}
\begin{pspicture}(0,0)(101,143)
\rput(28,87){$\Psi_{H_{l_x}}(j_x)$}
\psline[fillstyle=solid]{-|}(42.83,101.83)(40,99)
\pspolygon[](46,103)(44,101)(42,103)(44,105)
\psline[fillstyle=solid](51,110)(44.84,103.84)
\rput(40,104){$\scr{R_x}$}
\psline[fillstyle=solid]{|*-}(16.83,75.83)(14,73)
\rput(10,74){$\scr{R^{-1}_x}$}
\rput(11,77){$\scr{j_x}$}
\psline[fillstyle=solid]{-|}(51,110)(51,120)
\psline[fillstyle=solid]{|*-}(51,139)(51,143)
\rput(46,141){$\scr{j_z}$}
\psline[fillstyle=solid](12,71)(6,65)
\pspolygon[](15,72)(13,70)(11,72)(13,74)
\psline[fillstyle=solid]{-|}(59.17,101.83)(62,99)
\pspolygon[](58,105)(60,103)(58,101)(56,103)
\rput(62,104){$\scr{R_y}$}
\psline[fillstyle=solid]{|*-}(85.17,75.83)(88,73)
\rput(95,73){$\scr{R^{-1}_y}$}
\rput(90,77){$\scr{j_y}$}
\psline[fillstyle=solid](90,71)(96,65)
\pspolygon[](89,74)(91,72)(89,70)(87,72)
\rput(14,60){$\scr{j_l}$}
\rput(80,64){$\scr{j_l}$}
\psline[border=0.3,fillstyle=solid](51,110)(57.16,103.84)
\rput(47,120){$\scr{j_z}$}
\psline[fillstyle=solid]{-|}(18.17,62.83)(21,60)
\pspolygon[](17,66)(19,64)(17,62)(15,64)
\rput(21,65){$\scr{R_y}$}
\psline[fillstyle=solid]{|*-}(44.17,36.83)(47,34)
\rput(44,30){$\scr{R^{-1}_y}$}
\rput(49,38){$\scr{j_l}$}
\psline[fillstyle=solid](49,32)(51.83,29.17)
\pspolygon[](48,35)(50,33)(48,31)(46,33)
\psline[fillstyle=solid](11,70)(16.16,64.84)
\psline[fillstyle=solid]{-|}(56.17,33.17)(59,36)
\pspolygon[](53,32)(55,34)(57,32)(55,30)
\psline[fillstyle=solid](52,29)(54.16,31.16)
\rput(58,28){$\scr{R_x}$}
\psline[fillstyle=solid]{|*-}(82.17,59.17)(85,62)
\psline[fillstyle=solid](87,64)(92,69)
\pspolygon[](84,63)(86,65)(88,63)(86,61)
\rput(41,108){$\scr{j_x}$}
\rput(58,108){$\scr{j_y}$}
\rput(101,67){$\scr{j'_y}$}
\rput(3,68){$\scr{j'_x}$}
\rput(73,87){$\Psi_{H_{l_y}}(j_y)$}
\rput(33,48){$\Psi_{H_{l_{l_{y}}}}(j_l)$}
\rput(71,48){$\Psi_{H_{l_{l_{x}}}}(j_l)$}
\rput(51,130){$\Psi_{H_{l_z}}(j_z)$}
\rput(54,113){{\large{*}}}
\rput(84,68){{\large{*}}}
\rput(20,70){{\large{*}}}
\rput(52,25){{\large{*}}}
\end{pspicture}

\ea
\quad
\Bigg|
\ba
\ifx\JPicScale\undefined\def\JPicScale{0.55}\fi
\psset{unit=\JPicScale mm}
\psset{linewidth=0.3,dotsep=1,hatchwidth=0.3,hatchsep=1.5,shadowsize=1,dimen=middle}
\psset{dotsize=0.7 2.5,dotscale=1 1,fillcolor=black}
\psset{arrowsize=1 2,arrowlength=1,arrowinset=0.25,tbarsize=0.7 5,bracketlength=0.15,rbracketlength=0.15}
\begin{pspicture}(0,0)(101,143)
\psline{|*-}(19,78)(21.83,80.83)
\rput(17,81){$\scr{j_x}$}
\rput{45}(28.5,87.5){\psellipse[](0,0)(2.21,-2.12)}
\rput(24,91){$\scr{h_{x}}$}
\psline[fillstyle=solid]{-|}(42.83,101.83)(40,99)
\pspolygon[](46,103)(44,101)(42,103)(44,105)
\psline[fillstyle=solid](51,110)(44.84,103.84)
\rput(40,104){$\scr{R_x}$}
\psline[fillstyle=solid]{|*-}(16.83,75.83)(14,73)
\rput(10,74){$\scr{R^{-1}_x}$}
\rput(21,85){$\scr{R_x}$}
\psline[fillstyle=solid](26.83,85.83)(24,83)
\psline(30,89)(32.82,91.82)
\pspolygon[](36,93)(34,91)(32,93)(34,95)
\rput(32,97){$\scr{R^{-1}_x}$}
\psline[fillstyle=solid]{|-}(38,97)(35.17,94.17)
\rput(11,77){$\scr{j_x}$}
\psline{|*-}(51,137)(51,133)
\rput(44,129){$\scr{j_z}$}
\rput{0}(51,130){\psellipse[](0,0)(3,-3)}
\rput(51,130){$\scr{h_{z}}$}
\psline[fillstyle=solid]{-|}(51,110)(51,120)
\psline[fillstyle=solid]{|*-}(51,139)(51,143)
\psline{-|}(51,127)(51,123)
\rput(46,141){$\scr{j_z}$}
\pspolygon[](25,82)(23,80)(21,82)(23,84)
\psline[fillstyle=solid](12,71)(6,65)
\pspolygon[](15,72)(13,70)(11,72)(13,74)
\psline{|*-}(83,78)(80.17,80.83)
\rput(84,82){$\scr{j_y}$}
\rput(76,91){$\scr{h_{y}}$}
\psline[fillstyle=solid]{-|}(59.17,101.83)(62,99)
\pspolygon[](58,105)(60,103)(58,101)(56,103)
\rput(62,104){$\scr{R_y}$}
\psline[fillstyle=solid]{|*-}(85.17,75.83)(88,73)
\rput(95,73){$\scr{R^{-1}_y}$}
\rput(79,86){$\scr{R_y}$}
\psline[fillstyle=solid](75.17,85.83)(78,83)
\psline(72,89)(69.18,91.82)
\pspolygon[](68,95)(70,93)(68,91)(66,93)
\rput(72,96){$\scr{R^{-1}_y}$}
\psline[fillstyle=solid]{|-}(64,97)(66.83,94.17)
\rput(90,77){$\scr{j_y}$}
\pspolygon[](79,84)(81,82)(79,80)(77,82)
\psline[fillstyle=solid](90,71)(96,65)
\pspolygon[](89,74)(91,72)(89,70)(87,72)
\rput{43.83}(73.44,87.44){\psellipse[](0,0)(2.21,-2.12)}
\rput(14,60){$\scr{j_l}$}
\rput(80,64){$\scr{j_l}$}
\psline[border=0.3,fillstyle=solid](51,110)(57.16,103.84)
\rput(47,120){$\scr{j_z}$}
\psline{|*-}(42,39)(39.17,41.83)
\rput(43,43){$\scr{j_l}$}
\rput(35,52){$\scr{h_{l_y}}$}
\psline[fillstyle=solid]{-|}(18.17,62.83)(21,60)
\pspolygon[](17,66)(19,64)(17,62)(15,64)
\rput(21,65){$\scr{R_y}$}
\psline[fillstyle=solid]{|*-}(44.17,36.83)(47,34)
\rput(44,30){$\scr{R^{-1}_y}$}
\rput(38,47){$\scr{R_y}$}
\psline[fillstyle=solid](34.17,46.83)(37,44)
\psline(31,50)(28.18,52.82)
\pspolygon[](27,56)(29,54)(27,52)(25,54)
\rput(31,57){$\scr{R^{-1}_y}$}
\psline[fillstyle=solid]{|-}(23,58)(25.83,55.17)
\rput(49,38){$\scr{j_l}$}
\pspolygon[](38,45)(40,43)(38,41)(36,43)
\psline[fillstyle=solid](49,32)(51.83,29.17)
\pspolygon[](48,35)(50,33)(48,31)(46,33)
\rput{43.83}(32.44,48.44){\psellipse[](0,0)(2.21,-2.12)}
\psline[fillstyle=solid](11,70)(16.16,64.84)
\psline{|*-}(80,57)(77.17,54.17)
\rput(64,38){$\scr{j_l}$}
\rput{45}(70.5,47.5){\psellipse[](0,0)(2.21,-2.12)}
\rput(75,46){$\scr{h^{-1}_{l_x}}$}
\psline[fillstyle=solid]{-|}(56.17,33.17)(59,36)
\pspolygon[](53,32)(55,34)(57,32)(55,30)
\psline[fillstyle=solid](52,29)(54.16,31.16)
\rput(58,28){$\scr{R_x}$}
\psline[fillstyle=solid]{|*-}(82.17,59.17)(85,62)
\rput(88,58){$\scr{R^{-1}_x}$}
\rput(80,52){$\scr{R_x}$}
\psline[fillstyle=solid](72.17,49.17)(75,52)
\psline(69,46)(66.18,43.18)
\pspolygon[](63,42)(65,44)(67,42)(65,40)
\rput(71,40){$\scr{R^{-1}_x}$}
\psline[fillstyle=solid]{|-}(61,38)(63.83,40.83)
\pspolygon[](74,53)(76,55)(78,53)(76,51)
\psline[fillstyle=solid](87,64)(92,69)
\pspolygon[](84,63)(86,65)(88,63)(86,61)
\rput(41,108){$\scr{j_x}$}
\rput(58,108){$\scr{j_y}$}
\rput(101,67){$\scr{j'_y}$}
\rput(3,68){$\scr{j'_x}$}
\end{pspicture}
\ea
\Bigg\rangle
\label{psi}
\ee

namely the states \eqref{semiclassici ridotti inv} where the graphs in the ket notation are the basis states and the graphs out of the brackets are just the product of the functions $\Psi_{H_l}(j_l)=\mathcal{N}\psi_{H_l}(j_l)$ with the  invariant intertwiners proper of our model. In particular, these functions are such that $\sum_{j_l}|\Psi_{H_l}(j_l)|^2=1$, {\it i.e.} obtained normalizing $\psi_{H_l}(j_l)$, which are peaked on classical values $H_{l}$, in the magnetic spin variables \eqref{finscqrlg2} .

We are interested in describing the dynamics on the simplest possible state on which the operator \eqref{Hridotto}  has non vanishing expectation value $\langle{}^{R}\hat{H}^{m}_{E\cube}\rangle$,
\be
\langle{}^{R}\hat{H}^{m}_{E\cube}\rangle=\frac{\langle\Psi_{\bf{H}} |{}^{R}\hat{H}^{m}_{E\cube} |\Psi_{\bf{H}} \rangle}{\langle\Psi_{\bf{H}} |\Psi_{\bf{H}} \rangle}\,.
\ee
%{\bf- IO TOGLIEREI TUTTO  To understand the anisotropic behavior of $\hat{H}$ it's enough to restrict to a state on which non trivial action of the holonomies describing the curvature in the three orthogonal planes is possible and gives non vanishing contributions, to this aim it's enough to consider different values of the spins along the three direction but also loops determining the three orthogonal planes.} 
The best choice is a state based on a lattice with cubic topology and $6$-valent nodes. However the computation in this case complicates and will be presented in future work. A symmetrization of the state \eqref{psi} namely 
\be
|\Psi_{H} \; \vgraph \rangle=\sum_k \frac{1}{\sqrt{3}}|\Psi_{H} \; \vgraph^k\rangle
\label{sc}
\ee
with $k=x,y,z$  is the simplest possible state on which the Hamiltonian has non trivial action and we will focus on it in the following. Hence, since the states $|\Psi_{H} \; \vgraph^k\rangle$ are orthogonal we need to evaluate
\be
\langle \Psi_{H} \; \vgraph | {}^{R}\hat{H}^m_{E\cube}[N] |\Psi_{H} \; \vgraph \rangle= \sum_k\langle \Psi_{H} \; \vgraph^k |{}^{R}\hat{H}^m_{E\cube}[N]  |\Psi_{H} \; \vgraph^k \rangle=\sum_k\langle \Psi_{H} \; \vgraph^k |{}^{R}\hat{H}^{m,k}_{E\cube}[N]  |\Psi_{H} \; \vgraph^k \rangle.
\ee
It is enough to understand the behavior of a single term in the sum. Restricting to $k=z$ we have 

\be
\begin{split}
&{}^{R}\hat{H}^{m,z}_{E\cube}[N] |\Psi_{H} \; \vgraph^z\rangle=-N(\vgraph) C(m)\\
& \sum_{j_x,j_y,j_z, j_l}
\ba
\ifx\JPicScale\undefined\def\JPicScale{0.6}\fi
\psset{unit=\JPicScale mm}
\psset{linewidth=0.3,dotsep=1,hatchwidth=0.3,hatchsep=1.5,shadowsize=1,dimen=middle}
\psset{dotsize=0.7 2.5,dotscale=1 1,fillcolor=black}
\psset{arrowsize=1 2,arrowlength=1,arrowinset=0.25,tbarsize=0.7 5,bracketlength=0.15,rbracketlength=0.15}
\begin{pspicture}(0,0)(101,143)
\rput(28,87){$\Psi_{H_{l_x}}(j_x)$}
\psline[fillstyle=solid]{-|}(42.83,101.83)(40,99)
\pspolygon[](46,103)(44,101)(42,103)(44,105)
\psline[fillstyle=solid](51,110)(44.84,103.84)
\rput(40,104){$\scr{R_x}$}
\psline[fillstyle=solid]{|*-}(16.83,75.83)(14,73)
\rput(10,74){$\scr{R^{-1}_x}$}
\rput(11,77){$\scr{j_x}$}
\psline[fillstyle=solid]{-|}(51,110)(51,120)
\psline[fillstyle=solid]{|*-}(51,139)(51,143)
\rput(46,141){$\scr{j_z}$}
\psline[fillstyle=solid](12,71)(6,65)
\pspolygon[](15,72)(13,70)(11,72)(13,74)
\psline[fillstyle=solid]{-|}(59.17,101.83)(62,99)
\pspolygon[](58,105)(60,103)(58,101)(56,103)
\rput(62,104){$\scr{R_y}$}
\psline[fillstyle=solid]{|*-}(85.17,75.83)(88,73)
\rput(95,73){$\scr{R^{-1}_y}$}
\rput(90,77){$\scr{j_y}$}
\psline[fillstyle=solid](90,71)(96,65)
\pspolygon[](89,74)(91,72)(89,70)(87,72)
\rput(14,60){$\scr{j_l}$}
\rput(80,64){$\scr{j_l}$}
\psline[border=0.3,fillstyle=solid](51,110)(57.16,103.84)
\rput(47,120){$\scr{j_z}$}
\psline[fillstyle=solid]{-|}(18.17,62.83)(21,60)
\pspolygon[](17,66)(19,64)(17,62)(15,64)
\rput(21,65){$\scr{R_y}$}
\psline[fillstyle=solid]{|*-}(44.17,36.83)(47,34)
\rput(44,30){$\scr{R^{-1}_y}$}
\rput(49,38){$\scr{j_l}$}
\psline[fillstyle=solid](49,32)(51.83,29.17)
\pspolygon[](48,35)(50,33)(48,31)(46,33)
\psline[fillstyle=solid](11,70)(16.16,64.84)
\psline[fillstyle=solid]{-|}(56.17,33.17)(59,36)
\pspolygon[](53,32)(55,34)(57,32)(55,30)
\psline[fillstyle=solid](52,29)(54.16,31.16)
\rput(58,28){$\scr{R_x}$}
\psline[fillstyle=solid]{|*-}(82.17,59.17)(85,62)
\psline[fillstyle=solid](87,64)(92,69)
\pspolygon[](84,63)(86,65)(88,63)(86,61)
\rput(41,108){$\scr{j_x}$}
\rput(58,108){$\scr{j_y}$}
\rput(101,67){$\scr{j'_y}$}
\rput(3,68){$\scr{j'_x}$}
\rput(73,87){$\Psi_{H_{l_y}}(j_y)$}
\rput(33,48){$\Psi_{H_{l_l{y}}}(j_l)$}
\rput(71,48){$\Psi_{H_{l_l{x}}}(j_l)$}
\rput(51,130){$\Psi_{H_{l_z}}(j_z)$}
\rput(54,113){{\large{*}}}
\rput(84,68){{\large{*}}}
\rput(20,70){{\large{*}}}
\rput(52,25){{\large{*}}}
\end{pspicture}
\ea
\sum_{\mu'_x \mu'_y\mu_x,\mu_y}
H^{m\;\,j_xj'_xj_yj'_y}_{\mu_x\mu'_x\mu_y\mu'_y}(j_z,j_l)
\quad
\Bigg|\;
\ba
\ifx\JPicScale\undefined\def\JPicScale{0.55}\fi
\psset{unit=\JPicScale mm}
\psset{linewidth=0.3,dotsep=1,hatchwidth=0.3,hatchsep=1.5,shadowsize=1,dimen=middle}
\psset{dotsize=0.7 2.5,dotscale=1 1,fillcolor=black}
\psset{arrowsize=1 2,arrowlength=1,arrowinset=0.25,tbarsize=0.7 5,bracketlength=0.15,rbracketlength=0.15}
\begin{pspicture}(0,0)(104,143)
\psline{|*-}(19,78)(21.83,80.83)
\rput(14,78){$\scr{j_x+\mu_x}$}
\rput{45}(28.5,87.5){\psellipse[](0,0)(2.21,-2.12)}
\rput(24,91){$\scr{h_{x}}$}
\psline[fillstyle=solid]{-|}(42.83,101.83)(40,99)
\pspolygon[](46,103)(44,101)(42,103)(44,105)
\psline[fillstyle=solid](51,110)(44.84,103.84)
\rput(40,104){$\scr{R_x}$}
\psline[fillstyle=solid]{|*-}(16.83,75.83)(14,73)
\rput(10,74){$\scr{R^{-1}_x}$}
\rput(21,85){$\scr{R_x}$}
\psline[fillstyle=solid](26.83,85.83)(24,83)
\psline(30,89)(32.82,91.82)
\pspolygon[](36,93)(34,91)(32,93)(34,95)
\rput(32,97){$\scr{R^{-1}_x}$}
\psline[fillstyle=solid]{|-}(38,97)(35.17,94.17)
\psline{|*-}(51,137)(51,133)
\rput(44,129){$\scr{j_z}$}
\rput{0}(51,130){\psellipse[](0,0)(3,-3)}
\rput(51,130){$\scr{h_{z}}$}
\psline[fillstyle=solid]{-|}(51,110)(51,120)
\psline[fillstyle=solid]{|*-}(51,139)(51,143)
\psline{-|}(51,127)(51,123)
\rput(46,141){$\scr{j_z}$}
\pspolygon[](25,82)(23,80)(21,82)(23,84)
\psline[fillstyle=solid](12,71)(-2,57)
\pspolygon[](15,72)(13,70)(11,72)(13,74)
\psline{|*-}(83,78)(80.17,80.83)
\rput(76,91){$\scr{h_{y}}$}
\psline[fillstyle=solid]{-|}(59.17,101.83)(62,99)
\pspolygon[](58,105)(60,103)(58,101)(56,103)
\rput(62,104){$\scr{R_y}$}
\psline[fillstyle=solid]{|*-}(85.17,75.83)(88,73)
\rput(95,73){$\scr{R^{-1}_y}$}
\rput(79,86){$\scr{R_y}$}
\psline[fillstyle=solid](75.17,85.83)(78,83)
\psline(72,89)(69.18,91.82)
\pspolygon[](68,95)(70,93)(68,91)(66,93)
\rput(72,96){$\scr{R^{-1}_y}$}
\psline[fillstyle=solid]{|-}(64,97)(66.83,94.17)
\rput(90,79){$\scr{j_y-\mu_y}$}
\pspolygon[](79,84)(81,82)(79,80)(77,82)
\psline[fillstyle=solid](90,71)(104,57)
\pspolygon[](89,74)(91,72)(89,70)(87,72)
\rput{43.83}(73.44,87.44){\psellipse[](0,0)(2.21,-2.12)}
\rput(14,44){$\scr{j_l+\mu'_y}$}
\psline[border=0.3,fillstyle=solid](51,110)(57.16,103.84)
\rput(47,120){$\scr{j_z}$}
\psline{|*-}(42,26)(39.17,28.83)
\rput(38,23){$\scr{j_l+\mu'_y}$}
\rput(26,34){$\scr{h_{l_y}}$}
\psline[fillstyle=solid]{-|}(18.17,49.83)(21,47)
\pspolygon[](17,53)(19,51)(17,49)(15,51)
\rput(9,50){$\scr{R_y}$}
\psline[fillstyle=solid]{|*-}(44.17,23.83)(47,21)
\rput(44,17){$\scr{R^{-1}_y}$}
\rput(33,29){$\scr{R_y}$}
\psline[fillstyle=solid](34.17,33.83)(37,31)
\psline(31,37)(28.18,39.82)
\pspolygon[](27,43)(29,41)(27,39)(25,41)
\rput(20,39){$\scr{R^{-1}_y}$}
\psline[fillstyle=solid]{|-}(23,45)(25.83,42.17)
\pspolygon[](38,32)(40,30)(38,28)(36,30)
\psline[fillstyle=solid](49,19)(51.83,16.17)
\pspolygon[](48,22)(50,20)(48,18)(46,20)
\rput{43.83}(32.44,35.44){\psellipse[](0,0)(2.21,-2.12)}
\psline[fillstyle=solid](4,63)(16.16,51.84)
\psline{|*-}(80,44)(77.17,41.17)
\rput(66,22){$\scr{j_l+\mu'_x}$}
\rput{45}(70.5,34.5){\psellipse[](0,0)(2.21,-2.12)}
\rput(78,33){$\scr{h^{-1}_{l_x}}$}
\psline[fillstyle=solid]{-|}(56.17,20.17)(59,23)
\pspolygon[](53,19)(55,21)(57,19)(55,17)
\psline[fillstyle=solid](52,16)(54.16,18.16)
\rput(58,15){$\scr{R_x}$}
\psline[fillstyle=solid]{|*-}(82.17,46.17)(85,49)
\rput(88,45){$\scr{R^{-1}_x}$}
\rput(80,39){$\scr{R_x}$}
\psline[fillstyle=solid](72.17,36.17)(75,39)
\psline(69,33)(66.18,30.18)
\pspolygon[](63,29)(65,31)(67,29)(65,27)
\rput(71,27){$\scr{R^{-1}_x}$}
\psline[fillstyle=solid]{|-}(61,25)(63.83,27.83)
\pspolygon[](74,40)(76,42)(78,40)(76,38)
\psline[fillstyle=solid](87,51)(99,62)
\pspolygon[](84,50)(86,52)(88,50)(86,48)
\rput(39,108){$\scr{j_x}+\mu_x$}
\rput(61,108){$\scr{j_y-\mu_y}$}
\rput(104,62){$\scr{j'_y}$}
\rput(-2,62){$\scr{j'_x}$}
\rput(87,42){$\scr{j_l+\mu'_x}$}
\psline(51,116)(59,116)
\rput(58,118){$\scr\tilde{{m}}$}
\rput(60,115){$\scr{0}$}
\rput(47,114){$\scr{k}$}
\end{pspicture}
\ea
\Bigg\rangle
\end{split}
\label{Hpsi}
\ee

To proceed note that the state \eqref{v3} is not normalized in the $SU(2)$ scalar product as shown in \eqref{not orto}; to normalize it, it's enough to divide each three valent node by

\be
\sqrt{|<{\bf j_l}, {\bf x_{n_3}}|{\bf n_l}, \vec{{\bf u}}_l >|^2}=\sqrt{
\left(
\begin{array} {c}
\ifx\JPicScale\undefined\def\JPicScale{0.7}\fi
\psset{unit=\JPicScale mm}
\psset{linewidth=0.3,dotsep=1,hatchwidth=0.3,hatchsep=1.5,shadowsize=1,dimen=middle}
\psset{dotsize=0.7 2.5,dotscale=1 1,fillcolor=black}
\psset{arrowsize=1 2,arrowlength=1,arrowinset=0.25,tbarsize=0.7 5,bracketlength=0.15,rbracketlength=0.15}
\begin{pspicture}(7,0)(32,27)
\psline[fillstyle=solid]{-|}(12.83,8.83)(10,6)
\pspolygon[](16,10)(14,8)(12,10)(14,12)
\psline[fillstyle=solid](21,17)(14.84,10.84)
\rput(10,11){$\scr{R_1}$}
\psline[fillstyle=solid]{-|}(21,24)(21,27)
\psline[fillstyle=solid]{-|}(29.17,8.83)(32,6)
\pspolygon[](28,12)(30,10)(28,8)(26,10)
\rput(32,11){$\scr{R_2}$}
\psline[fillstyle=solid](21,17)(27.16,10.84)
\rput(17,27){$\scr{j_3}$}
\rput(11,15){$\scr{j_1}$}
\rput(28,15){$\scr{j_2}$}
\pspolygon[](19.5,24)(22.5,24)(22.5,21)(19.5,21)
\psline(21,21)(21,17)
\rput(25,22){$\scr{R_3}$}
\end{pspicture}
\end{array}
\right)^{*}
\begin{array} {c}
\ifx\JPicScale\undefined\def\JPicScale{0.7}\fi
\psset{unit=\JPicScale mm}
\psset{linewidth=0.3,dotsep=1,hatchwidth=0.3,hatchsep=1.5,shadowsize=1,dimen=middle}
\psset{dotsize=0.7 2.5,dotscale=1 1,fillcolor=black}
\psset{arrowsize=1 2,arrowlength=1,arrowinset=0.25,tbarsize=0.7 5,bracketlength=0.15,rbracketlength=0.15}
\begin{pspicture}(0,0)(32,27)
\psline[fillstyle=solid]{-|}(12.83,8.83)(10,6)
\pspolygon[](16,10)(14,8)(12,10)(14,12)
\psline[fillstyle=solid](21,17)(14.84,10.84)
\rput(10,11){$\scr{R_1}$}
\psline[fillstyle=solid]{-|}(21,24)(21,27)
\psline[fillstyle=solid]{-|}(29.17,8.83)(32,6)
\pspolygon[](28,12)(30,10)(28,8)(26,10)
\rput(32,11){$\scr{R_2}$}
\psline[fillstyle=solid](21,17)(27.16,10.84)
\rput(17,27){$\scr{j_3}$}
\rput(11,15){$\scr{j_1}$}
\rput(28,15){$\scr{j_2}$}
\pspolygon[](19.5,24)(22.5,24)(22.5,21)(19.5,21)
\psline(21,21)(21,17)
\rput(25,22){$\scr{R_3}$}
\end{pspicture}
\end{array}
}
\ee 
In the coherent states \eqref{psi}, this normalization must be done twice: for both the intertwiners in the basis elements and the intertwiners in the coefficients (since the latter are dual to the former). This corresponds to use a normalized intertwiners basis for which each intertwiner is just a phase and the expression above is equal to 1.  
Having normalized intertwiners, the full state $|\Psi_{H} \; \vgraph \rangle$ is normalized too, {\it i.e.}
\be
\langle \Psi_{H} \; \vgraph\,  |\Psi_{H} \; \vgraph \,\rangle\,=1\,.
\ee
We have then
\be
\begin{split}
&\langle\Psi_{H} \; \vgraph^z |{}^{R}\hat{H}^{m,z}_{E\cube} |\Psi_{H} \; \vgraph^z\rangle=
-N(\vgraph) C(m)\sum_{j_x,j_y,j_z, j_l}\;\sum_{\mu'_x \mu'_y\mu_x,\mu_y=\pm m}
\ba
\ifx\JPicScale\undefined\def\JPicScale{1}\fi
\psset{unit=\JPicScale mm}
\psset{linewidth=0.3,dotsep=1,hatchwidth=0.3,hatchsep=1.5,shadowsize=1,dimen=middle}
\psset{dotsize=0.7 2.5,dotscale=1 1,fillcolor=black}
\psset{arrowsize=1 2,arrowlength=1,arrowinset=0.25,tbarsize=0.7 5,bracketlength=0.15,rbracketlength=0.15}
\begin{pspicture}(0,0)(38,27)
\psline[fillstyle=solid]{-|}(18.83,8.83)(16,6)
\pspolygon[](22,10)(20,8)(18,10)(20,12)
\psline[fillstyle=solid](27,17)(20.84,10.84)
\rput(16,11){$\scr{R_x}$}
\psline[fillstyle=solid]{-|}(27,17)(27,27)
\psline[fillstyle=solid]{-|}(35.17,8.83)(38,6)
\pspolygon[](34,12)(36,10)(34,8)(32,10)
\rput(38,11){$\scr{R_y}$}
\psline[border=0.3,fillstyle=solid](27,17)(33.16,10.84)
\rput(23,27){$\scr{j_z}$}
\rput(17,15){$\scr{j_x+\mu_x}$}
\rput(34,15){$\scr{j_y-\mu_y}$}
\rput(31,21){{\large{*}}}
\end{pspicture}
\ea
\hspace{-1cm}
\ba
\ifx\JPicScale\undefined\def\JPicScale{1}\fi
\psset{unit=\JPicScale mm}
\psset{linewidth=0.3,dotsep=1,hatchwidth=0.3,hatchsep=1.5,shadowsize=1,dimen=middle}
\psset{dotsize=0.7 2.5,dotscale=1 1,fillcolor=black}
\psset{arrowsize=1 2,arrowlength=1,arrowinset=0.25,tbarsize=0.7 5,bracketlength=0.15,rbracketlength=0.15}
\begin{pspicture}(0,0)(50,27)
\psline[fillstyle=solid]{-|}(25.83,8.83)(23,6)
\pspolygon[](29,10)(27,8)(25,10)(27,12)
\psline[fillstyle=solid](34,17)(27.84,10.84)
\rput(23,11){$\scr{R_x}$}
\psline[fillstyle=solid]{-|}(34,17)(34,27)
\rput(30,21){$\scr{k}$}
\psline(34,25)(42,25)
\psline[fillstyle=solid]{-|}(42.17,8.83)(45,6)
\pspolygon[](41,12)(43,10)(41,8)(39,10)
\rput(45,11){$\scr{R_y}$}
\rput(30,27){$\scr{j_z}$}
\rput(18,7){$\scr{j_x+\mu_x}$}
\rput(50,8){$\scr{j_y-\mu_y}$}
\psline[border=0.3,fillstyle=solid](34,17)(40,11)
\rput(41,27){$\scr\tilde{{m}}$}
\rput(43,24){$\scr{0}$}
\end{pspicture}
\ea
\\
&
\left(
\ba
\ifx\JPicScale\undefined\def\JPicScale{0.6}\fi
\psset{unit=\JPicScale mm}
\psset{linewidth=0.3,dotsep=1,hatchwidth=0.3,hatchsep=1.5,shadowsize=1,dimen=middle}
\psset{dotsize=0.7 2.5,dotscale=1 1,fillcolor=black}
\psset{arrowsize=1 2,arrowlength=1,arrowinset=0.25,tbarsize=0.7 5,bracketlength=0.15,rbracketlength=0.15}
\begin{pspicture}(0,0)(101,143)
\rput(28,87){$\Psi_{H_{l_x}}(j_x+\mu_x)$}
\psline[fillstyle=solid]{-|}(42.83,101.83)(40,99)
\pspolygon[](46,103)(44,101)(42,103)(44,105)
\psline[fillstyle=solid](51,110)(44.84,103.84)
\rput(40,104){$\scr{R_x}$}
\psline[fillstyle=solid]{|*-}(16.83,75.83)(14,73)
\rput(10,74){$\scr{R^{-1}_x}$}
\rput(11,77){$\scr{j_x+\mu_x}$}
\psline[fillstyle=solid]{-|}(51,110)(51,120)
\psline[fillstyle=solid]{|*-}(51,139)(51,143)
\rput(46,141){$\scr{j_z}$}
\psline[fillstyle=solid](12,71)(6,65)
\pspolygon[](15,72)(13,70)(11,72)(13,74)
\psline[fillstyle=solid]{-|}(59.17,101.83)(62,99)
\pspolygon[](58,105)(60,103)(58,101)(56,103)
\rput(62,104){$\scr{R_y}$}
\psline[fillstyle=solid]{|*-}(85.17,75.83)(88,73)
\rput(95,73){$\scr{R^{-1}_y}$}
\rput(90,77){$\scr{j_y-\mu_y}$}
\psline[fillstyle=solid](90,71)(96,65)
\pspolygon[](89,74)(91,72)(89,70)(87,72)
\rput(14,60){$\scr{j_l+\mu'_y}$}
\rput(80,64){$\scr{j_l+\mu'_x}$}
\psline[border=0.3,fillstyle=solid](51,110)(57.16,103.84)
\rput(47,120){$\scr{j_z}$}
\psline[fillstyle=solid]{-|}(18.17,62.83)(21,60)
\pspolygon[](17,66)(19,64)(17,62)(15,64)
\rput(21,65){$\scr{R_y}$}
\psline[fillstyle=solid]{|*-}(44.17,36.83)(47,34)
\rput(44,30){$\scr{R^{-1}_y}$}
\rput(45,40){$\scr{j_l+\mu'_y}$}
\rput(63,40){$\scr{j_l+\mu'_x}$}
\psline[fillstyle=solid](49,32)(51.83,29.17)
\pspolygon[](48,35)(50,33)(48,31)(46,33)
\psline[fillstyle=solid](11,70)(16.16,64.84)
\psline[fillstyle=solid]{-|}(56.17,33.17)(59,36)
\pspolygon[](53,32)(55,34)(57,32)(55,30)
\psline[fillstyle=solid](52,29)(54.16,31.16)
\rput(58,28){$\scr{R_x}$}
\psline[fillstyle=solid]{|*-}(82.17,59.17)(85,62)
\psline[fillstyle=solid](87,64)(92,69)
\pspolygon[](84,63)(86,65)(88,63)(86,61)
\rput(41,108){$\scr{j_x+\mu_x}$}
\rput(58,108){$\scr{j_y-\mu_y}$}
\rput(101,67){$\scr{j'_y}$}
\rput(3,68){$\scr{j'_x}$}
\rput(73,87){$\Psi_{H_{l_y}}(j_y-\mu_y)$}
\rput(33,48){$\Psi_{H_{l_{l_{y}}}}(j_l+\mu_y')$}
\rput(71,48){$\Psi_{H_{l_{l_{x}}}}(j_l+\mu_x')$}
\rput(51,130){$\Psi_{H_{l_z}}(j_z)$}
\rput(54,113){{\large{*}}}
\rput(84,68){{\large{*}}}
\rput(20,70){{\large{*}}}
\rput(52,25){{\large{*}}}
\end{pspicture}
\ea
\right)^{{\huge{*}}}
H^{m\;\,j_xj'_xj_yj'_y}_{\mu_x\mu'_x\mu_y\mu'_y}(j_z,j_l)
\left(\ba
\ifx\JPicScale\undefined\def\JPicScale{0.6}\fi
\psset{unit=\JPicScale mm}
\psset{linewidth=0.3,dotsep=1,hatchwidth=0.3,hatchsep=1.5,shadowsize=1,dimen=middle}
\psset{dotsize=0.7 2.5,dotscale=1 1,fillcolor=black}
\psset{arrowsize=1 2,arrowlength=1,arrowinset=0.25,tbarsize=0.7 5,bracketlength=0.15,rbracketlength=0.15}
\begin{pspicture}(0,0)(101,143)
\rput(28,87){$\Psi_{H_{l_x}}(j_x)$}
\psline[fillstyle=solid]{-|}(42.83,101.83)(40,99)
\pspolygon[](46,103)(44,101)(42,103)(44,105)
\psline[fillstyle=solid](51,110)(44.84,103.84)
\rput(40,104){$\scr{R_x}$}
\psline[fillstyle=solid]{|*-}(16.83,75.83)(14,73)
\rput(10,74){$\scr{R^{-1}_x}$}
\rput(11,77){$\scr{j_x}$}
\psline[fillstyle=solid]{-|}(51,110)(51,120)
\psline[fillstyle=solid]{|*-}(51,139)(51,143)
\rput(46,141){$\scr{j_z}$}
\psline[fillstyle=solid](12,71)(6,65)
\pspolygon[](15,72)(13,70)(11,72)(13,74)
\psline[fillstyle=solid]{-|}(59.17,101.83)(62,99)
\pspolygon[](58,105)(60,103)(58,101)(56,103)
\rput(62,104){$\scr{R_y}$}
\psline[fillstyle=solid]{|*-}(85.17,75.83)(88,73)
\rput(95,73){$\scr{R^{-1}_y}$}
\rput(90,77){$\scr{j_y}$}
\psline[fillstyle=solid](90,71)(96,65)
\pspolygon[](89,74)(91,72)(89,70)(87,72)
\rput(14,60){$\scr{j_l}$}
\rput(80,64){$\scr{j_l}$}
\psline[border=0.3,fillstyle=solid](51,110)(57.16,103.84)
\rput(47,120){$\scr{j_z}$}
\psline[fillstyle=solid]{-|}(18.17,62.83)(21,60)
\pspolygon[](17,66)(19,64)(17,62)(15,64)
\rput(21,65){$\scr{R_y}$}
\psline[fillstyle=solid]{|*-}(44.17,36.83)(47,34)
\rput(44,30){$\scr{R^{-1}_y}$}
\rput(49,38){$\scr{j_l}$}
\psline[fillstyle=solid](49,32)(51.83,29.17)
\pspolygon[](48,35)(50,33)(48,31)(46,33)
\psline[fillstyle=solid](11,70)(16.16,64.84)
\psline[fillstyle=solid]{-|}(56.17,33.17)(59,36)
\pspolygon[](53,32)(55,34)(57,32)(55,30)
\psline[fillstyle=solid](52,29)(54.16,31.16)
\rput(58,28){$\scr{R_x}$}
\psline[fillstyle=solid]{|*-}(82.17,59.17)(85,62)
\psline[fillstyle=solid](87,64)(92,69)
\pspolygon[](84,63)(86,65)(88,63)(86,61)
\rput(41,108){$\scr{j_x}$}
\rput(58,108){$\scr{j_y}$}
\rput(101,67){$\scr{j'_y}$}
\rput(3,68){$\scr{j'_x}$}
\rput(73,87){$\Psi_{H_{l_y}}(j_y)$}
\rput(33,48){$\Psi_{H_{l_l{y}}}(j_l)$}
\rput(71,48){$\Psi_{H_{l_l{x}}}(j_l)$}
\rput(51,130){$\Psi_{H_{l_z}}(j_z)$}
\rput(54,113){{\large{*}}}
\rput(84,68){{\large{*}}}
\rput(20,70){{\large{*}}}
\rput(52,25){{\large{*}}}
\end{pspicture}
\ea\right)
\end{split}
\label{azione H esatta}
\ee
where the coefficients in the first line are the only remnants of the scalar product between the basis elements in the expression \eqref{Hpsi} and the dual basis elements in $\langle \Psi_H \vgraph^z|$.
This complicated expression can be greatly simplified using the explicit form of the coherent states \eqref{finscqrlg}. In fact for large mean values one has
\be
\Psi^*_{H_{l}}(j_l+\mu) \Psi_{H_{l}}(j_l)\approx \mathcal{N}^2 e^{-\alpha\left(j_{l}-\bar{j}_l\right)^2} e^{-i\theta_l \mu}\qquad \bar{j}_l\gg\mu,
\ee
$\mathcal{N}$ being the factor normalizing $\Psi_{H_{l}}$ and for $\bar{j}_l\gg\mu$ also the gaussian $e^{-\alpha\left(j_{l}-\bar{j}_l\right)^2}$. Hence, to compute \eqref{azione H esatta} we only need to understand the role of the reduced intertwiners. To this aim we note that:
\be
\ba
\ifx\JPicScale\undefined\def\JPicScale{1}\fi
\psset{unit=\JPicScale mm}
\psset{linewidth=0.3,dotsep=1,hatchwidth=0.3,hatchsep=1.5,shadowsize=1,dimen=middle}
\psset{dotsize=0.7 2.5,dotscale=1 1,fillcolor=black}
\psset{arrowsize=1 2,arrowlength=1,arrowinset=0.25,tbarsize=0.7 5,bracketlength=0.15,rbracketlength=0.15}
\begin{pspicture}(0,0)(34,32)
\psline[fillstyle=solid]{-|}(12.83,13.83)(10,11)
\pspolygon[](16,15)(14,13)(12,15)(14,17)
\psline[fillstyle=solid](21,22)(14.84,15.84)
\rput(10,16){$\scr{R_1}$}
\psline[fillstyle=solid]{-|}(21,29)(21,32)
\psline[fillstyle=solid]{-|}(29.17,13.83)(32,11)
\pspolygon[](28,17)(30,15)(28,13)(26,15)
\rput(32,16){$\scr{R_2}$}
\psline[fillstyle=solid](21,22)(27.16,15.84)
\rput(17,32){$\scr{j_3}$}
\rput(11,20){$\scr{j_1}$}
\rput(28,20){$\scr{j_2}$}
\pspolygon[](19.5,29)(22.5,29)(22.5,26)(19.5,26)
\psline(21,26)(21,22)
\rput(25,27){$\scr{R_3}$}
\psline[fillstyle=solid]{-|}(12.83,7.83)(10,5)
\pspolygon[](16,9)(14,7)(12,9)(14,11)
\psline[fillstyle=solid](21,16)(14.84,9.84)
\rput(18,8){$\scr{R_1}$}
\psline[fillstyle=solid]{-|}(29.17,7.83)(32,5)
\pspolygon[](28,11)(30,9)(28,7)(26,9)
\rput(23,8){$\scr{R_2}$}
\psline[fillstyle=solid](21,16)(27.16,9.84)
\rput(8,3){$\scr{\mu_1}$}
\rput(34,3){$\scr{\mu_2}$}
\rput(21,13){$\scr{m}$}
\end{pspicture}
\ea
\quad
=\sum_{k1, k2} 
\begin{array} {c}
\ifx\JPicScale\undefined\def\JPicScale{1}\fi
\psset{unit=\JPicScale mm}
\psset{linewidth=0.3,dotsep=1,hatchwidth=0.3,hatchsep=1.5,shadowsize=1,dimen=middle}
\psset{dotsize=0.7 2.5,dotscale=1 1,fillcolor=black}
\psset{arrowsize=1 2,arrowlength=1,arrowinset=0.25,tbarsize=0.7 5,bracketlength=0.15,rbracketlength=0.15}
\begin{pspicture}(0,0)(39,27)
\psline[fillstyle=solid]{-|}(10.83,6.83)(6,2)
\pspolygon[](14,8)(12,6)(10,8)(12,10)
\psline[fillstyle=solid](21,17)(12.84,8.84)
\rput(8,9){$\scr{R_1}$}
\psline[fillstyle=solid]{-|}(21,24)(21,27)
\psline[fillstyle=solid]{-|}(31.17,6.83)(36,2)
\pspolygon[](30,10)(32,8)(30,6)(28,8)
\rput(34,9){$\scr{R_2}$}
\psline[fillstyle=solid](21,17)(29.16,8.84)
\rput(17,27){$\scr{j_3}$}
\rput(14,17){$\scr{j_1}$}
\rput(26,17){$\scr{j_2}$}
\pspolygon[](19.5,24)(22.5,24)(22.5,21)(19.5,21)
\psline(21,21)(21,17)
\rput(25,22){$\scr{R_3}$}
\psline[fillstyle=solid]{-|}(9,5)(10,1)
\psline[fillstyle=solid](21,12)(19,10)
\psline[fillstyle=solid]{-|}(33,5)(31.83,1.17)
\psline[fillstyle=solid](21,12)(23,10)
\rput(13,1){$\scr{\mu_1}$}
\rput(29,1){$\scr{\mu_2}$}
\psline[fillstyle=solid](16,12)(19,10)
\psline[fillstyle=solid](26,12)(23,10)
\rput(3,2){$\scr{j_1}$}
\rput(39,2){$\scr{j_2}$}
\rput(31,12){$\scr{k_2}$}
\rput(11,12){$\scr{k_1}$}
\rput(21,9){$\scr{m}$}
\end{pspicture}
\end{array}
=\sum_{k_1 k_2} \sixj{j_1}{j_2}{j_3}{k_2}{k_1}{m}
\begin{array} {c}
\ifx\JPicScale\undefined\def\JPicScale{1}\fi
\psset{unit=\JPicScale mm}
\psset{linewidth=0.3,dotsep=1,hatchwidth=0.3,hatchsep=1.5,shadowsize=1,dimen=middle}
\psset{dotsize=0.7 2.5,dotscale=1 1,fillcolor=black}
\psset{arrowsize=1 2,arrowlength=1,arrowinset=0.25,tbarsize=0.7 5,bracketlength=0.15,rbracketlength=0.15}
\begin{pspicture}(0,0)(39,27)
\psline[fillstyle=solid]{-|}(10.83,6.83)(6,2)
\pspolygon[](14,8)(12,6)(10,8)(12,10)
\psline[fillstyle=solid](21,17)(12.84,8.84)
\rput(8,9){$\scr{R_1}$}
\psline[fillstyle=solid]{-|}(21,24)(21,27)
\psline[fillstyle=solid]{-|}(31.17,6.83)(36,2)
\pspolygon[](30,10)(32,8)(30,6)(28,8)
\rput(34,9){$\scr{R_2}$}
\psline[fillstyle=solid](21,17)(29.16,8.84)
\rput(17,27){$\scr{j_3}$}
\pspolygon[](19.5,24)(22.5,24)(22.5,21)(19.5,21)
\psline(21,21)(21,17)
\rput(25,22){$\scr{R_3}$}
\psline[fillstyle=solid]{-|}(9,5)(10,1)
\psline[fillstyle=solid]{-|}(33,5)(31.83,1.17)
\rput(13,1){$\scr{\mu_1}$}
\rput(29,1){$\scr{\mu_2}$}
\rput(3,2){$\scr{j_1}$}
\rput(39,2){$\scr{j_2}$}
\rput(31,12){$\scr{k_2}$}
\rput(11,12){$\scr{k_1}$}
\end{pspicture}
\end{array}
\label{trick angolo}
\ee

and 

\be
\ba
\ifx\JPicScale\undefined\def\JPicScale{1}\fi
\psset{unit=\JPicScale mm}
\psset{linewidth=0.3,dotsep=1,hatchwidth=0.3,hatchsep=1.5,shadowsize=1,dimen=middle}
\psset{dotsize=0.7 2.5,dotscale=1 1,fillcolor=black}
\psset{arrowsize=1 2,arrowlength=1,arrowinset=0.25,tbarsize=0.7 5,bracketlength=0.15,rbracketlength=0.15}
\begin{pspicture}(0,0)(52,31)
\psline[fillstyle=solid]{-|}(26.83,12.83)(18,4)
\pspolygon[](30,14)(28,12)(26,14)(28,16)
\rput(24,15){$\scr{R_x}$}
\rput(27,19){$\scr{j_x}$}
\psline[fillstyle=solid]{-|}(35,21)(35,31)
\psline[fillstyle=solid]{-|}(43.17,12.83)(52,4)
\rput(46,15){$\scr{R_y}$}
\rput(41,18){$\scr{j_y}$}
\psline[fillstyle=solid]{|*-}(24,2)(31,9)
\rput(32,5){$\scr{R_x}$}
\rput(35,15){$\scr{m}$}
\psline[fillstyle=solid](33,11)(37,15)
\pspolygon[](30,10)(32,12)(34,10)(32,8)
\psline[fillstyle=solid]{|*-}(46,2)(39,9)
\rput(39,5){$\scr{R_y^{-1}}$}
\rput(44,22){$\scr{\tilde{m}}$}
\pspolygon[](38,8)(36,10)(38,12)(40,10)
\psline[fillstyle=solid](37,15)(45,26)
\psline[fillstyle=solid](38,12)(37,15)
\rput(33,23){$\scr{j_z}$}
\rput(31,31){$\scr{j_z}$}
\psline[border=0.3,fillstyle=solid](35,21)(41,15)
\psline[fillstyle=solid](35,21)(28.84,14.84)
\pspolygon[](42,16)(44,14)(42,12)(40,14)
\rput(46,27){$\scr{0}$}
\rput(29,3){$\scr{m}$}
\rput(39,3){$\scr{m}$}
\rput(22,0){$\scr{\mu_x}$}
\rput(47,0){$\scr{\mu_y}$}
\end{pspicture}
\ea
=
\sum_{i,l,k}
\ba
\ifx\JPicScale\undefined\def\JPicScale{1}\fi
\psset{unit=\JPicScale mm}
\psset{linewidth=0.3,dotsep=1,hatchwidth=0.3,hatchsep=1.5,shadowsize=1,dimen=middle}
\psset{dotsize=0.7 2.5,dotscale=1 1,fillcolor=black}
\psset{arrowsize=1 2,arrowlength=1,arrowinset=0.25,tbarsize=0.7 5,bracketlength=0.15,rbracketlength=0.15}
\begin{pspicture}(0,0)(42,35)
\psline[fillstyle=solid](14.83,16.83)(12,14)
\pspolygon[](18,18)(16,16)(14,18)(16,20)
\psline[fillstyle=solid](23,25)(16.84,18.84)
\rput(12,19){$\scr{R_x}$}
\rput(15,23){$\scr{j_x}$}
\psline[fillstyle=solid]{-|}(23,25)(23,35)
\rput(19,31){$\scr{k}$}
\psline(23,29)(25,29)
\psline(23,33)(31,33)
\psline[fillstyle=solid](31.17,16.83)(34,14)
\pspolygon[](30,20)(32,18)(30,16)(28,18)
\rput(34,19){$\scr{R_y}$}
\rput(29,22){$\scr{j_y}$}
\psline[fillstyle=solid](13,15)(19,13)
\rput(18,11){$\scr{R_x}$}
\rput(23,19){$\scr{m}$}
\psline[fillstyle=solid](21,15)(25,19)
\pspolygon[](18,14)(20,16)(22,14)(20,12)
\psline[fillstyle=solid](33,15)(27,13)
\rput(27,25){$\scr\tilde{{m}}$}
\pspolygon[](26,12)(24,14)(26,16)(28,14)
\psline[fillstyle=solid](25,19)(25,29)
\rput(21,27){$\scr{j_z}$}
\rput(19,35){$\scr{j_z}$}
\rput(10,15){$\scr{l}$}
\rput(36,15){$\scr{i}$}
\rput(28,11){$\scr{R^{-1}_y}$}
\psline[fillstyle=solid](26,16)(25,19)
\psline[border=0.3,fillstyle=solid](23,25)(29,19)
\rput(30,35){$\scr\tilde{{m}}$}
\rput(32,32){$\scr{0}$}
\psline[fillstyle=solid]{-|}(11.83,13.83)(7,9)
\psline[fillstyle=solid]{-|}(34.17,13.83)(39,9)
\psline[fillstyle=solid]{-|}(10,12)(11,8)
\psline[fillstyle=solid]{-|}(36,12)(34.83,8.17)
\rput(14,8){$\scr{\mu_x}$}
\rput(32,8){$\scr{\mu_y}$}
\rput(4,9){$\scr{j_x}$}
\rput(42,9){$\scr{j_y}$}
\end{pspicture}
\ea
=\sum_{i,l,k}
\tinyninej {k}{\tilde{m}}{j_z} {l} {m} {j_y}  {i} {m}  {j_x}
\ba
\ifx\JPicScale\undefined\def\JPicScale{1}\fi
\psset{unit=\JPicScale mm}
\psset{linewidth=0.3,dotsep=1,hatchwidth=0.3,hatchsep=1.5,shadowsize=1,dimen=middle}
\psset{dotsize=0.7 2.5,dotscale=1 1,fillcolor=black}
\psset{arrowsize=1 2,arrowlength=1,arrowinset=0.25,tbarsize=0.7 5,bracketlength=0.15,rbracketlength=0.15}
\begin{pspicture}(0,0)(42,35)
\psline[fillstyle=solid](14.83,16.83)(12,14)
\pspolygon[](18,18)(16,16)(14,18)(16,20)
\psline[fillstyle=solid](23,25)(16.84,18.84)
\rput(12,19){$\scr{R_x}$}
\psline[fillstyle=solid]{-|}(23,25)(23,35)
\rput(19,31){$\scr{k}$}
\psline(23,33)(31,33)
\psline[fillstyle=solid](31.17,16.83)(34,14)
\pspolygon[](30,20)(32,18)(30,16)(28,18)
\rput(34,19){$\scr{R_y}$}
%\rput(23,19){$\scr{m}$}
\rput(19,35){$\scr{j_z}$}
\rput(10,15){$\scr{l}$}
\rput(36,15){$\scr{i}$}
\psline[border=0.3,fillstyle=solid](23,25)(29,19)
\rput(30,35){$\scr\tilde{{m}}$}
\rput(32,32){$\scr{0}$}
\psline[fillstyle=solid]{-|}(11.83,13.83)(7,9)
\psline[fillstyle=solid]{-|}(34.17,13.83)(39,9)
\psline[fillstyle=solid]{-|}(10,12)(11,8)
\psline[fillstyle=solid]{-|}(36,12)(34.83,8.17)
\rput(14,8){$\scr{\mu_x}$}
\rput(32,8){$\scr{\mu_y}$}
\rput(4,9){$\scr{j_1}$}
\rput(42,9){$\scr{j_2}$}
\end{pspicture}
\ea
\label{trick vertice}
\ee

We can thus proceed observing that the gaussians peak the $j_l$ around large values $\bar{j}_l$'s and this allow us to use a fundamental approximation for the Clebsh Gordan coefficients viable when $a,c>>b$ \cite{brussaard}:
\be\label{rel cb}
C^{cc}_{a\alpha b\beta}\approx\delta_{\beta,c-\alpha}  \delta_{\beta,c-a}\,.
\ee
The coefficients appearing in the previous formula \eqref{trick angolo} and \eqref{trick vertice} are of the form:
\be
C^{k\kappa}_{jj m\mu}=(-1)^{j+m-k} C^{k\kappa}_{ m\mu jj}=(-1)^{j+m-k}\sqrt{\frac{d_k}{d_j}}(-1)^{m-\mu}C^{jj}_{ k\kappa m\,-\mu},
\ee
and using \eqref{rel cb} we have:
\be
C^{k\kappa}_{jj m\mu}\approx (-1)^{j+m-k} (-1)^{m-\mu} \sqrt{\frac{d_k}{d_j}} \delta_{-\mu,j-\kappa}  \delta_{-\mu,j-k}=(-1)^{j+m-k} (-1)^{m-\mu} \sqrt{\frac{d_k}{d_j}} \delta_{\kappa ,j+\mu}  \delta_{k,j+\mu}\,.
\label{key approx}
\ee
Hence, equation \eqref{trick angolo} can be approximated as
\be
\begin{split}
\ba
\ifx\JPicScale\undefined\def\JPicScale{1}\fi
\psset{unit=\JPicScale mm}
\psset{linewidth=0.3,dotsep=1,hatchwidth=0.3,hatchsep=1.5,shadowsize=1,dimen=middle}
\psset{dotsize=0.7 2.5,dotscale=1 1,fillcolor=black}
\psset{arrowsize=1 2,arrowlength=1,arrowinset=0.25,tbarsize=0.7 5,bracketlength=0.15,rbracketlength=0.15}
\begin{pspicture}(0,0)(34,32)
\psline[fillstyle=solid]{-|}(12.83,13.83)(10,11)
\pspolygon[](16,15)(14,13)(12,15)(14,17)
\psline[fillstyle=solid](21,22)(14.84,15.84)
\rput(10,16){$\scr{R_1}$}
\psline[fillstyle=solid]{-|}(21,29)(21,32)
\psline[fillstyle=solid]{-|}(29.17,13.83)(32,11)
\pspolygon[](28,17)(30,15)(28,13)(26,15)
\rput(32,16){$\scr{R_2}$}
\psline[fillstyle=solid](21,22)(27.16,15.84)
\rput(17,32){$\scr{j_3}$}
\rput(11,20){$\scr{j_1}$}
\rput(28,20){$\scr{j_2}$}
\pspolygon[](19.5,29)(22.5,29)(22.5,26)(19.5,26)
\psline(21,26)(21,22)
\rput(25,27){$\scr{R_3}$}
\psline[fillstyle=solid]{-|}(12.83,7.83)(10,5)
\pspolygon[](16,9)(14,7)(12,9)(14,11)
\psline[fillstyle=solid](21,16)(14.84,9.84)
\rput(18,8){$\scr{R_1}$}
\psline[fillstyle=solid]{-|}(29.17,7.83)(32,5)
\pspolygon[](28,11)(30,9)(28,7)(26,9)
\rput(23,8){$\scr{R_2}$}
\psline[fillstyle=solid](21,16)(27.16,9.84)
\rput(8,3){$\scr{\mu_1}$}
\rput(34,3){$\scr{\mu_2}$}
\rput(21,13){$\scr{m}$}
\end{pspicture}
\ea
\,
&=
 \sum_{k_1,k_2} \sum_{\kappa_1 \kappa_2}\sixj{j_1}{j_2}{j_3}{k_2}{k_1}{m}
\ba
\ifx\JPicScale\undefined\def\JPicScale{1}\fi
\psset{unit=\JPicScale mm}
\psset{linewidth=0.3,dotsep=1,hatchwidth=0.3,hatchsep=1.5,shadowsize=1,dimen=middle}
\psset{dotsize=0.7 2.5,dotscale=1 1,fillcolor=black}
\psset{arrowsize=1 2,arrowlength=1,arrowinset=0.25,tbarsize=0.7 5,bracketlength=0.15,rbracketlength=0.15}
\begin{pspicture}(0,0)(51,34)
\psline[fillstyle=solid]{-|}(12.83,8.83)(8,4)
\pspolygon[](21,15)(19,13)(17,15)(19,17)
\psline[fillstyle=solid](28,24)(19.84,15.84)
\rput(15,16){$\scr{R_1}$}
\psline[fillstyle=solid]{-|}(28,31)(28,34)
\psline[fillstyle=solid]{-|}(43.17,8.83)(48,4)
\pspolygon[](37,17)(39,15)(37,13)(35,15)
\rput(41,16){$\scr{R_2}$}
\psline[fillstyle=solid](28,24)(36.16,15.84)
\rput(24,34){$\scr{j_3}$}
\pspolygon[](26.5,31)(29.5,31)(29.5,28)(26.5,28)
\psline(28,28)(28,24)
\rput(32,29){$\scr{R_3}$}
\psline[fillstyle=solid]{-|}(11,7)(12,3)
\psline[fillstyle=solid]{-|}(45,7)(43.83,3.17)
\rput(15,3){$\scr{\mu_1}$}
\rput(41,3){$\scr{\mu_2}$}
\rput(5,4){$\scr{j_1}$}
\rput(51,4){$\scr{j_2}$}
\rput(38,19){$\scr{k_2}$}
\rput(18,19){$\scr{k_1}$}
\psline(18,14)(15,11)
\rput(14,10){$\scr{\kappa_1}$}
\psline(38,14)(41,11)
\rput(42,10){$\scr{\kappa_2}$}
\end{pspicture}
\ea
\\
&\approx
\sqrt{\frac{d_{j_1+\mu_1}}{d_{j_1}}} \sqrt{\frac{d_{j_1+\mu_2}}{d_{j_2}}}
\sixj{j_1}{j_2}{j_3}{j_2+\mu_2}{j_1+\mu_1}{m}
\ba
\ifx\JPicScale\undefined\def\JPicScale{1}\fi
\psset{unit=\JPicScale mm}
\psset{linewidth=0.3,dotsep=1,hatchwidth=0.3,hatchsep=1.5,shadowsize=1,dimen=middle}
\psset{dotsize=0.7 2.5,dotscale=1 1,fillcolor=black}
\psset{arrowsize=1 2,arrowlength=1,arrowinset=0.25,tbarsize=0.7 5,bracketlength=0.15,rbracketlength=0.15}
\begin{pspicture}(0,0)(41,34)
\pspolygon[](21,15)(19,13)(17,15)(19,17)
\psline[fillstyle=solid](28,24)(19.84,15.84)
\rput(15,16){$\scr{R_1}$}
\psline[fillstyle=solid]{-|}(28,31)(28,34)
\pspolygon[](37,17)(39,15)(37,13)(35,15)
\rput(41,16){$\scr{R_2}$}
\psline[fillstyle=solid](28,24)(36.16,15.84)
\rput(24,34){$\scr{j_3}$}
\pspolygon[](26.5,31)(29.5,31)(29.5,28)(26.5,28)
\psline(28,28)(28,24)
\rput(32,29){$\scr{R_3}$}
\rput(38,19){$\scr{j_2+\mu_2}$}
\rput(18,19){$\scr{j_1+\mu_1}$}
\psline{-|*}(18,14)(15,11)
\psline{-|*}(38,14)(41,11)
\rput(14,9){$\scr{j_1+\mu_1}$}
\rput(43,9){$\scr{j_2+\mu_2}$}
\end{pspicture}
\ea
+
O\left(\frac{1}{\sqrt{j_1}}\right)
+
O(\frac{1}{\sqrt{j_2}})
\end{split}
\label{trick angolo approx}
\ee
where in the first line we have explicitly reintroduced the magnetic indexes in the inferior legs and we used  \eqref{key approx} to evaluate the two inferior Clebsh coefficients.
Proceeding in the same way for \eqref{trick vertice} we find
\be
\ba
\ifx\JPicScale\undefined\def\JPicScale{1}\fi
\psset{unit=\JPicScale mm}
\psset{linewidth=0.3,dotsep=1,hatchwidth=0.3,hatchsep=1.5,shadowsize=1,dimen=middle}
\psset{dotsize=0.7 2.5,dotscale=1 1,fillcolor=black}
\psset{arrowsize=1 2,arrowlength=1,arrowinset=0.25,tbarsize=0.7 5,bracketlength=0.15,rbracketlength=0.15}
\begin{pspicture}(0,0)(52,31)
\psline[fillstyle=solid]{-|}(26.83,12.83)(18,4)
\pspolygon[](30,14)(28,12)(26,14)(28,16)
\rput(24,15){$\scr{R_x}$}
\rput(27,19){$\scr{j_x}$}
\psline[fillstyle=solid]{-|}(35,21)(35,31)
\psline[fillstyle=solid]{-|}(43.17,12.83)(52,4)
\rput(46,15){$\scr{R_y}$}
\rput(41,18){$\scr{j_y}$}
\psline[fillstyle=solid]{|*-}(24,2)(31,9)
\rput(32,5){$\scr{R_x}$}
\rput(35,15){$\scr{m}$}
\psline[fillstyle=solid](33,11)(37,15)
\pspolygon[](30,10)(32,12)(34,10)(32,8)
\psline[fillstyle=solid]{|*-}(46,2)(39,9)
\rput(39,5){$\scr{R_y^{-1}}$}
\rput(44,22){$\scr{\tilde{m}}$}
\pspolygon[](38,8)(36,10)(38,12)(40,10)
\psline[fillstyle=solid](37,15)(45,26)
\psline[fillstyle=solid](38,12)(37,15)
\rput(33,23){$\scr{j_z}$}
\rput(31,31){$\scr{j_z}$}
\psline[border=0.3,fillstyle=solid](35,21)(41,15)
\psline[fillstyle=solid](35,21)(28.84,14.84)
\pspolygon[](42,16)(44,14)(42,12)(40,14)
\rput(46,27){$\scr{0}$}
\rput(29,3){$\scr{m}$}
\rput(39,3){$\scr{m}$}
\rput(22,0){$\scr{\mu_x}$}
\rput(47,0){$\scr{\mu_y}$}
\end{pspicture}
\ea
\approx \sum_k \sqrt{\frac{d_{j_x+\mu_x}}{d_{j_x}}}
\tinyninej {k}{\tilde{m}}{j_z} {j_x+\mu_x} {m} {j_y}  {j_y-\mu_y} {m}  {j_x}
\ba
\ifx\JPicScale\undefined\def\JPicScale{1}\fi
\psset{unit=\JPicScale mm}
\psset{linewidth=0.3,dotsep=1,hatchwidth=0.3,hatchsep=1.5,shadowsize=1,dimen=middle}
\psset{dotsize=0.7 2.5,dotscale=1 1,fillcolor=black}
\psset{arrowsize=1 2,arrowlength=1,arrowinset=0.25,tbarsize=0.7 5,bracketlength=0.15,rbracketlength=0.15}
\begin{pspicture}(0,0)(50,27)
\psline[fillstyle=solid]{-|}(25.83,8.83)(23,6)
\pspolygon[](29,10)(27,8)(25,10)(27,12)
\psline[fillstyle=solid](34,17)(27.84,10.84)
\rput(23,11){$\scr{R_x}$}
\psline[fillstyle=solid]{-|}(34,17)(34,27)
\rput(30,21){$\scr{k}$}
\psline(34,25)(42,25)
\psline[fillstyle=solid]{-|}(42.17,8.83)(45,6)
\pspolygon[](41,12)(43,10)(41,8)(39,10)
\rput(45,11){$\scr{R_y}$}
\rput(30,27){$\scr{j_z}$}
\rput(18,7){$\scr{j_x+\mu_x}$}
\rput(50,8){$\scr{j_y-\mu_y}$}
\psline[border=0.3,fillstyle=solid](34,17)(40,11)
\rput(41,27){$\scr\tilde{{m}}$}
\rput(43,24){$\scr{0}$}
\end{pspicture}
\ea
+O\left(\frac{1}{\sqrt{j_x}}\right)+O\left(\frac{1}{\sqrt{j_y}}\right)
\label{trick vertice approx}
\ee
Up to subleading corrections we can use \eqref{trick angolo approx} and \eqref{trick vertice approx} to simplify \eqref{azione H esatta}.
We note that the expression \eqref{azione H esatta} is made by the product of the following four kind of terms:

\begin{enumerate}
	\item two factors in the first line, namely the remnant of the scalar product between the basis elements,
	\item the coefficients in the big conjugate parenthesis $()^*$ left from the coherent state coefficients of $\langle \Psi_{H}\vgraph^z|$,
	\item the coefficients in the big $()$ parenthesis left from the coherent state coefficients of $| \Psi_{H}\vgraph^z\rangle$,
	\item the matrix elements of the Hamiltonian operator ${}^{R}\hat{H}^{m}_{E\cube}$.
\end{enumerate}
The terms of the kind $2)$ and $3)$ are disposed according to their original position with respect to the state, {\it i.e.} node, left corner, corner opposite to the node and right corner.
The matrix elements of ${}^{R}\hat{H}^{m}_{E\cube}$ too consist of a $9j$ and two $6j$-s associated respectively to the node and the two corners.
We illustrate the simplification looking at the coefficient involving the node.

The first coefficient in the first line of \eqref{azione H esatta} times the node coefficient in the complex conjugate parenthesis $()^*$ simplifies due to the normalization. The node contribution left is then the node coefficient in $()$, the second factor in the first line of the \eqref{azione H esatta} and the $9j$ in $H^{m\ldots}_{\ldots}$. The product of the latter two terms is the left hand side of \eqref{trick vertice approx} at the leading order. This expression is in turn made of two factors: a 3-valent reduced intertwiner in the $j$-s representations and a second in the $m$ and $1$ representation. The first is just the dual of the one appearing in $()$ and their product gives $1$ according to the normalization. Proceeding in the same way for the corners we obtain the leading order contribution:
\be
\begin{split}
&\langle\Psi_{H} \; \vgraph^z |{}^{R}\hat{H}^{m}_{E\cube} |\Psi_{H} \; \vgraph^z\rangle\approx -N(\vgraph)C(m)(8\pi\gamma l_P^2)^{3/2}\\
&\approx\sum_{\tilde{m}}\sum_{\mu=\pm m}\sum_{\mu_x,\mu_y=\pm m}\sum_{\mu'_x,\mu'_y=\pm m}\sum_{j_x,j_y,j_z,j_l} \sqrt{j_x\;j_y\;(j_z+\mu)}\; s(\mu)  C^{mm}_{mm \; \tilde{m}0}
\ba
\ifx\JPicScale\undefined\def\JPicScale{0.6}\fi
\psset{unit=\JPicScale mm}
\psset{linewidth=0.3,dotsep=1,hatchwidth=0.3,hatchsep=1.5,shadowsize=1,dimen=middle}
\psset{dotsize=0.7 2.5,dotscale=1 1,fillcolor=black}
\psset{arrowsize=1 2,arrowlength=1,arrowinset=0.25,tbarsize=0.7 5,bracketlength=0.15,rbracketlength=0.15}
\begin{pspicture}(0,0)(101,143)
\rput(18,87){$\Psi^*_{H_{l_x}}(j_x+\mu_x)\Psi_{H_{l_x}}(j_x)$}
\psline[fillstyle=solid]{-|}(42.83,101.83)(40,99)
\pspolygon[](46,103)(44,101)(42,103)(44,105)
\psline[fillstyle=solid](51,110)(44.84,103.84)
\rput(40,104){$\scr{R_x}$}
\psline[fillstyle=solid]{|*-}(16.83,75.83)(14,73)
\rput(10,74){$\scr{R^{-1}_x}$}
\rput(19,78){$\scr{\mu_x}$}
\psline[fillstyle=solid]{-|}(51,110)(51,120)
%\psline[fillstyle=solid]{|*-}(51,139)(51,143)
%\rput(46,141){$\scr{j_z}$}
\psline[fillstyle=solid](12,71)(11,70)
\pspolygon[](15,72)(13,70)(11,72)(13,74)
\psline[fillstyle=solid]{-|}(59.17,101.83)(62,99)
\pspolygon[](58,105)(60,103)(58,101)(56,103)
\rput(62,104){$\scr{R_y}$}
\psline[fillstyle=solid]{|*-}(85.17,75.83)(88,73)
\rput(95,73){$\scr{R^{-1}_y}$}
\rput(84,78){$\scr{\mu_y}$}
\psline[fillstyle=solid](90,71)(92,69)
\pspolygon[](89,74)(91,72)(89,70)(87,72)
\rput(11,64){$\scr{m}$}
\rput(22,57){$\scr{\mu'_y}$}
\rput(93,66){$\scr{m}$}
\rput(81,56){$\scr{\mu'_x}$}
\psline[border=0.3,fillstyle=solid](51,110)(57.16,103.84)
\rput(51,122){$\scr{0}$}
\psline[fillstyle=solid]{-|}(18.17,62.83)(21,60)
\pspolygon[](17,66)(19,64)(17,62)(15,64)
\rput(21,65){$\scr{R_y}$}
\psline[fillstyle=solid]{|*-}(44.17,36.83)(47,34)
\rput(44,30){$\scr{R^{-1}_y}$}
\rput(53,27){$\scr{m}$}
\rput(40,38){$\scr{\mu'_y}$}
\rput(61,38){$\scr{\mu'_x}$}
\psline[fillstyle=solid](49,32)(51.83,29.17)
\pspolygon[](48,35)(50,33)(48,31)(46,33)
\psline[fillstyle=solid](11,70)(16.16,64.84)
\psline[fillstyle=solid]{-|}(56.17,33.17)(59,36)
\pspolygon[](53,32)(55,34)(57,32)(55,30)
\psline[fillstyle=solid](52,29)(54.16,31.16)
\rput(58,28){$\scr{R_x}$}
\rput(92,60){$\scr{R_x}^{-1}$}
\psline[fillstyle=solid]{|*-}(82.17,59.17)(85,62)
\psline[fillstyle=solid](87,64)(92,69)
\pspolygon[](84,63)(86,65)(88,63)(86,61)
\rput(48,115){$\scr{\tilde{m}}$}
\rput(45,108){$\scr{m}$}
\rput(38,97){$\scr{\mu_x}$}
\rput(57,108){$\scr{m}$}
\rput(65,97){$\scr{\mu_y}$}
%\rput(101,67){$\scr{j'_y}$}
%\rput(3,68){$\scr{j'_x}$}
\rput(88,87){$\Psi^*_{H_{l_y}}(j_y-\mu_y)\Psi_{H_{l_y}}(j_y)$}
\rput(18,48){$\Psi^*_{H_{l_{l_{y}}}}(j_l+\mu'_y)\Psi_{H_{l_{l_{y}}}}(j_l)$}
\rput(88,48){$\Psi^*_{H_{l_{l_{x}}}}(j_l+\mu'_x)\Psi_{H_{l_{l_{x}}}}(j_l)$}
\rput(51,130){$\Psi^*_{H_{l_z}}(j_z)\Psi_{H_{l_z}}(j_z)$}
\end{pspicture}
\ea
\end{split}
\label{LO1}
\ee

We see how each link of the plaquette provides a contribution $\Psi^*_{H_{l}}(j_l+\mu)\Psi_{H_{l}}(j_l)$ for $\mu=\pm m$, which gives a phase terms and the product of two gaussian centered around different values, {\it i.e.} 
\be
\Psi^*_{H_{l}}(j_l+\mu) \Psi_{H_{l}}(j_l)\propto \mathcal{N}^2\, e^{-\frac{\alpha_l}2\left(j_{l}-\bar{j}_l+\mu\right)^2} e^{-\frac{\alpha_l} 2\left(j_{l}-\bar{j}_l\right)^2} e^{-i\theta \mu},
\ee
which can be rewritten as
\be
\Psi^*_{H_{l}}(j_l+\mu) \Psi_{H_{l}}(j_l)\propto \mathcal{N}^2\, e^{-\alpha_l\mu \left(j_{l}-\bar{j}_l\right)-\frac{\alpha_l}2\mu^2} e^{-\frac{\alpha_l}2\left(j_{l}-\bar{j}_l\right)^2} e^{-i\theta \mu},
\ee
The sum over the spin numbers $j_x$, $j_y$, $j_z$ and $j_l$ can be approximated with an integral over continuous variables as they go to infinity, such that the expression \eqref{LO1} can be evaluated via a saddle point expansion around $\bar{j}_l$, so finding 
\be
\begin{split}
&\langle\Psi_{H} \; \vgraph^z |{}^{R}\hat{H}^{m}_{E\cube} |\Psi_{H} \; \vgraph^z\rangle\approx -N(\vgraph)C(m)(8\pi\gamma l_P^2)^{3/2}\\
&\sum_{\tilde{m}}\sum_{\mu=\pm m}\sum_{\mu_x,\mu_y=\pm m}\sum_{\mu'_x,\mu'_y=\pm m}\sqrt{\bar{j}_x\;\bar{j}_y\;(\bar{j}_z+\mu)}\; s(\mu)  C^{mm}_{mm \; \tilde{m}0}
\ba
\ifx\JPicScale\undefined\def\JPicScale{0.6}\fi
\psset{unit=\JPicScale mm}
\psset{linewidth=0.3,dotsep=1,hatchwidth=0.3,hatchsep=1.5,shadowsize=1,dimen=middle}
\psset{dotsize=0.7 2.5,dotscale=1 1,fillcolor=black}
\psset{arrowsize=1 2,arrowlength=1,arrowinset=0.25,tbarsize=0.7 5,bracketlength=0.15,rbracketlength=0.15}
\begin{pspicture}(0,0)(101,143)
\rput(28,87){$e^{-i\theta_{l_x}\mu_x}$}
\psline[fillstyle=solid]{-|}(42.83,101.83)(40,99)
\pspolygon[](46,103)(44,101)(42,103)(44,105)
\psline[fillstyle=solid](51,110)(44.84,103.84)
\rput(40,104){$\scr{R_x}$}
\psline[fillstyle=solid]{|*-}(16.83,75.83)(14,73)
\rput(10,74){$\scr{R^{-1}_x}$}
\rput(19,78){$\scr{\mu_x}$}
\psline[fillstyle=solid]{-|}(51,110)(51,120)
%\psline[fillstyle=solid]{|*-}(51,139)(51,143)
%\rput(46,141){$\scr{j_z}$}
\psline[fillstyle=solid](12,71)(11,70)
\pspolygon[](15,72)(13,70)(11,72)(13,74)
\psline[fillstyle=solid]{-|}(59.17,101.83)(62,99)
\pspolygon[](58,105)(60,103)(58,101)(56,103)
\rput(62,104){$\scr{R_y}$}
\psline[fillstyle=solid]{|*-}(85.17,75.83)(88,73)
\rput(95,73){$\scr{R^{-1}_y}$}
\rput(84,78){$\scr{\mu_y}$}
\psline[fillstyle=solid](90,71)(92,69)
\pspolygon[](89,74)(91,72)(89,70)(87,72)
\rput(11,64){$\scr{m}$}
\rput(22,57){$\scr{\mu'_y}$}
\rput(93,66){$\scr{m}$}
\rput(81,56){$\scr{\mu'_x}$}
\psline[border=0.3,fillstyle=solid](51,110)(57.16,103.84)
\rput(51,122){$\scr{0}$}
\psline[fillstyle=solid]{-|}(18.17,62.83)(21,60)
\pspolygon[](17,66)(19,64)(17,62)(15,64)
\rput(21,65){$\scr{R_y}$}
\psline[fillstyle=solid]{|*-}(44.17,36.83)(47,34)
\rput(44,30){$\scr{R^{-1}_y}$}
\rput(53,27){$\scr{m}$}
\rput(40,38){$\scr{\mu'_y}$}
\rput(61,38){$\scr{\mu'_x}$}
\psline[fillstyle=solid](49,32)(51.83,29.17)
\pspolygon[](48,35)(50,33)(48,31)(46,33)
\psline[fillstyle=solid](11,70)(16.16,64.84)
\psline[fillstyle=solid]{-|}(56.17,33.17)(59,36)
\pspolygon[](53,32)(55,34)(57,32)(55,30)
\psline[fillstyle=solid](52,29)(54.16,31.16)
\rput(58,28){$\scr{R_x}$}
\rput(92,60){$\scr{R_x}^{-1}$}
\psline[fillstyle=solid]{|*-}(82.17,59.17)(85,62)
\psline[fillstyle=solid](87,64)(92,69)
\pspolygon[](84,63)(86,65)(88,63)(86,61)
\rput(48,115){$\scr{\tilde{m}}$}
\rput(45,108){$\scr{m}$}
\rput(38,97){$\scr{\mu_x}$}
\rput(57,108){$\scr{m}$}
\rput(65,97){$\scr{\mu_y}$}
%\rput(101,67){$\scr{j'_y}$}
%\rput(3,68){$\scr{j'_x}$}
\rput(78,87){$e^{i\theta_{l_y}\mu_y}$}
\rput(28,48){$e^{-i\theta_{l_{l_{y}}}\mu'_y}$}
\rput(78,48){$e^{-i\theta_{l_{l_{x}}}\mu'_x}$}
\end{pspicture}
\ea\label{eq fasi}
\end{split}
\ee
whose leading order corrections are $O(\alpha_l)$ and since, as discussed in \cite{Bianchi:2009ky}, $\alpha_l=1/(\bar{j}_l)^k$ with $k>1$, they are negligible in the limit $\bar{j}_l\rightarrow\infty$.

Let us now fix $m=1/2$, which implies $\tilde{m}=1$ and $C^{mm}_{mm\;\tilde{m}0}=C^{\frac{1}{2}\frac{1}{2}}_{\frac{1}{2} \frac{1}{2}\; 1 0}=\frac{1}{\sqrt{3}}$. The sums over $\mu$'s in the plaquette are now actually sums over all the components of the $SU(2)$ fundamental representations $m=1/2$ and we have 
\be
\sum_{\mu=\pm 1/2}R_{i\mu'\mu}e^{-i\theta\mu}R^{-1}_{i\mu\mu''}=(e^{-\frac{i}2\theta\sigma_i})_{\mu'\mu''}\equiv h_{\mu'\mu''}(\theta_{l_i}),\quad i=x,y,z\,,
\ee
$\sigma_i$ being Pauli matrices. Hence, we can represent the expression \eqref{eq fasi} as follows 
\be
\langle\Psi_{H} \; \vgraph^z|{}^{R}\hat{H}^{1/2}_{E\cube} |\Psi_{H} \; \vgraph^z\rangle\approx
-N(\vgraph)(8\pi\gamma l_P^2)^{1/2}\frac{2i}{3\sqrt{3}} \sum_{\mu=\pm 1/2}\sqrt{\bar{j}_x\;\bar{j}_y\;(\bar{j}_z+\mu)}\; s(\mu)  
\ba
\ifx\JPicScale\undefined\def\JPicScale{1}\fi
\psset{unit=\JPicScale mm}
\psset{linewidth=0.3,dotsep=1,hatchwidth=0.3,hatchsep=1.5,shadowsize=1,dimen=middle}
\psset{dotsize=0.7 2.5,dotscale=1 1,fillcolor=black}
\psset{arrowsize=1 2,arrowlength=1,arrowinset=0.25,tbarsize=0.7 5,bracketlength=0.15,rbracketlength=0.15}
\begin{pspicture}(0,0)(67,64)
\rput(21,43){$\scr{h(\theta_{l_x})}$}
\psline[fillstyle=solid](26,38)(18,30)
\psline(31,43)(40,52)
\psline[fillstyle=solid]{-|}(40,52)(40,62)
\rput(37,57){$\scr{1}$}
\rput(40,64){$\scr{0}$}
\rput{0}(28.5,40.5){\psellipse[](0,0)(3.5,-3.5)}
\psline(31,43)(40,52)
\psline[fillstyle=solid](27,21)(18,30)
\rput{0}(29.5,18.5){\psellipse[](0,0)(3.5,-3.5)}
\psline[fillstyle=solid](41,7)(32,16)
\psline[fillstyle=solid](55,21)(63,29)
\psline(50,16)(41,7)
\rput(41,4){$\scr{\frac{1}{2}}$}
\rput{0}(52.5,18.5){\psellipse[](0,0)(3.5,-3.5)}
\psline(50,16)(41,7)
\psline[fillstyle=solid](54,38)(63,29)
\rput{0}(51.5,40.5){\psellipse[](0,0)(3.5,-3.5)}
\psline[fillstyle=solid](40,52)(49,43)
\rput(60,43){$\scr{h^{-1}(\theta_{l_y})}$}
\rput(63,15){$\scr{h(\theta_{l_{l_x}})}$}
\rput(14,30){$\scr{\frac{1}{2}}$}
\rput(67,29){$\scr{\frac{1}{2}}$}
\rput(32,49){$\scr{\frac{1}{2}}$}
\rput(49,49){$\scr{\frac{1}{2}}$}
\rput(19,15){$\scr{h(\theta_{l_{l_y}})}$}
\end{pspicture}
\ea.
\label{quasi finale}
\ee
We can reverse the orientation of $h^{-1}(\theta_y)$ such that the 3-valent intertwiner projected on $0$ coincides with the Pauli matrix $\sigma_3$ (modulo a factor $1/\sqrt{3}$) and we can rewrite Eq.\eqref{quasi finale} as
\be
\langle\Psi_{H} \; \vgraph^z|{}^{R}\hat{H}^{1/2}_{E\cube} |\Psi_{H} \; \vgraph^z\rangle\approx -\frac{2i}{9}N(\vgraph)(8\pi\gamma l_P^2)^{1/2}
\sum_{\mu=\pm 1/2}\sqrt{\bar{j}_x\;\bar{j}_y\;(\bar{j}_z+\mu)}\; s(\mu)  Tr\{\sigma_3 h(\theta_{l_x})h(\theta_{l_{l_y}})h(\theta_{l_{l_x}})h(\theta_{-l_y})\}
\ee 
We have seen how $\theta(l_i)=\pm\bar{c}_i\epsilon_l$, $\epsilon_l$ being the length of the link $l$ and the sign depends on the orientation, while $\bar{c}_i$ denote locally constant connections around which the semiclassical state is peaked. The expression above becomes
\be
\langle\Psi_{H} \; \vgraph^z|{}^{R}\hat{H}^{1/2}_{E\cube} |\Psi_{H} \; \vgraph^z\rangle\approx
\frac{2}{9}N(\vgraph)(8\pi\gamma l_P^2)^{1/2}\sum_{\mu=\pm 1/2}\sqrt{\bar{j}_x\;\bar{j}_y\;(\bar{j}_z+\mu)}\; s(\mu) \sin{(\epsilon_{l_x}\bar{c}_x)} \sin{(\epsilon_{l_y}\bar{c}_y)},
\ee 
and by expanding $\sqrt{\bar{j}_z+\mu}$ and making the sum we get
\be
\langle\Psi_{H} \; \vgraph^z|{}^{R}\hat{H}^{1/2}_{E\cube} |\Psi_{H} \; \vgraph^z\rangle\approx
\frac{2}{9}N(\vgraph)(8\pi\gamma l_P^2)^{1/2}\sqrt{\frac{\bar{j}_x\;\bar{j}_y}{\bar{j}_z}}\;  \sin{(\epsilon_{l_x}\bar{c}_x)} \sin{(\epsilon_{l_y}\bar{c}_y)}.
\ee 
The full semiclassical state is the sum over the directions $x,y,z$ \eqref{sc}
%\be
%|\Psi^{sc}\rangle=|\Psi_{H} \; v^x_1\rangle+|\Psi_{H} \; v^y_2\rangle+|\Psi_{H} \; v^z_3\rangle,
%\ee
and remembering the relations \eqref{delta} and \eqref{hamEc}, we can write the expectation value of the scalar constraint as
\be
\label{expH}
\langle\, {}^{R}\hat{H}^{1/2}_{\cube}\,\rangle_{\vgraph} \,\approx
\frac{2}{9}\frac{1}{\gamma^2}N(\vgraph)\delta\bigg(\sqrt{\frac{\bar{p}^x\;\bar{p}^y}{\bar{p}^z}}\;  \sin{(\epsilon_{l_x}\bar{c}_x)} \sin{(\epsilon_{l_y}\bar{c}_y)}+\sqrt{\frac{\bar{p}^y\;\bar{p}^z}{\bar{p}^x}}\;  \sin{(\epsilon_{l_y}\bar{c}_y)} \sin{(\epsilon_{l_z}\bar{c}_z)}+\sqrt{\frac{\bar{p}^z\;\bar{p}^x}{\bar{p}^y}}\;  \sin{(\epsilon_{l_z}\bar{c}_{z})} \sin{(\epsilon_{l_x}\bar{c}_x)}\bigg),
\ee
where we assumed $\delta_x=\delta_y=\delta_z=\delta$.
In the continuum limit $\epsilon,\delta\rightarrow 0$, the scalar constraint describing a local Bianchi I dynamics comes out (the term within square brackets into equation (\ref{HbI})) if we also assume $\epsilon_{l_x}=\epsilon_{l_y}=\epsilon_{l_z}=\epsilon$:
\be
\langle\, {}^{R}\hat{H}^{1/2}_{\cube}\,\rangle_{\vgraph} \,\rightarrow
\frac{2}{9}\frac{1}{\gamma^2}N(\vgraph)\delta\epsilon^2\left(\sqrt{\frac{\bar{p}^x\;\bar{p}^y}{\bar{p}^z}}\; \bar{c}_x \bar{c}_y+\sqrt{\frac{\bar{p}^y\;\bar{p}^z}{\bar{p}^x}}\;  \bar{c}_y \bar{c}_z+\sqrt{\frac{\bar{p}^z\;\bar{p}^x}{\bar{p}^y}}\;  \bar{c}_{z}\bar{c}_x\right)\,,\label{somma nodi}
\ee
which means that the model has the proper semiclassical limit (\ref{localBI}), $\frac{2}{9}\delta\epsilon^2$ playing the role of the volume element $V(\vgraph)$ of the homogeneous patch around the node $\vgraph$ (this result has been foreseen in \cite{Cianfrani:2010ji}). Generically, we have arbitrary values for $\delta$'s
and $\epsilon$'s, in which case the proper semiclassical limit is achieved in the continuum limit for
\be\label{delta finale}
\delta_x=\frac{9\,V(\vgraph)}{2\epsilon_{l_x}\sqrt{\epsilon_{l_y}\epsilon_{l_z}}},\quad
\delta_y=\frac{9\,V(\vgraph)}{2\epsilon_{l_y}\sqrt{\epsilon_{l_z}\epsilon_{l_x}}},\quad
\delta_z=\frac{9\,V(\vgraph)}{2\epsilon_{l_z}\sqrt{\epsilon_{l_x}\epsilon_{l_y}}}.
\ee
If instead we fix non vanishing values for $\epsilon,\delta$, the expectation value of the scalar constraint 
is given by the expression (\ref{expH}). By using Eqs. (\ref{delta finale}) this expression becomes 
\begin{align}
\langle\, {}^{R}\hat{H}^{1/2}_{\cube}\,\rangle_{\vgraph} \,\approx
\frac{1}{\gamma^2}N(\vgraph)V(\vgraph)\bigg(&\sqrt{\frac{\bar{p}^x\;\bar{p}^y}{\bar{p}^z}}\;  \frac{\sin{(\epsilon_{l_x}\bar{c}_x)}}{\epsilon_{l_x}} \frac{\sin{(\epsilon_{l_y}\bar{c}_y)}}{\epsilon_{l_y}}+\sqrt{\frac{\bar{p}^y\;\bar{p}^z}{\bar{p}^x}}\;  \frac{\sin{(\epsilon_{l_y}\bar{c}_y)}}{\epsilon_{l_y}} \frac{\sin{(\epsilon_{l_z}\bar{c}_z)}}{\epsilon_{l_z}}+\nonumber\\
&+\sqrt{\frac{\bar{p}^z\;\bar{p}^x}{\bar{p}^y}}\;  \frac{\sin{(\epsilon_{l_z}\bar{c}_z)}}{\epsilon_{l_z}} \frac{\sin{(\epsilon_{l_x}\bar{c}_x)}}{\epsilon_{l_x}}\bigg),
\end{align}
and it coincides with the expectation value of the Bianchi I scalar constraint 
in LQC \cite{MartinBenito:2008wx,Ashtekar:2009vc} at the leading order in the semiclassical expansion as far as one identifies $\epsilon_{l_i}$ with the regulator $\bar\mu_i$ adopted in LQC.  

\section{Conclusions}\label{concl}
We discussed the semiclassical limit of the scalar constraint operator acting on a three-valent node in QRLG. In order to get a nontrivial result we had to ``dress'' the node by adding a loop and summing over all the permutations of the three fiducial directions. This procedure allowed us to construct semiclassical states in the kinematical Hilbert space of QRLG by mimicking the procedure adopted in Loop Quantum Gravity \cite{Thiemann:2002vj}. 

We evaluated explicitly the expectation value of the euclidean part of the (non graph-changing) scalar constraint on such states. With respect to the previous works on QRLG we admit also the presence of states projected on the minimum magnetic number of the $SU(2)$ representation. These states enter the construction of the scalar constraint operator.
In the limit $\bar{j}>>1$, $\bar{j}$ denoting the spin quantum numbers around which the semiclassical states are peaked, we could approximate the expectation value of the scalar constraint using the asymptotic forms of the Clebsch-Gordan coefficients involved. 

This way, we demostrated how the expectation value of the scalar constraint acting on the coherent states based at dressed nodes reproduces the local Bianchi I dynamics for high occupation numbers, {\it i.e.} $\bar{j}>>1$, and in the continuum limit, which means sending the area of the dressing loop to zero. Therefore, \emph{the classical limit of QRLG coincides with a local Bianchi I dynamics}, {\it i.e.} it reproduces General Relativity in the proper (BKL) approximation scheme. This result makes the whole QRLG a viable scenario to investigate the quantum corrections to the early Universe dynamics.     

Furthermore, by taking only the limit of high occupation numbers for spins, while retaining a nonvanishing loop, then we reproduced the leading order term of the scalar constraint in LQC. \emph{The length of the edges into the loop plays the role of the regulator in LQC}. Therefore, we can trace back the origin of the LQC regulator as entering the definition of semiclassical states in QRLG. However, from this analysis we get no indication on how to fix such a parameter or on its dependence from the spins (as in the $\bar{\mu}$ scheme).   

The next step is to investigate the semiclassical corrections to the classical dynamics. These are of two kinds: the corrections coming from the expansion in $\epsilon$ and those due to the expansion around $\bar{j}$. While the latter are expected to provides (at least qualitatively) the same corrections as in LQC, the former will provides new contributions which \emph{survive in the continuous limit}. These will be determined by considering the next-to-leading order expansion of the 3j, 6j and 9j symbols entering the expression \eqref{azione H esatta}. The order of magnitude of these corrections will tell us whether they can be discussed in the QRLG paradigm or if the full LQG theory is needed.     
Moreover, it remains to investigate the Lorentzian part of the constraint, which in the classical limit is proportional to the Euclidean one. It will be discussed elsewhere. However, we gave in this work all the necessary tools to make such an analysis and we expect it to be pursued straightforwardly.  

Furthermore, we discussed only the case of a three-valent node. In order to realize a realistic description of a quantum Universe we must consider a generic three-dimensional reduced graph, whose nodes are up to six-valent. We expect that the approximation scheme adopted here is still suitable to provide a proper semiclassical limit, the only difficulty being that more complicated $n$-j symbols appears in calculations.  

Finally, the semiclassical techniques we developed are expected to be useful also with respect to the quantization of a generic metric in the diagonal form, in which case a combination of the scalar and the vector constraints generates the dynamics.

\appendix

%%%%%%%%%%%%%%%%%%%%%%%%

\section{Reduced Recoupling }\label{appendix}
 
 The standard multiplication of $SU(2)$ holonomies and their recoupling, {\it i.e.}

\be
D^{j_1}_{m_1n_1}(g) D^{j_2}_{m_2n_2}(g)=\sum_k C^{km}_{j_1m_1j_2 m_2} D^k_{mn}(g)\;  C^{kn}_{j_1n_1 j_2 n_2} \label{couplingCG}
\ee  
using the graphical calculus, introduced in \cite{io e antonia}  and based on 3j-symbols related to Clebsch-Gordan coefficients by
\be
C^{j_3m_3}_{j_1m_1 j_2m_2}=(-1)^{j_1-j_2+m_3} \sqrt{d_{j_3}} 
\;
\big(
\begin{array}{ccc}
j_1  & j_2 & j_3 \\
m_1 & m_2 & -m_3
\end{array}
\big),
\ee
can be written as

\be
\begin{array}{c}
\ifx\JPicScale\undefined\def\JPicScale{1}\fi
\psset{unit=\JPicScale mm}
\psset{linewidth=0.3,dotsep=1,hatchwidth=0.3,hatchsep=1.5,shadowsize=1,dimen=middle}
\psset{dotsize=0.7 2.5,dotscale=1 1,fillcolor=black}
\psset{arrowsize=1 2,arrowlength=1,arrowinset=0.25,tbarsize=0.7 5,bracketlength=0.15,rbracketlength=0.15}
\begin{pspicture}(0,0)(25,20)
\psline(12,9)(12,1)
\psline{-<<}(12,5)(5,5)
\psline{<-}(25,5)(18,5)
\psline(12,9)(18,5)
\psline(18,5)(12,1)
\psline(12,20)(12,12)
\psline{-<<}(12,16)(5,16)
\psline{<-}(25,16)(18,16)
\psline(12,20)(18,16)
\psline(18,16)(12,12)
\rput(22,3){$j_1$}
\rput(22,14){$j_2$}
\end{pspicture}
\end{array}
=
\sum_k d_k
\begin{array}{c}
\ifx\JPicScale\undefined\def\JPicScale{1}\fi
\psset{unit=\JPicScale mm}
\psset{linewidth=0.3,dotsep=1,hatchwidth=0.3,hatchsep=1.5,shadowsize=1,dimen=middle}
\psset{dotsize=0.7 2.5,dotscale=1 1,fillcolor=black}
\psset{arrowsize=1 2,arrowlength=1,arrowinset=0.25,tbarsize=0.7 5,bracketlength=0.15,rbracketlength=0.15}
\begin{pspicture}(0,0)(26,10)
\psline(12,9)(12,1)
\psline{-<<}(9,5)(2,2)
\psline{<-}(26,8)(21,5)
\psline(12,9)(18,5)
\psline(18,5)(12,1)
\psline{-<<}(9,5)(2,8)
\psline{<-}(26,2)(21,5)
\rput(3,0){$j_1$}
\rput(2,10){$j_2$}
\psline(9,5)(12,5)
\psline(18,5)(21,5)
\rput(24,9){$j_2$}
\rput(24,1){$j_1$}
\rput(10,3){$k$}
\end{pspicture}
\end{array}
\ee
where the triangle denotes a generic $SU(2)$ group element and the notation with the two kind of arrows is used to distinguish indexes belonging to the vector space $\mathcal{H}^j$ or the dual vector space $\mathcal{H}^{j*}$. 
The expression \eqref{couplingCG} in the quantum reduced case \cite{Alesci:2013xd} becomes
\be
D^{|n_1|}_{n_1n_1}(g) D^{|n_2|}_{n_2n_2}(g)= C^{|n_1+n_2|\,n_1+n_2}_{j_1n_1j_2 n_2} D^{|n_1+n_2|}_{n_1+n_2\,n_1+n_2}(g)\;  C^{|n_1+n_2|\,n_1+n_2}_{j_1n_1j_2 n_2} \label{reduced couplingCG}
\ee
If in the graphical notation we use 3-valent nodes to represent Clebsh-Gordan coefficients instead of 3-j symbols, the graphical transposition of the previous formula, using the label $n$ to denote the magnetic number of a link in representation $|n|$, is just:
\be
 \begin {split}
\begin{array}{c}
\ifx\JPicScale\undefined\def\JPicScale{1}\fi
\psset{unit=\JPicScale mm}
\psset{linewidth=0.3,dotsep=1,hatchwidth=0.3,hatchsep=1.5,shadowsize=1,dimen=middle}
\psset{dotsize=0.7 2.5,dotscale=1 1,fillcolor=black}
\psset{arrowsize=1 2,arrowlength=1,arrowinset=0.25,tbarsize=0.7 5,bracketlength=0.15,rbracketlength=0.15}
\begin{pspicture}(0,0)(42,19)
\psline(19,8)(19,0)
\psline{-<<}(19,4)(12,4)
\psline{<-}(32,4)(25,4)
\psline(19,8)(25,4)
\psline(25,4)(19,0)
\psline(19,19)(19,11)
\psline{-<<}(19,15)(12,15)
\psline{<-}(32,15)(25,15)
\psline(19,19)(25,15)
\psline(25,15)(19,11)
\rput(29,2){$j_1$}
\rput(28,13){$j_2$}
\psline{<-|}(42,4)(35,4)
\psline{<-|}(42,15)(35,15)
\psline{|-<<}(9,4)(2,4)
\psline{|-<<}(9,15)(2,15)
\psline(12,16)(12,14)
\psline(32,16)(32,14)
\psline(32,5)(32,3)
\psline(12,5)(12,3)
\rput(6,2){$j_1$}
\rput(6,13){$j_2$}
\rput(38,13){$j_2$}
\rput(38,2){$j_1$}
\end{pspicture}
\end{array}
=
\begin{array}{c}
\ifx\JPicScale\undefined\def\JPicScale{1}\fi
\psset{unit=\JPicScale mm}
\psset{linewidth=0.3,dotsep=1,hatchwidth=0.3,hatchsep=1.5,shadowsize=1,dimen=middle}
\psset{dotsize=0.7 2.5,dotscale=1 1,fillcolor=black}
\psset{arrowsize=1 2,arrowlength=1,arrowinset=0.25,tbarsize=0.7 5,bracketlength=0.15,rbracketlength=0.15}
\begin{pspicture}(0,0)(58,9)
\psline(28,8)(28,0)
\psline(28,8)(34,4)
\psline(34,4)(28,0)
\rput(22,2){$\scr{n_1+n_2}$}
\psline{-<<}(7,4)(0,1)
\psline{-<<}(7,4)(0,7)
\rput(1,-1){$n_1$}
\rput(0,9){$n_2$}
\psline{-|}(7,4)(14,4)
\rput(10,2){$\scr{n_1+n_2}$}
\psline{<-}(58,7)(53,4)
\psline{<-}(58,1)(53,4)
\psline{|-}(46,4)(53,4)
\rput(56,8){$n_2$}
\rput(56,0){$n_1$}
\rput(48,2){$\scr{n_1+n_2}$}
\psline{-<<}(28,4)(18,4)
\psline(18,5)(18,3)
\psline{<-}(42,4)(34,4)
\psline(42,5)(42,3)
\end{pspicture}
\end{array}
\end{split}
 \ee
where
the projection on the reduced Hilbert space forces the magnetic number $n_1+n_2$ of the recoupled group element to be equal to the spin admitting only the channel $K=|n_1+n_2|$.

{\acknowledgments
The authors wish to thank T. Thiemann, K. Giesel, T. Pawlowski for useful discussions.
This work of FC was supported by funds provided by the National Science Center under the agreement DEC12
2011/02/A/ST2/00294.
The work of E.A. was supported by the grant of Polish Narodowe Centrum Nauki nr 2011/02/A/ST2/00300.}

\end{document}